\documentclass[preprints,review,accept,moreauthors]{Definitions/mdpi}
\usepackage{siunitx}

\usepackage{amsmath,slashed,amssymb,amsfonts,epsfig,amsthm,bm,graphicx}
\usepackage{tensor}
\usepackage{caption}
\usepackage{subcaption}
\usepackage{slashed}
\usepackage{siunitx}
\usepackage{color}
\definecolor{darkgreen}{rgb}{0,0.35,0}

\def\be{\begin{equation}}
\def\ee{\end{equation}}

\def\bea{\begin{eqnarray}}
\def\eea{\end{eqnarray}}

\usepackage{tikz}
\usetikzlibrary{matrix}
\def\1{{_{1}}}\def\2{{_{2}}}

\def\noHe0{:\;\!\!\;\!\!:H_e(0):\;\!\!\;\!\!:}
\def\noHm0{:\;\!\!\;\!\!:H_\mu(0):\;\!\!\;\!\!:}

\def\1{{_{1}}}\def\2{{_{2}}}
\usepackage{tikz}
\usetikzlibrary{matrix}
\usepackage{enumitem}
\usepackage{comment}
\def\ii{\mathrm{i}}  % imaginary number

\newcommand{\christoffel}[3]{\left\{
\begin{array}{c}
#1\\ #2 \, #3
\end{array}
\right\}}
\firstpage{1}
\pubvolume{1}
\issuenum{1}
\articlenumber{0}
\pubyear{2022}
\copyrightyear{2022}
\datereceived{}
\dateaccepted{}
\datepublished{}
\hreflink{https://doi.org/} % If needed use \linebreak
\Title{Torsion at different scales: from materials to the Universe}

\Author{Nick E. Mavromatos$^{1,2}$\orcidB{}, Pablo Pais $^{3,4}$\orcidC{},Alfredo Iorio $^{4}$\orcidA{}}

\AuthorNames{Nick E. Mavromatos, Pablo Pais and Alfredo Iorio}

\address{
$^{1}$ \quad Physics Department, School of Applied Mathematical and Physical Sciences, National Technical University of Athens, Athens 157 80, Greece \\
$^{2}$ \quad Theoretical Particle Physics and Cosmology Group, Department of Physics, King’s College London, Strand, London WC2R 2LS, United Kingdom \\
$^{3}$ \quad Instituto de Ciencias F\'isicas y Matem\'aticas, Universidad Austral de Chile, Casilla 567, 5090000, Valdivia, Chile \\
$^{4}$ \quad Faculty of Mathematics and Physics, Charles University, V Hole\v{s}ovi\v{c}k\'ach 2, 18000 Prague 8, Czech Republic}
\corres{Correspondence: nikolaos.mavromatos@kcl.ac.uk}

\abstract{The concept of torsion  in geometry, although known for a long time, has not gained considerable attention by the physics community until relatively recently, due to its diverse and potentially important applications to a plethora of contexts of physical interest. These range from novel materials, such as graphene and graphene-like materials, to advanced theoretical ideas, such as string theory and supersymmetry/supergravity and applications thereof in understanding the dark sector of our Universe. This work reviews such applications of torsion at different physical scales.}

\begin{document}

%%%%%%%%%%%%%%%%%%%%%%%%%%%%%%%%%%%%%%%%%%%%%%%%%%%%%%%%%%%%%%%%%%%%%%%%%%
\section{Introduction}\label{SecIntroduction}
%%%%%%%%%%%%%%%%%%%%%%%%%%%%%%%%%%%%%%%%%%%%%%%%%%%%%%%%%%%%%%%%%%%%%%%%%%

Torsion is as important a concept of differential geometry as curvature~\cite{cartan,hehl,shapiro}. The latter plays a key role in General Relativity (GR), but the former plays no role at all there. Nonetheless, torsion enters in various contexts and formulations, directing to diverse physical predictions and realizations that span a huge range of length scales: from cosmology to condensed matter and particle physics.
Therefore, the related literature is huge, and it is not possible to cover it all in the restricted space of this review.

Here, we focus on specific aspects of torsion, either in the emergent geometric description of the physics of various materials of great interest to condensed matter physic---mainly graphene---or in the spacetime geometry itself, in particular in the early Universe. These two situations correspond to scales that are separated by a huge amount, yet the mathematical properties of torsion appear to be universal. Torsion has important physical effects, in principle experimentally testable, in both scenarios.

Graphene and related materials provide a tabletop realization of some high-energy scenarios where torsion is associated with (the continuum limit of) the appropriate dislocations in the material. A way to represent the effect of dislocations, in the long wave-length regime, through torsion tensor is to consider a continuum field-theoretic fermionic system in a (2 + 1)-dimensional space with a torsion-full spin-connection.

In the case of fundamental physics, torsion is associated with supergravity (SUGRA) theories or with the geometry of the early Universe (cosmology). We discuss physical aspects of torsion that may affect particle physics phenomenology.
In such cases, the (totally antisymmetric component of) torsion corresponds to a dynamical pseudoscalar (axion-like) degree of freedom, which is responsible for giving the vacuum a form encountered in the so-called running vacuum model (RVM) cosmology, characterised by a dynamical inflation without external inflaton fields, but rather due to non-linearities of the underlying gravitational dynamics. Moreover, under some circumstances, the torsion-associated axions can lead to background configurations that spontaneously violate  Lorentz (and CPT) symmetry, leading, in some models with right-handed neutrinos, to lepton asymmetry in the early radiation epoch, that succeeds  the exit from inflation.

The structure of the review is as follows. First, in Section \ref{SecTorsion}, we extensively discuss the concept of the torsion tensor in general geometric terms. This has the double scope of introducing our notations but also, and more importantly, of elucidating as many details as possible of the geometry and physics of torsion. The following Section \ref{sec:TorsQED} is dedicated to an important illustration of how torsion may affect well known theories, such as quantum electrodynamics (QED), while Section \ref{sec:immir} focuses on some ambiguities of the Einstein-Cartan gravity theories and on the Barbero-Immirzi (BI) parameter. In Section \ref{SecGraphene} we discuss how torsion can be practically realized in a tabletop system, that is graphene. Then, after recalling in Section \ref{sec:sugra}, how standard SUGRAs necessarily include torsion, we discuss in Section \ref{SecUSUSY} a novel type of local supersymmetry (SUSY), without superpartners, whose natural realization is in graphene. The rather extended Section \ref{SecCosmology} is dedicated to the important and hot topic of torsion in cosmology. Our concluding remarks and some brief description of other applications of torsion, which are not covered in this review, are given in the last Section \ref{SecConclusions}.

%%%%%%%%%%%%%%%%%%%%%%%%%%%%%%%%%%%%%%%%%%%%%%%%%%%%%%%%%%%%%%%%%%%%%%%%%%

\section{Properties of Torsion}\label{SecTorsion}

\begin{figure}[ht]
\begin{center}
\includegraphics[width=5.5cm]{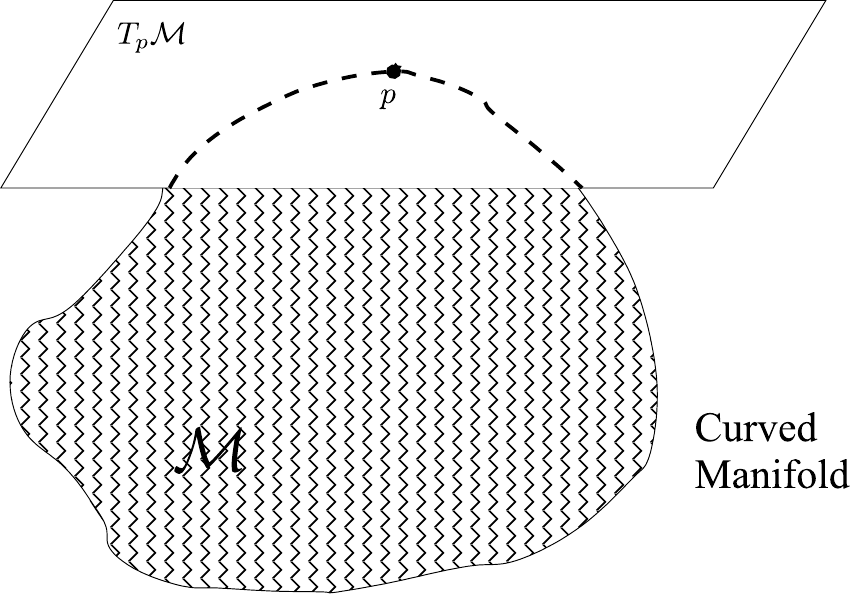}
\end{center}
\caption{Tangent hyperplane $T_{p}\mathcal{M}$ at a point $p$ of a curved $(d+1)$-dimensional manifold $\mathcal{M}$, used in the first order formalism of GR to define the vielbein $\tensor{e}{^{a}}{_{\mu}}$ map $\mathcal{M} \ \rightarrow \ T_{p}\mathcal{M}$.}
\label{fig:tangent}
\end{figure}

As already mentioned, torsion is an old subject~\cite{cartan,hehl,shapiro} that goes beyond GR, as it constitutes a more general formalism in the sense that, to obtain Einstein's GR, one needs to impose a constraint to guarantee the absence (vanishing) of torsion tensor in a Riemannian spacetime. Specifically, let $\mathcal M$ be a (3+1)-dimensional Minkowski-signature curved world-manifold~\footnote{Although the contorted geometry formalism can be generic and valid in $(d+1)-$dimensional spacetime, nonetheless for the sake of concreteness, in this work  we shall present the analysis for $d=3$, and, in the case of graphene,  for $d=2$.}, parametrized by coordinates $x^\mu$,
where Greek indices $\mu, \nu =0, \dots 3$ are spacetime volume indices, raised and lowered by the curved metric $g_{\mu\nu} = \eta_{ab}\, \tensor{e}{^{a}}{_{\mu}}\, \tensor{e}{^{b}}{_{\nu}}$, with $e_{\,\mu}^{a}$ the vielbein (we also define the inverse vielbein as $\tensor{E}{^{\mu}}{_{a}}\,\tensor{e}{^{a}}{_{\nu}}=\delta^{\mu}_{\nu}$, and $\tensor{E}{^{\mu}}{_{a}}\,\tensor{e}{^{b}}{_{\mu}}=\delta^{b}_{a}$, such that $g^{\mu\nu} = \eta^{ab} \, \tensor{E}{^{\mu}}{_{a}} \, \tensor{E}{^{\nu}}{_{b}}$ gives the inverse metric tensor). In the above formulae,
Latin indices $a,b, \dots = 0, \dots 3$ are (Lorentz) indices on the tangent hyperplane of $\mathcal M$ at a given point $p$ ({\it cf.} Fig.~\ref{fig:tangent}), and are raised and lowered by the Minkowski metric $\eta_{ab}$ (and its inverse $\eta^{ab}$), which is the metric of the tangent space $T_{p}\mathcal{M}$.

In differential form language~\cite{Eguchi,Nakahara}, which we use here often for notational convenience, the torsion two form is defined as~\cite{cartan,hehl,olivetorsion,shapiro}:
\begin{align}\label{tordef}
T^{a} = \frac{1}{2} \, \tensor{T}{^a_{\mu\nu}} \, dx^\mu \wedge dx^\nu \equiv  d e^a + {\omega}^a_{\,\,b} \wedge \, e^b\,,
\end{align}
where in the first equality we used the definition of a differential two form~\cite{Eguchi}, and the $\wedge$ denotes the exterior product,\footnote{The action of $\wedge$ on forms is expressed as~\cite{Eguchi,Nakahara}: $f^{(k)} \wedge g^{(\ell)} = (-1)^{k\, \ell}\, g^{(\ell)} \wedge f^{(k)}$, where $f^{(k)}$ and $g^{(\ell)}$ are $k$-forms and $\ell$-forms, respectively.}  and $\tensor{\omega}{^a}{_{b\,\mu}} $ is the generalized (contorted) spin connection one form, which can
can be split into a part that is torsion-free, $\tensor{\mathring{\omega}}{^a}{_{b\,\mu}} $, and related to the standard Christoffel symbols of GR, and another part that involves the \emph{contorsion one-form}\footnote{Some references refer to it as \emph{contortion tensor} \cite{hehl}. However, as we are following closer the terminology of \cite{Nakahara}, we keep the name \emph{contorsion}. As far as we know, there is no consensus yet about the name of this quantity.} $\mathcal{K}^a_{\,\,b\,\mu}$~\cite{hehl,shapiro}:
\begin{align}\label{spinconn}
\tensor{\omega}{^a}{_{b\,\mu}} = \tensor{\mathring{\omega}}{^a}{_{b\,\mu}} + \tensor{\mathcal{K}}{^a}{_{b\,\mu}} \;.
\end{align}
We can use the one-form $\tensor{\omega}{^a}{_{b}}$ to define the covariant derivative  $D(\omega)$ acting on $q$-forms $Q^{a\dots}_{\,\,\,b \dots}$ in this contorted spacetime~\cite{olivetorsion}:
\begin{align}\label{covderdef}
D(\omega) \, Q^{a\dots}_{\,\,\,b \dots} = d  \,  Q^{a\dots}_{\,\,\,b \dots} +
\tensor{\omega}{^a}{_{c}} \wedge Q^{c\dots}_{\,\,\,b \dots}  + \dots -  (-1)^q \, Q^{a\dots}_{\,\,\,d \dots}  \wedge \tensor{\omega}{^d}{_{b}} - \dots
\end{align}
It can be readily seen, using the covariant constancy of the Minkowski tangent space metric $\eta^{ab}$
\begin{align}\label{MetricCompatib}
D (\omega) \, \eta_{ab} = 0 \,,
\end{align}
that the spin connection \eqref{spinconn} is antisymmetric in its Lorentz indices
\begin{align}\label{connantisym}
\omega_{ab} = - \, \omega_{ba}\;.
\end{align}
We also have covariant constancy for the totally antisymmetric Levi-Civita tensor $\epsilon_{abcd}$:
 \begin{align}\label{leviciv}
D (\omega) \, \epsilon_{abcd}= 0\;.
\end{align}

In this Section, we discuss the generalization of Einstein-Hilbert action for spacetime geometries with torsion. To this end, we first note that the \textit{generalized} Riemann curvature, or Lorentz curvature, two-form is defined as:
\begin{align}\label{curv2form}
\tensor{R}{^a}{_{b}} = d\, \tensor{\omega}{^a}{_{b}} +
\tensor{\omega}{^a}{_{c}}\, \wedge \, \tensor{\omega}{^c}{_{b}}\,.
\end{align}
We can write the components of the Lorentz curvature in terms of the Riemann curvature two-form $\tensor{\mathring{R}}{^a}{_{b}}$, defined only by the torsion-less spin-connection, i.e., $\tensor{\mathring{R}}{^a}{_{b}} = d\, \tensor{\mathring{\omega}}{^a}{_{b}} + \tensor{\mathring{\omega}}{^a}{_{c}}\, \wedge \, \tensor{\mathring{\omega}}{^c}{_{b}}$, and the contorsion $\tensor{\mathcal{K}}{^{a}_{b}}$,
\begin{equation}
\tensor{R}{^a}{_{b}} = \tensor{\mathring{R}}{^a}{_{b}} + D (\mathring{\omega})\,\tensor{\mathcal{K}}{^{a}_{b}} + \tensor{\mathcal{K}}{^{a}_{c}}\,\tensor{\mathcal{K}}{^{c}_{b}} \;,
\end{equation}
where the quantity $D (\mathring{\omega})$ denotes the diffeomorphic covariant derivative of GR. From the definition of the covariant derivative \eqref{covderdef}, we therefore have that the torsion two form is just the covariant derivative of the vielbein
\begin{equation}\label{cderviel}
T^{a} = d\, e^{a} +
\tensor{\omega}{^a}{_{b}}\, \wedge \, e^{b} \;,
\end{equation}
and~\cite{olivetorsion}
\begin{align}\label{properties}
& D(\omega) \ T^a = \tensor{R}{^a}{_{b}} \wedge e^{b} \,, \nonumber \\
&D(\omega) \, \tensor{R}{^a}{_{b}} = 0 \;,
\end{align}
where the equations \eqref{properties} are the generalization of the usual Bianchi identity. The two equations, (\ref{curv2form}) and (\ref{cderviel}), are known as \emph{Cartan structure equations} \cite{Nakahara}.

Taking into account that the full diffeomorphic covariant derivative on the vielbein is zero, $D (\omega, \Gamma) \, e^a = 0$, we obtain a relation between the affine connection, $\tensor{\Gamma}{^{\lambda}_{\mu \nu}}$, and the spin connection \eqref{spinconn}, $\tensor{\omega}{_{\mu}^{a}_{b}}$. In components~\cite{olivetorsion}:
\begin{align}\label{propcovder}
D_\mu(\Gamma) \, e^a_{\, \nu} = \partial_\mu \, e_{\, \nu}^a - \tensor{\Gamma}{^{\lambda}_{\nu \mu}}\, e^a_{\,\lambda}
= -\tensor{\omega}{_{\mu}^{a}_{b}}\, e^b_{\,\nu} \;.
\end{align}
From \eqref{propcovder}, \eqref{tordef} and \eqref{curv2form}, we easily obtain
\begin{align}\label{antisymmaffine}
T^a_{\,\,\,\mu\nu} = e^a_{\,\lambda} \,\Big(\Gamma^\lambda_{\,\,\,\mu\nu} - \Gamma^\lambda_{\,\,\,\nu\mu}\Big) \equiv 2 \, e^a_{\,\lambda} \,\Gamma^\lambda_{\,\,\,[\mu\nu]} \;,
\end{align}
where $[\mu \nu]$ indicates antisymmetrization of the indices.

The relation \eqref{antisymmaffine} expresses the essence of torsion, namely that in its presence the affine connection loses its symmetry in its lower indices. The torsion \textit{tensor} is associated with the antisymmetric part (in the lower indices) of the affine connection, which is its only part that transforms as a tensor under general coordinate transformations.

The spin connection then, in general, is torsion-full. If we want a torsion-free connection (that is the case of GR) we need to impose
\begin{equation}\label{torsionzero}
d \,  e^a + \omega^a_{\,b} \wedge e^b = 0 \,,
\end{equation}
and we have that the antisymmetric (because of \eqref{MetricCompatib}) connection is $\omega_{a b} = \mathring{\omega}_{a b}$. In other words, covariant constancy of the metric is a separate request from zero torsion. In fact, in Riemann-Cartan spaces the metric is compatible, hence $\omega_{a b}$ is antisymmetric, but torsion is nonzero.

We next remark that the contorsion one-form coefficients $\tensor{\mathcal{K}}{^{a}_{b}_{c}}=\tensor{\mathcal{K}}{^{a}_{b}_{\mu}}\,E^{\mu}_{c}$ satisfy $\mathcal{K}^c_{\,\,ab}=-\mathcal{K}^c_{\,\,ba}$ and are related to the torsion tensor coefficients $\tensor{T}{^{a}_{b}_{c}}=\tensor{T}{^{a}_{\mu}_{\nu}}\,E^{\mu}_{b}\,E^{\nu}_{c}$ via~\cite{Nakahara}
\begin{equation}\label{torcontor}
\quad \mathcal{K}_{abc} =- \frac{1}{2}( T_{cab} - T_{abc} - T_{bca})
\Rightarrow \, T_{[abc]}=-2\mathcal{K}_{[abc]} \;,
\end{equation}
where $[abc]$ denotes total antisymmetrization. From (\ref{torcontor}) we can write the torsion tensor in term of contorsion as
\begin{align}\label{contortor}
T^a_{\,\,\,bc}=-2\mathcal{K}^a_{[bc]}\;.
\end{align}
Equations (\ref{torcontor}) and (\ref{contortor}) tell us that both torsion and contorsion tensors carry the same information.

\subsection{Geometric Interpretation}\label{sec:phystors}

Let us now discuss the geometric interpretation of torsion, by parallel transporting the vector $v^{\mu}$ along the direction $dx^{\nu}$, using the connection ${\Gamma^{\lambda}}_{\mu \nu}$ that appears in (\ref{antisymmaffine})
\begin{equation*}
\delta_{\|}v^{\mu} = v^{\mu}_{\|}(x + dx) - v^{\mu}(x) = \Gamma{^{\mu}}{_{\rho\nu}}\,v^{\rho}\,dx^{\nu} \;.
\end{equation*}
Then, the covariant derivative $D(\Gamma)\,v$ can be written, in its components, as the difference
\begin{align}
v^{\mu}(x + dx) - v^{\mu}_{\|} (x + dx) & =  v^{\mu}(x+dx) - v^{\mu}(x) - \delta_{\|}\,v^{\mu}
 = \left(\partial_{\nu} v^{\mu} - \Gamma{^{\mu}}{_{\rho\nu}} v^{\rho}\right)\,dx^{\nu}  \nonumber \\ & \equiv D_{\nu}(\Gamma) \, v^{\mu}\,dx^{\nu}  \;.
\end{align}
Both curvature and torsion measure the noncommutativity of covariant derivatives of a vector taken along two directions  \cite{Santalo}
\begin{equation*}
 D_{\nu}(\Gamma) \, D_{\rho}(\Gamma) \, v^{\mu} -  D_{\rho}(\Gamma) \,  D_{\nu}(\Gamma) \, v^{\mu} = R{^{\mu}}{_{\sigma\rho\nu}}\,v^{\sigma} + \tensor{T}{^{\sigma}_{\rho\nu}} D_{\sigma}(\Gamma) \, v^{\mu} \;,
\end{equation*}
where the torsion components have been defined already and the curvature components can be written as
\begin{equation}\label{curvature_def}
\tensor{R}{^{\mu}}{_{\sigma\rho\nu}} =  \partial_{\rho}\Gamma{^{\mu}}{_{\sigma\nu}} - \partial_{\nu}\Gamma{^{\mu}}{_{\sigma\rho}}  + \Gamma{^{\mu}}{_{\lambda\nu}}\,\Gamma{^{\lambda}}{_{\sigma\rho}} - \Gamma{^{\mu}}{_{\lambda\rho}}\,\Gamma{^{\lambda}}{_{\sigma\nu}} \;.
\end{equation}
It is remarkable that for a scalar field, $\varphi$, such noncommutativity is entirely due to torsion
\begin{equation*}
 D_{\nu}(\Gamma) \,  D_{\rho}(\Gamma) \, \varphi -  D_{\rho}(\Gamma) \,  D_{\nu}(\Gamma) \, \varphi = \tensor{T}{^{\sigma}}{_{\rho\nu}} \partial_{\sigma} \varphi \;.
\end{equation*}
\begin{figure}
\begin{center}
\includegraphics[width=0.40\textwidth,angle=0]{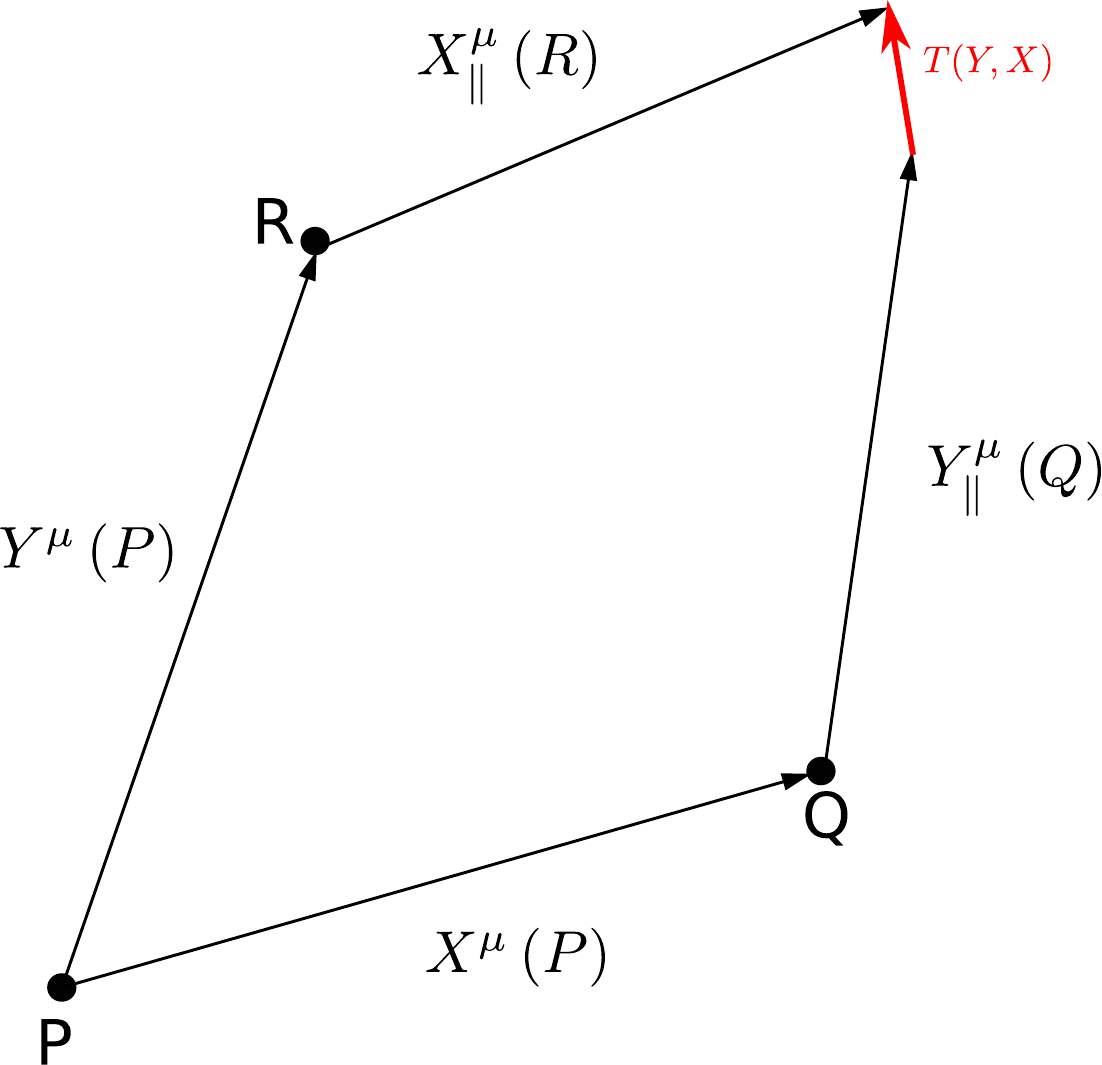}
\caption{A geometric interpretation of torsion in Riemann-Cartan spaces. Consider two vector fields, $X$ and $Y$, at a point $P$. First, parallel-transport $X$ along $Y$ to the infinitesimally close point $R$. Then, again from $P$, parallel-transport $Y$ along $X$ to reach a point $Q$. The failure of the closure of the parallelogram is the geometrical signal of torsion, and its value is the difference $T(X,Y)$ between the two resulting vectors, here in red. An $n$-dimensional manifold $M$ with a linear connection preserving local distances, i.e. fulfilling condition (\ref{metric_compatibility_condition}), is called a \emph{Riemann-Cartan space}, denoted by $U_{n}$. In Riemannian spaces, $V_{n}$, this tensor is assumed to be zero. The picture was inspired by \cite{HehlObukhov2007} but with the notation of \cite{Nakahara}, and was taken from \cite{DICEtorsion}.}\label{FigTorsion}
\end{center}
\end{figure}
We introduce the \emph{metric tensor} $g_{\mu\nu}$ when it is necessary to measure angles and distances between events in a spacetime manifold. The line element is
\begin{equation}\label{metric_def}
ds^{2} = g_{\mu\nu}\,dx^{\mu}\,dx^{\nu} \;.
\end{equation}
We can define the longitude of any curve on $\mathcal{M}$ by integrating (\ref{metric_def}).

A very reasonable assumption usually taken is that local distances do not change under parallel transport, i.e.,
\begin{equation}\label{metric_compatibility_condition}
 D_{\rho}(\Gamma) \, g_{\mu\nu} = 0 \;.
\end{equation}
The condition (\ref{metric_compatibility_condition}) for a linear connection $\Gamma$ is called \emph{metric compatibility}, which leads to the antisymmetry of the spin-connection (\ref{connantisym}) \cite{Nakahara}. Fig. \ref{FigTorsion} illustrates a geometric interpretation of torsion, with details in the caption.

The unique linear metric-compatible torsionless connection, called the \emph{Levi-Civita connection}, can then be obtained from the metric $g_{\mu\nu}$ (see \cite{Nakahara} for details)
\begin{equation}\label{Christoffel_symbols}
\christoffel{\mu}{\nu}{\rho} = \frac{1}{2} \, g^{\mu\sigma} \, ( \partial_{\nu} g_{\rho\sigma} + \partial_{\rho} g_{\nu\sigma} -\partial_{\sigma} g_{\nu\rho} ) \;.
\end{equation}
The quantities (\ref{Christoffel_symbols}) are called \emph{Christoffel symbols}, and the curvature associated with the Levi-Civita connection is the \emph{Riemannian curvature tensor}, denoted by $\tensor{\mathring{R}}{^{\mu}_{\nu\rho\sigma}}$. In this way, the linear connection in a Riemann-Cartan space can be written as
\begin{equation*}
\Gamma{^{\mu}}{_{\nu\rho}} = \christoffel{\mu}{\nu}{\rho} + \tensor{K}{^{\mu}_{\nu\rho}} \;.
\end{equation*}

%Notice that,  contrary to the torsion tensor, $\tensor{K}{^{\mu}_{\nu\rho}}$ is not necessarily antisymmetric in the last two indices, unless %torsion is totally antisymmetric.

\subsection{Gravitational Dynamics in presence of Torsion}\label{sec:gravdyn}

The Einstein-Hilbert scalar curvature term corresponding to the generalized contorted Riemann tensor is given by
\begin{align}\label{gravact}
 \mathcal S_{\rm grav} &= \frac{1}{2\kappa^2}\, \int d^4x \, \sqrt{-g} \, R = \frac{1}{2\kappa^2}\,\int \,  R_{ab}\, \wedge \, \star(e^a \, \wedge \, e^b)  \nonumber \\
& = \frac{1}{2\kappa^2}\,\int \, ( \mathring{R}_{ab} + D (\mathring{\omega}) \, \mathcal{K}_{ab} +  \mathcal{K}_{ac} \, \wedge \,  \mathcal{K}^c_{\,\,\,b} )\, \wedge \, \star(e^a \, \wedge \, e^b)  \nonumber \\
& = \frac{1}{2\kappa^2}\,\int \, (\mathring{R}_{ab} + \mathcal{K}_{ac} \, \wedge \, \mathcal{K}^c_{\,\,\,b} )\, \wedge \, \star(e^a \, \wedge \, e^b) \;,
\end{align}
where in the last two equalities we used form language and took into account the definition of the generalized curvature two form \eqref{curv2form} in terms of the contorted spin connection \eqref{spinconn}. In \eqref{gravact}, %\footnote{Our conventions and definitions used throughout this work are: signature of metric $(-,+,+,+ )$, Riemann Curvature tensor
%$R^\lambda_{\,\,\,\,\mu \nu \sigma} = \partial_\nu \, \Gamma^\lambda_{\,\,\mu\sigma} + \Gamma^\rho_{\,\, \mu\sigma} \, \Gamma^\lambda_{\,\, \rho\nu} - (\nu \leftrightarrow \sigma)$, Ricci tensor $R_{\mu\nu} = R^\lambda_{\,\,\,\,\mu \lambda \nu}$, and Ricci scalar $R = R_{\mu\nu}g^{\mu\nu}$.},
$\kappa^2=8\pi {\rm G} = M_{\rm Pl}^{-2}$ is the gravitational constant in four dimensions, which is the inverse of the square of the reduced Planck mass $M_{\rm Pl}$ in units $\hbar=c=1$ we work throughout. In passing from the second to the third equality we used the fact that the term $D (\mathring{\omega}) \, \mathcal{K}_{ab}\wedge\star\,(e^{a}\,\wedge\,e^{b})$ is a total derivative and thus yields, by means of Stoke theorem, a boundary term that we assume to be zero (we used also the metric compatibility of the spin-connection (\ref{connantisym})).

For completeness, we give below the component form of the gravitational action (in the notation of \cite{olivetorsion}):
\begin{align}\label{delta}
\mathcal S_{\rm grav} &= \frac{1}{2\kappa^2} \, \int d^4x \, \sqrt{-g}\, \Big( \mathring{R} + \Delta\Big)\,,\nonumber \\  \Delta &\equiv K^\lambda_{\quad\mu\nu}\, K^{\nu\mu}_{\quad \lambda} - K^{\mu\nu}_{\quad \nu}\, K_{\mu\lambda}^{\quad \lambda} = T^\nu_{\quad \nu\mu}\, T^{\lambda \quad \mu}_{\quad \lambda} - \frac{1}{2} \, T^\mu_{\quad \nu \lambda}\, T^{\nu \lambda}_{\quad \mu} + \frac{1}{4} \, T_{\mu\nu\lambda}\, T^{\mu\nu\lambda} \;.
\end{align}
Next, we decompose the torsion tensor in its irreducible parts under the Lorentz group~\cite{olivetorsion,Capozziello2001,shapiro},
\begin{equation}\label{decom}
 T_{\mu\nu\rho} = \frac{1}{3}\left(T_\nu g_{\mu\rho} - T_\rho g_{\mu\nu} \right) - \frac{1}{6} \epsilon_{\mu\nu\rho\sigma} S^\sigma + q_{\mu\nu\rho} \;,
 \end{equation}
 where
 \begin{equation}\label{p1}
 T_\mu \equiv T^\nu_{\,\,\mu\nu} \;,
 \end{equation}
is the \emph{torsion trace vector}, transforming like a vector,
 \begin{equation}\label{p2}
 S_\mu \equiv \epsilon_{\mu\nu\rho\sigma} T^{\nu\rho\sigma} \;,
 \end{equation}
is the \emph{pseudotrace axial vector } and the antisymmetric tensor $q_{\mu\nu\rho}$ satisfies
\begin{equation}\label{p3}
 q^\nu_{\,\,\rho\nu} = 0 = \epsilon^{\sigma\mu\nu\rho} q_{\mu\nu\rho}~. \end{equation}
Thus, we may write the contorsion tensor components as:
\begin{align}\label{K}
\mathcal{K}_{abc} = \frac{1}{2} \epsilon_{abcd}\, S^d + \widehat{\mathcal{K}}_{abc}\;,
\end{align}
being $ \widehat{\mathcal{K}}_{abc}$, by definition, the difference of $\mathcal{K}_{abc}$ and the first term of (\ref{K}).
This yields for the quantity $\Delta$ in \eqref{delta}:
\begin{align}
 \Delta = \frac{3}{2} S_d\, S^d + \widehat{\Delta}\,,
\end{align}
with $\widehat{\Delta}$ being given by the combination appearing in the expression for $\Delta$ in \eqref{delta} in terms of the contorsion tensor, but with $K$ replaced by $\widehat K$~\cite{olivetorsion}.

Using the decomposition (\ref{decom}) and the relations (\ref{p1}), (\ref{p2}), (\ref{p3}) and discarding total derivative terms,
the gravitational part of the action can be written as:
\begin{align}\label{torsiondecompaction}
S_{\rm grav} & =\frac{1}{2\kappa^2} \int d^4 x\, \sqrt{-g}\, \left(E^\mu_{\, a} \, E^\nu_{\, b} \, R^{a b}_{\,\,\,\mu\nu} (\omega)
+ \frac{1}{24} S_\mu S^\mu - \frac{2}{3} T_\mu T^\mu + \frac{1}{2} q_{\mu\nu\rho} q^{\mu\nu\rho}\right) \nonumber \\
&\equiv \frac{1}{2\kappa^2} \int d^4 x \, \sqrt{-g} \Big( R + \widehat \Delta \Big) + \frac{3}{4\kappa^2} \int \, S \, \wedge \, \star S \,,
\end{align}
where in the last line we used mixed notation components/form, following \cite{olivetorsion}, as this will be more convenient for the discussion that follows. For future use, the reader should notice that $\widehat \Delta$ is independent of the pseudovector $S^d$.

An important part of our review will deal with fermionic torsion, that is torsion induced by fermion fields in the theory. Such a feature arises either in certain materials, such as graphene, to be discussed in Section \ref{SecGraphene}, or in fundamental theories, which may play a r\^ole in particle physics, such as SUGRA  (local SUSY, Section \ref{sec:sugra}), unconventional supersymmetry (USUSY) (Section \ref{SecUSUSY}), and string theory (with applications to cosmology, Section \ref{sec:string}). In the next Section we review such a (quantum) torsion in a fermionic theory corresponding to QED, as an instructive example, which can be generalized to non-Abelian gauge fields as well.

\section{(Quantum) Torsion, Axions and Anomalies in Einstein-Cartan Quantum Electrodynamics}\label{sec:TorsQED}

%***Links with our graphene minimal action***

Our starting point is a $(3+1)$-dimensional QED with torsion (termed, from now on, ``contorted QED''), describing the dynamics of a massless Dirac fermion field $\psi (x)$, coupled to a gauged (electromagnetic) U(1) field $A_\mu$, in a curved spacetime with torsion\footnote{For a recent study of the massive case, where the focus is on neutrino mixing and oscillations, see \cite{capolupo2023quantum}.}. The action of the model reads~\cite{olivetorsion}:
\begin{align}\label{actorsqed}
    \mathcal S_{\rm TorsQED} = \frac{\ii}{2} \int d^4x \, \sqrt{-g}\, \Big[ \overline \psi (x) \, \gamma^{\mu} \, \mathcal{D}_\mu (\omega,A) \, \psi (x) - \overline{\mathcal D_\mu(\omega,A) \psi(x)} \gamma^{\mu} \, \psi(x) \Big]\;,
\end{align}
where $\mathcal{D}_\mu (\omega,A) \equiv D_\mu(\omega) - \ii \, e\, A_\mu$ is the diffeomorphic \textit{and} gauge covariant derivative and~\cite{hehl,shapiro}:
\begin{align}\label{covder}
D_\mu(\omega) \, = \, \partial_\mu + \ii \, \omega^a_{\,\,\,b\,\mu} \, \tensor{\sigma}{^{b}_{a}}\; , \quad \sigma^{ab} \equiv \frac{\ii}{4} [\gamma^{a}\,, \, \gamma^{b}]\;.
\end{align}
The quantities $\gamma^{a}$ and $\gamma^{\mu}$ denote the $4\times 4$ Dirac matrices in the tangent space and in the manifold, respectively.
On account of \eqref{covder} and \eqref{spinconn} (discussed in Section \ref{SecTorsion}), the action \eqref{actorsqed} becomes:
\begin{align}\label{qedplustors}
S_{\rm TorQED} = S_{\rm QED}(\mathring{\omega}, A) + \frac{1}{8} \, \int d^4x \sqrt{-g} \,  \overline{\psi} (x) \{\gamma^{c}\,, \,
\sigma^{ab}\}\, \mathcal{K}_{abc} \, \psi (x) \,,
\end{align}
where $S_{\rm QED}(\mathring{\omega}, A)$ is the standard QED action in a torsion-free curved spacetime and $\{\,,\,\}$ denotes the standard anticommutator. Since the Dirac $\gamma$-matrices obey
$$\{\gamma^{c}\,, \sigma^{ab}\} = 2 \epsilon^{abc}_{\quad d}\, \gamma^{d}\, \gamma^{5}\;,$$
where $\epsilon^{abcd}$ is the Levi-Civita tensor in $(3+1)$-dimensions, one can prove that it is only the totally antisymmetric part of the torsion that couples to fermionic matter \cite{shapiro}. Indeed, on using \eqref{torcontor}, we may write \eqref{qedplustors} in the form
\begin{align}\label{axial}
S_{\rm TorQED} = S_{\rm QED}(\omega, A) - \frac{3}{4} \int d^4 x\, \sqrt{-g} \, S_\mu \, \overline \psi \, \gamma^{\mu}\, \gamma^{5} \, \psi \,,\end{align}
where $S_d = \frac{1}{3\!} \epsilon^{abc}_{\quad d} \, T_{abc}$ (or in form language $S = \star T$) is the dual pseudovector constructed out of the totally antisymmetric part of the torsion. From \eqref{axial} we thus observe that only the totally antisymmetric part of the torsion couples to the fermion axial current
\begin{align}\label{j5}
j^{5\mu}= \overline \psi \,\gamma^{\mu} \, \gamma^{5} \, \psi \;.
\end{align}
The $(2+1)$-dimensional version of (\ref{axial}) will be our starting point to describe the conductivity electrons in graphene-like materials in a fixed spacetime with torsion  (see Section \ref{SecGraphene}).
In contorted QED, the Maxwell tensor is defined with respect to the ordinary torsion-free geometry, $F_{\mu\nu} = \partial_\mu A_\nu - \partial_\nu A_\mu = D_{\mu}(\mathring{\omega}) A_{\nu} - D_{\nu}(\mathring{\omega}) A_{\mu}$. This way, the Maxwell tensor continues to satisfy the Bianchi identity (in form language $d\, F =0$) even in  the presence of torsion. Thus the standard Maxwell term, independent from torsion, is added to the action \eqref{actorsqed} to describe the dynamics of the photon field:
\begin{align}\label{qedaction}
S_{\rm Max} = -\frac{1}{4} \int d^4x \, \sqrt{-g}\, F_{\mu\nu} \, F^{\mu\nu} =
-\frac{1}{2} \int \, F \, \wedge \, \star F\;,
\end{align}
where $\star$ denotes the Hodge star~\cite{Eguchi,Nakahara}.

The dynamics of the gravitational field is described by adding Einstein-Hilbert scalar curvature action \eqref{gravact} (or, equivalently, \eqref{delta}, in component form) of Section \ref{SecTorsion} to the above actions. By adding \eqref{axial} to \eqref{delta}, so as to obtain the full gravitational action in a contorted geometry,  with QED as its matter content, we obtain from the graviton equations of motion the stress-energy tensor of the theory, which can be decomposed into various components gauge, fermion and torsion-$S$ (the reader should recall that only the totally antisymmetric part of the torsion $S$ couples to matter in the theory):
\begin{align}\label{stress}
T^A_{\mu\nu} &= F_{\mu\lambda}\, F^\lambda_{\,\,\,\nu} - \frac{1}{2}\, g_{\mu\nu} \, F_{\alpha\beta}\, F^{\alpha\beta} \,, \nonumber\\
T^\psi_{\mu\nu} &= -\left(\frac{i}{2} \overline \psi
\gamma_{(\mu} \, \mathcal{D}_{\nu)} \, \psi - (\mathcal{D}_{(\mu} \overline \psi )\gamma_{\nu)}\, \psi \right) + \frac{3}{4} \, S_{(\mu}\,\overline \psi \, \gamma_{\nu)}\, \gamma^{5} \, \psi\;, \nonumber \\
T^S_{\mu\nu} &= -\frac{3}{2\, \kappa^2} \Big(S_\mu\, S_\nu -\frac{1}{2}\, g_{\mu\nu}\, S_\alpha \, S^\alpha \Big)\;,
\end{align}
where $(\dots)$ denotes indices symmetrization.

Variation of the above gravitational action with respect to the torsion components $T^\mu$, $q^{\mu\nu\rho}$ and $S^\mu$ ({\it Cf.} (\ref{decom})), treated as {\it independent} field variables, leads to the equations of motion (in form language):
\begin{align}\label{sj5}
T^\mu = 0, \qquad q_{\mu\nu\rho} = 0\,,
\qquad S = \frac{\kappa^2}{2} j^5\,,
\end{align}
respectively, where $j^{5}$ is the axial fermion current one form, which in components is given by Eq.~\eqref{j5}. Thus, classically, only the totally antisymmetric component of the torsion is non vanishing in this Einstein-Cartan theory with fermions.
From \eqref{spinconn},\eqref{torcontor} and \eqref{decom},  we then obtain for the \textit{on-shell} torsion-full spin connection:
\begin{equation}\label{torsionaxcurr}
\omega^{ab}_\mu = \mathring{\omega}^{ab}_{\mu}  +  \frac{\kappa^2}{4} \epsilon^{ab}_{\,\,\,\,\,\,cd}\, e^c_{\,\mu} \, j^{5\,d}~,
\end{equation}
thereby associating the torsion part of the connection, induced by the fermions, with the spinor axial current.

We next remark that the  equations of motion for the fermion, stemming from \eqref{axial},
imply the gauged Dirac equation with the vector pseudovector $S_\mu$, corresponding to the totally antisymmetric torsion component, playing the r\^ole of an axial source:
\begin{align}\label{diraceq}
\ii \gamma^{\mu} \, \mathcal{D}_\mu (\omega,A) \, \psi = \frac{3}{4}   \, S_\mu \, \gamma^{\mu} \, \gamma^{5} \, \psi\;.
\end{align}
Classically, \eqref{sj5} implies a direct substitution of the torsion by the axial fermion current in \eqref{stress}, \eqref{diraceq}. Moreover, as a result of the Dirac equation \eqref{diraceq}, a classical conservation of the axial current follows,
$d \, \star \, j^5 =0$. In view of \eqref{sj5}, this, in turn, implies a classical conservation of the torsion pseudovector $S$ , that is:
\begin{equation}\label{torsioncons}
  d \, \star  S  = 0\;.
\end{equation}
Because the action is quadratic in  $S_\mu$ one could integrate it out exactly in a path integral, thus producing repulsive four fermion interactions
\begin{equation}\label{4f}
 -\frac{3\, \kappa^2}{16} \int \, j^5 \, \wedge \, \star \, j^5 \;,
\end{equation}
which are a characteristic feature of Einstein-Cartan theories.

However, this would {\it not} be a self consistent procedure in view of the fact that, due to chiral anomalies, the axial fermion current conservation is violated at a quantum level~\cite{adler,belljackiw,toranom1,anomalies1,anomalies2,anomalies3}. Specifically at one loop one obtains for the divergence of the axial fermion current in a curved spacetime with torsion:
\begin{align}\label{chianom}
 d \, \star \, j^5 = \frac{e^2}{8\pi^2} \, F \, \wedge \, F - \frac{1}{96 \pi^2} R^a_{\,\,b} \, \wedge \, R^b_{\,\,a} \equiv \mathcal G(\omega,A) \;.
\end{align}
It can be shown~\cite{toranom1,toranom2,toranom3} that by the addition of appropriate counterterms, the torsion contributions to $\mathcal G(\omega, A)$ can be removed, and hence one obtains
\begin{align}\label{anomcurr}
 d \, \star \, j^5 = \frac{e^2}{8\pi^2} \, F \, \wedge \, F - \frac{1}{96 \pi^2} \mathring{R}^a_{\,\,b} \, \wedge \, \mathring{R}^b_{\,\,a} \equiv \mathcal G(\mathring{\omega},A) \;,
\end{align}
where only torsion-free quantities appear in the anomaly equation.

Therefore, to consistently integrate over the torsion $S_\mu$ in the path integral of the contorted QED, we need to add appropriate counterterms order by order in perturbation theory. This will ensure the conservation law \eqref{torsioncons} in the quantum theory, despite the presence of the anomaly \eqref{anomcurr}.
This can be achieved~\cite{olivetorsion} by implementing \eqref{torsioncons} as a
$\delta$-functional constraint in the path integral, represented by means of a Lagrange multiplier pseudoscalar field $\Phi$:
\begin{align}
\delta (d \, \star S) = \int D\Phi \, \exp\Big(\ii\,\int \, \Phi \,   d\,  \star S\Big)\,,
\end{align}
thus writing for the $S$-path integral
\begin{align}
 \mathcal Z &\propto \int D \, S \,  \delta (d \star S) \, \exp\Big(\ii\,\int \Big[\frac{3}{4\,\kappa^2} \, S \, \wedge \, \star S - \frac{3}{4} \, S \, \wedge \, \star  j^5\Big]\Big) \nonumber \\
 &= \int D S \, D\Phi \,\exp\Big(\ii\,\int \Big[\frac{3}{4\,\kappa^2}  \, S \, \wedge \, \star S - \frac{3}{4} S \, \wedge \, \star j^5 + \Phi\, d\,  \star S\Big]\Big) \;.
\end{align}
The path integral over $  S$ can then be performed, making this way the field $\Phi$ dynamical. Normalising the kinetic term of $\Phi$, requires the rescaling $\Phi = (3/(2\kappa^2))^{1/2}\,b$. We may write then for the result of the $  S$ path-integration~\cite{olivetorsion}:
\begin{align}\label{axion}
 \mathcal Z &\propto \int Db \, \exp\Big[\ii\,\int \, \Big(-\frac{1}{2}   d b \, \wedge \, \star   d b -\frac{1}{f_b} \,   d b \, \wedge \, \star j^5 - \frac{1}{2f_b^2}\,   j^5 \, \wedge \star   j^5 \Big)\Big]\,, \nonumber \\
 & f_b \equiv (3\kappa^2/8)^{-1/2}\,,
\end{align}
which demonstrates the emergence of a massless axion-like degree of freedom $b(x)$ from torsion. The characteristic shift-symmetric coupling of the axion to the axial fermionic current with $f_b$ the corresponding coupling parameter~\cite{Kim}.  Using the anomaly equation \eqref{anomcurr} we may partially integrate this term to obtain:
\begin{align}\label{axionanom}
 \mathcal Z \propto \int Db \, \exp\Big[\ii\int \, \Big(-\frac{1}{2}   d b \, \wedge \, \star   d b + \frac{1}{f_b} \,  b \, \mathcal G(\omega,A) - \frac{1}{2f_b^2}\,   j^5 \, \wedge \star   j^5 \Big) \Big]\, .
\end{align}
The repulsive four fermion interactions in \eqref{axion} and \eqref{axionanom} are characteristic of Einstein-Cartan theories, as already mentioned, but as we see from \eqref{axionanom} this is not the only effect of torsion. One has also the coupling of torsion to anomalies, which induces a coupling of the axion to gauge and gravitational anomaly parts of the theory.
The emergence of axionic degrees of freedom from torsion is an important result which will play a crucial r\^ole in our cosmological considerations.  We have observed that, in the massless chiral QED case, torsion became dynamical, due to anomalies.
We stress that the effective field theory \eqref{axionanom} guarantees the conservation law \eqref{torsioncons}, and hence the conservation of the axion charge
\begin{align}
  Q_S=\int \star   S\,,
\end{align}
order by order in perturbation theory.

Viewed as a gravitational theory, \eqref{axionanom} corresponds to a Chern-Simons gravity~\cite{JackiwPiChernSimons,GURALNIK2003222,yunes}, due to the presence of the gravitational anomaly.
From a physical point of view, placing the theory on an expanding Universe Friedman-Lemaitre-Robertson-Walker (FLRW) background spacetime, we observe that the gravitational anomaly term vanishes~\cite{JackiwPiChernSimons,yunes}. However, the gauge chiral anomaly survives. This could have important consequences for the cosmology of the model.

In fact, although above we discussed QED, we could easily consider more general models, with several fermion species, some of which could couple to non-Abelian gauge fields, e.g., the SU(3) colour group of Quantum Chromodynamics (QCD).
In such a case, torsion, being gravitational in origin, couples to all fermion  species, in a similar way as in  the aforementioned QED case, \eqref{axial}, but now the axial current \eqref{j5} is generalized to include all the fermion species:
\begin{align}\label{totcurr}
    J^{5\, \mu}_{\rm tot} = \sum_{i={\rm fermion~species}} \overline \psi_{i} \,\gamma^{\mu} \, \gamma^{5} \, \psi_{i} \;.
\end{align}
Chiral anomalies of the axial fermion current as a result of (non-perturbative) instanton effects of the non-Abelian gauge group, e.g. SU(3), during the QCD cosmological era of the Universe,
will be responsible for inducing a
breaking of the axion shift-symmetry, by generating a potential for the axion $b$
of the generic form~\cite{Kim}
\begin{align}\label{inst}
V(b) = \int d^4x \sqrt{-g} \, \Lambda_{\rm QCD}^4 \,\Big[1-{\rm cos}\Big(\frac{b}{f_b}\Big) \Big]\, \;,
\end{align}
where $\Lambda_{\rm QCD}$ is the energy scale at which the instantons are dominant configurations. As we observe from \eqref{inst} one obtains this way a {\it mass} for the torsion-induced axion
$m_b = \frac{\Lambda_{\rm QCD}^2}{f_b}$, which can thus play a r\^ole of a dark matter component. In this way we can have a geometric origin of  the dark matter component in the Universe~\cite{philtrans}, which we discuss
in Section \ref{sec:string}, where we describe a more detailed scenario in which such cosmological aspects of torsion are realised in the context of string-inspired cosmologies.

\section{Ambiguities in the Einstein-Cartan Theory-The Barbero-Immirzi parameter.}\label{sec:immir}

The contorted gravitational actions discussed in the previous Section can be modified by the addition of total derivative topological terms, which do not affect the equations of motion, and hence the associated dynamics. One particular form of such total derivative terms plays an important r\^ole in Loop quantum gravity~\cite{loop1,loop2}, a non-perturbative  approach to the canonical quantization of gravity. Below, we shall briefly mention such modifications, which, as we shall see, introduce an extra (complex) parameter, $\beta$, in the connection, termed ``BI parameter'', due to its discoverer~\cite{immirzi1,immirzi2}. This is a free parameter of the theory and it may be thought of as the analogue of the instanton angle $\theta$  of non-Abelian gauge theories, such as QCD, associated with strong CP violation.

Let us commence our discussion by presenting the case of pure gravity in the first-order formalism.
In pure gravity, if torsion is absent, a term in the action linear in the \textit{dual} of the Riemann curvature tensor, $\tilde{\mathring{R}}^{ab} _{\mu\nu} \equiv \epsilon^{ab}_{\,\,\,cd}\mathring{R}^{cd}_{\,\,\,\mu\nu}$ , called the Holst term~\cite{holst}
\begin{equation}\label{holst}
S_{{\rm Holst}} = -\frac{\beta}{4\kappa^2} \int d^4 x \, e \, E^\mu_a E^\nu_b \tilde{\mathring{R}}^{ab} _{\mu\nu} \;,
\end{equation}
where $e=\sqrt{-g}$ is the vielbein determinant, vanishes identically, as a result of the corresponding Bianchi identity of the Riemann curvature tensor:
\begin{gather}\label{antisym}
\mathring{R}_{\alpha\mu\nu\rho} +
\mathring{R}_{\alpha\nu\rho\mu} + \mathring{R}_{\alpha\rho\mu\nu} =0.
\end{gather}

However, if torsion is present, such a term yields non-trivial contributions, since in that case the Bianchi identity (\ref{antisym}) is not valid. In the general case $\beta$ is a complex parameter, and the reader might worry that in order to guarantee the
reality of the effective action one should add the appropriate complex conjugate (\emph{i.e.} impose reality conditions). As we shall discuss below, however, the effective action contributions in the second-order formalism, obtained from (\ref{holst}) upon decomposing the connection into torsion and torsion-free parts, and using the solutions for the torsion obtained by varying the Holst modification of the general relativity action with respect the independent torsion components, as in the Einstein-Cartan theory discussed previously, are independent of the BI parameter $\beta$, which can thus take on any value.

We mention for completeness that the term (\ref{holst}) has been added by Holst~\cite{holst} to the standard first-order GR Einstein-Hilbert term in the action in order to derive  a Hamiltonian formulation of canonical general relativity suggested by Barbero~\cite{barbero1,barbero2}  from an action. This formulation made use of a real SU(2) connection in general relativity, as opposed to the complex connection introduced by Ashtekar in his canonical formulation of gravity~\cite{ashtekar1}. The link between the two approaches was provided by Immirzi~\cite{immirzi1,immirzi2} who, by means of a canonical transformation, introduced a finite complex number $\beta \ne 0$  (the \emph{BI parameter}, previously mentioned) in the definition of the connection. When the (otherwise free) parameter takes on the purely imaginary values $\beta = \pm i$,  the theory reduces to the self (or anti-self) dual  formulation of canonical quantum gravity proposed by Ashtekar~\cite{ashtekar1,ashtekar2} and Ashtekar-Romano-Tate~\cite{art}. The values $\beta = \pm 1$ lead to Barbero's real Hamiltonian formulation of canonical gravity.
The Holst modification (\ref{holst}), can then be used to derive these formulations from an effective action, with the coefficient $\beta$ in (\ref{holst}) playing the r\^ole of the complex BI parameter\footnote{In the original formulation of Barbero and Immirzi, the BI parameter is $\gamma = 1/\beta$, but this is not important for our purposes.}.

In the presence of fermions, the Holst modification (\ref{holst}) is \emph{not }a total derivative. Therefore, if added
it will lead to the false prediction of ``observable effects'' of the BI parameter. In particular, following exactly the same procedure as for the Einstein-Cartan theory of the previous Section, and
using the decomposition (\ref{decom}) of the torsion in the Holst modification of the Einstein action,
obtained by adding (\ref{holst}) to the combined actions (\ref{torsiondecompaction}) and (\ref{axial}), \eqref{qedaction},
one
can derive the following extra contributions in the action (up to total derivatives)~\cite{rovelli,freidel,mercuri,calcagni}
\begin{eqnarray}\label{holsttorsion}
S_{\rm Holst} = -\frac{1}{2\kappa^2} \int d^4 x e\,\left(\frac{\beta}{3}T_\mu S^\mu + \frac{\beta}{2} \epsilon_{\mu\nu\rho\sigma}q_\lambda^{\,\,\,\,\mu\rho} \, q^{\lambda \nu \sigma}\right)~.
\end{eqnarray}

By varying independently the combined actions (\ref{torsiondecompaction}), (\ref{axial}) and (\ref{holsttorsion})
with respect to the torsion components, as in the Einstein-Cartan theory, one arrives at the
equations:
\begin{eqnarray}\label{variation}
&& \frac{1}{24\kappa^2} S^\mu + \frac{\beta}{6\kappa^2} T^\mu - \frac{1}{8}j^{5\mu} =0~, \nonumber \\
&& - 4T^\mu + \beta S^\mu = 0 ~, \nonumber \\
&& q_{\mu\nu\rho} + \beta \epsilon_{\nu\sigma\rho\lambda} q_\mu^{\,\sigma\lambda} = 0~.
\end{eqnarray}

The solutions of (\ref{variation}) are~\cite{rovelli,mercuri}
\begin{eqnarray}\label{false}
T^\mu = \frac{3\kappa^2}{4} \frac{\beta}{\beta^2 + 1} j^{5\mu} ~, \quad
S^\mu = \frac{3\kappa^2 }{\beta^2 + 1}\, j^{5\mu}~, \quad q_{\mu\nu\rho} = 0~.
\end{eqnarray}
Substituting these back into the action, following the steps of what done for the Einstein-Cartan theory, this would lead to a four-fermion induced interaction term of the form~\cite{rovelli}
\begin{equation}\label{4fermi}
S_{\rm j^5-j^5} = -\frac{3 }{16 (\beta^2 + 1)}\kappa^2 \int d^4 x e\, j^{5\mu}\, j^{5}_{\,\mu}~.
\end{equation}
The coupling of this term depends on the BI parameter $\beta$, which is in contradiction to its r\^ole in the canonical formulation of gravity~\cite{immirzi1,immirzi2}, as a free parameter, being implemented by a canonical transformation in the connection field. Moreover, for purely imaginary values of $\beta$, such that $|\beta|^2 > 1$,
the four fermion interaction is \emph{attractive}. For values of $\beta \to \pm \ii$ (which corresponds to the well-defined Ashtekar-Romano-Tate theory~\cite{art}) the interaction diverges, which presents a puzzle. Furthermore, for values of $|\beta| \to 1^+$ the coupling of the four-fermion interaction is strong. Such strong couplings can lead to the formation of fermion condensates in flat spacetimes, given that the attractive four-fermion effective coupling of (\ref{4fermi}) in this case is much stronger than the weak gravitational coupling $\kappa^2 \propto G_N$. These features are all in contradiction with the allegedly topological nature of the BI parameter.

The above are indeed pathologies related to the mere addition of a Holst term in a theory with fermions. Such an addition is inconsistent with the first-order formalism, for the simple reason that the Holst term (\ref{holst}) alone is not a total derivative in the presence of fermions, and thus there is no surprise that its addition leads to ``observable'' effects (\ref{4fermi})  in the effective action.
In addition, as observed in \cite{mercuri}, the solution (\ref{false}) of (\ref{variation}) is mathematically inconsistent, given that the first line of (\ref{false}) equates a proper vector ($T^\mu$) with an axial one (the axial spinor current $j^{5\mu}$).

The only consistent cases are those for which either  $\beta \to 0 $  or $\beta \to \infty$. The first is the Einstein-Cartan theory. The second means no torsion, in the sense that in a path integral formalism, where one integrates over all spin connection configurations, only the zero torsion contributions survive in the partition function, so as to compensate the divergent coefficient. In either case,  $T^\mu \to 0$, and the solution (\ref{false}) reduces to that of the Einstein-Cartan theory (\ref{axial}), \eqref{4f}. However, this is in sharp contradiction with the arbitrariness of the BI parameter $\beta$ of the canonical formulation of gravity, which is consistent for every (complex in general) $\beta$.

The resolution of the problem was provided by Mercuri \cite{mercuri}, who noticed that
an appropriate Holst-like modification of a gravity theory in the presence of fermions is possible, if the Holst modification contains additional fermionic-field dependent terms so as to become a total derivative and thus retains its topological nature that characterises such modifications in the torsion-free pure gravity case.
The proposed Holst-like term for the torsion-full case of gravity in the presence of fermions
contains the Holst term (\ref{holst}) and an \emph{additional} fermion-piece of the form~\cite{mercuri} (we ignore the electromagnetic interactions from now on, for brevity, as they do not play an essential r\^ole in our arguments):
\begin{equation}\label{fermiholst}
S_{\rm Holst-fermi} = \frac{\alpha}{2}\int d^4x  e\, \left(\overline{\psi} \gamma^\mu \gamma_5 D_\mu (\omega) \psi
+ \overline{D_\mu (\omega) \psi} \gamma^\mu  \gamma_5 \psi \right)~, \quad \alpha = {\rm const.}~,
\end{equation}
so that the total Holst-like modification is given by the sum $S_{\rm Holst-total} \equiv S_{{\rm Holst}} + S_{\rm Holst-fermi}$.

We next note that the fermionic Holst contributions (\ref{fermiholst}) when combined with the Dirac kinetic terms of the QED action, yield terms of the form (in our relative normalization with respect the Einstein terms in the total action):
\begin{equation}\label{kineticholstfermi}
S_{\rm Dirac-Holst-fermi} = \frac{\ii}{2} \int d^4 x e \,\left[\overline{\psi} \gamma^\mu \left(1 - \ii \alpha \gamma_5 \right)D_\mu(\omega) \psi + \overline{D_\mu(\omega) \psi}  \gamma^\mu \left(1 - {i \alpha}\gamma_5 \right) \psi \right]~.
\end{equation}
We thus observe that in the Ashtekar limit~\cite{ashtekar1,ashtekar2} $\beta = \pm \ii$, the terms in the parentheses in (\ref{kineticholstfermi}) containing the constant $\alpha$ become the chirality matrices
$\left( 1 \pm \gamma_5 \right)/2$ and this is why the specific theory is chiral.

In general, the (complex) parameter $\alpha$ is to be fixed by the requirement that the integrand in $S_{\rm Holst-total}$ is a \emph{total derivative}, so that it does not contribute to the equations of motion. It can be readily seen that this is achieved when
\begin{equation}\label{todercond}
\alpha = \beta~.
\end{equation}

In that case one recovers the results of the Einstein-Cartan theory, as far as the torsion decomposition and the second-order final form of the effective action are concerned.\footnote{Indeed, by applying the decomposition
(\ref{decom}) onto (\ref{fermiholst}), prior to imposing (\ref{todercond}),  we obtain the following extra contribution in the effective action, as compared to
the terms discussed previously in the case $\alpha = 0$~\cite{mercuri}:
\begin{equation}\label{alphadecomp}
\int d^4 x e \, \frac{\alpha}{2} T_\mu j^{5\mu}~.
\end{equation}
Including such contributions and considering the vanishing variations of the total action with respect to the (independent)  torsion components,  $T^\mu, S^\mu$ and $q^{\mu\nu\rho}$, we obtain the solution
\begin{align}\label{correctdecomp}
T^\mu = \frac{ 3\, \kappa^2 }{4} \left(\frac{\beta - \alpha}{\beta^2 + 1}\right) j^{5\mu}\,, \qquad
S^\mu = 3\kappa^2 \frac{1 + \alpha \beta}{1 + \beta^2 } \, j^{5\mu} \;, \qedsymbol{} \,  q_{\mu\nu\rho} = 0\,.
\end{align}
Clearly, as we discussed above, the first equation is  problematic from the point of view of leading to a proportionality
relation between a vector and a pseudovector, except in the Einstein-Cartan  case $\beta = 0$ and the
limit $\alpha = \beta$, where the situation is reduced again to the Einstein-Cartan theory, given that in such a case the Holst-like modification is a total derivative.}

\subsection{Holst actions for fermions and topological invariants.}\label{sec:holst1}

A final comment concerns the precise expression of the total derivative term that amounts to the total Holst-like modification
$S_{\rm Holst-total}$. As discussed in \cite{mercuri}, this action can be cast in a form involving (in the integrand) a  \emph{topologically invariant density},  the so-called Nieh-Yan topological density~\cite{Nieh}, which is the only exact form invariant under local Lorentz transformations associated with torsion:
\begin{equation}\label{holsttotalterm}
S_{\rm Holst-total} = -\ii\,\frac{\beta}{2} \int d^4 x \left[I_{\rm NY} + \partial_\mu j^{5\mu} \right]~,
\end{equation}
with $I_{\rm NY}$ the Nieh-Yan invariant density~\cite{Nieh}:
\begin{equation}\label{niehinvterm}
I_{\rm NY} \equiv \epsilon^{\mu\nu\rho\sigma} \left(T_{\mu\nu}^{\,\,\,\, a} T_{\rho\sigma \, a} - \frac{1}{2} e^a_\mu e^b_\nu R_{\rho\sigma a b}(\omega)\right).
\end{equation}
Taking into account that in our case the  torsion-full connection has the form (\ref{torsionaxcurr}),
we observe that the first term in $I_{\rm NY}$, quadratic in the torsion $T$, vanishes identically, as a result of
appropriate Fierz identities. Thus, upon taking into account (\ref{torsionaxcurr}), the Holst-like modification of the gravitational action in this case becomes a total derivative of the form\cite{kaul}:
\begin{equation}\label{holsttot}
S_{\rm Holst-total} = \frac{\ii\,\beta }{4}\int d^4 x \partial_\mu j^{5\mu} =
-\frac{\ii\,\beta}{6} \int d^4 x \epsilon^{\mu\nu\rho\sigma} \partial_\mu T_{\nu\rho\sigma} (\psi) ~,
\end{equation}
where the last equality stems from the specific form of torsion in terms of the axial fermion current, implying
$2 \epsilon^{\mu\nu\rho\sigma} T_{\nu\rho\sigma} (\psi) + 3j^{5\mu} = 0$.
In general, the Nieh-Yan density is just the divergence of the pseudotrace axial vector associated with torsion,
$I_{\rm NY} = \epsilon^{\mu\nu\rho\sigma} \partial_\mu T_{\nu\rho\sigma} ~.$

The alert reader can notice that if the axial fermion current is conserved in a theory, then the Holst action (\ref{holsttot}) vanishes trivially. However, in the case of chiral anomalies, examined above, the axial current is not conserved but its divergence yields the mixed anomaly (\ref{chianom}). In that case, by promoting the BI parameter to a canonical pseudoscalar field $\beta \to \beta (x)$~\cite{calcagni}, the Holst term \eqref{holsttot} becomes equivalent to the
torsion-axion-$j^{5\mu}$ interaction term in \eqref{axion}, upon identifying $\beta(x)= \frac{b(x)}{f_b}$. In this case, the field-prompted BI parameter plays a r\^ole analogous to the QCD CP violating parameter~\cite{calcagni}. As we have discussed in Section \ref{sec:TorsQED}. Therefore, this is consistent with the association of torsion with an axion-like dynamical degree of freedom, and thus the works of \cite{calcagni} and \cite{olivetorsion} lead to equivalent physical results from this point of view~\cite{torsionmav}.

Before closing this Subection we remark that Holst modifications, along the lines discussed for spin $1/2$ fermions above, are known to exist for higher spin 3/2 fermions, $\psi_\mu$, like gravitinos of SUGRA  theories ~\cite{castel,kaul}.
In fact, Holst-like modifications, including fermionic contributions,  have been constructed in \cite{tsuda,kaul} for various supergravities  ({\it e.g.} N=1,2,4) ,  non-trivially extending the spin 1/2 case discussed above. The total derivative nature of these Holst-like actions implies no modifications to the equations of motion. On-shell (local and global) supersymmetries are then preserved. We discuss such issues in Section \ref{sec:sugra}.

\subsection{Barbero-Immirzi Parameter as an axion field}\label{sec:immiraxion}

The classical models described until now in this Section \ref{sec:immir} lack the presence of a dynamical pseudoscalar (axion-like) degree of freedom, which, as we have seen in Section \ref{sec:TorsQED}, is associated with quantum torsion.

Such a pseudoscalar degree of freedom arises in \cite{castel,taveras}, which were the first works to promote the BI parameter  to a dynamical field, the starting point is the so-called Holst action \eqref{holst}, which by itself is {\it not} a topological invariant, in contrast to the Nieh-Yan term \eqref{niehinvterm}. The work of \cite{castel,taveras} deals with matter free cases. If $\gamma (x)$ represents the BI field, the Holst term now reads (in form language)
\begin{align}\label{holstform}
S_{\rm Holst} = \frac{1}{2\, \kappa^2} \,  \int \overline \gamma(x) \, e^a \wedge e^b \wedge R_{ab},
\end{align}
where $R_{ab}$ is the curvature two-form, in the presence of torsion, and we used the notation of \cite{taveras} for the inverse of the BI field $\overline \gamma(x) = 1 / \gamma(x)$, to distinguish this case from the Kalab-Ramond (KR) axion $b(x)$ in our string-inspired one.
The analysis of \cite{castel,taveras} showed that the gravitational sector results in the action
\begin{align}\label{gravactholst}
\mathcal S^{\rm eff}_{\rm grav+ Holst+BI-field}
&=\; \int d^{4}x\sqrt{-g}\left[ -\dfrac{1}{2\kappa^{2}}\, R + \frac{3}{4 \kappa^2 \, ( \overline \gamma^2 + 1) }\, \partial_\mu \overline \gamma \, \partial^\mu \overline \gamma \right] \;.
\end{align}

Coupling the theory to fermionic matter~\cite{freidel,rovelli,holstferm} can be achieved by introducing a rather generic non-minimal coupling parameter $\alpha$, for massless Dirac fermions in the form
\begin{align}\label{fermion}
\mathcal S_F = \frac{i}{12} \int \epsilon_{abcd} e^a \wedge e^b \wedge e^c \wedge \Big[(1 - i \alpha) \, \overline \psi \gamma^d D(\omega) \, \psi - (1 + i  \alpha) \overline{( D (\omega) \, \psi)}\, \gamma^d \, \psi \Big] \;,
\end{align}
where $\alpha \in \mathbb R$ is a constant parameter.  The case of constant $\overline \gamma$ has been discussed in \cite{freidel,rovelli} (in fact,  Ref.~\cite{rovelli} deals with minimally-coupled fermions, {\it i.e}. the limit $\alpha = 0$),
whilst the work of \cite{mercuri} extended the analysis to coordinate-dependent BI, $\overline \gamma (x)$.

The extension of the BI to a coordinate dependent quantity, which is assumed to be a {\it pseudoscalar field}, implies:  \\
{\bf (i)} Consistency of \eqref{false}, given that now the BI parameter being a pseudoscalar field, reinstates the validity of the first of the equations \eqref{false}, since the product of its right-hand side is now parity even, and thus transforms as a vector, in agreement with the nature of the left-hand side of the equation.\\
{\bf (ii)} Additional terms of interaction of the fermions with the derivative of the BI field $\partial_\mu \overline \gamma$:
\begin{align}\label{fdg}
\mathcal S_{F\, \partial \overline \gamma} = \frac{1}{2} \int \sqrt{-g} \,  \Big( \frac{3}{2 ( \overline \gamma^2 + 1)} \, \partial^\mu \overline \gamma \, \Big[ -j^5_\mu + \alpha \, \overline \gamma (x) \, j_\mu \Big]\Big)\,.
\end{align}
with $j^5_\mu$ the axial current \eqref{j5}
and
\begin{align}\label{vector}
j_\mu = \overline \psi \, \gamma_\mu \, \psi~,
\end{align}
the vector current.\\
{\bf (iii)} Interaction terms of fermions with non-derivative $\overline \gamma (x)$ terms:
\begin{align}\label{fermionnonder}
\mathcal S_{F - \textrm{non-deriv} \overline{\gamma}} &= \frac{i}{2} \int \sqrt{-g} \Big[
 \left[(1 - \ii \alpha) \, \overline \psi \gamma^d D(\mathring{\Gamma}) \, \psi - (1 + \ii \alpha) \overline{(D(\mathring{\Gamma}) \psi)}\, \gamma^d \, \psi \right]  \nonumber \\
 &- \int \sqrt{-g} \, \frac{3}{16(\overline \gamma^2 + 1)} \left[ j_\mu^5 \, j^{5\mu} - 2\alpha \, \overline \gamma \, j_\mu^5 \, j^{\mu} - \alpha^2 \, j_\mu \, j^\mu \right] \;,
\end{align}
with $D (\mathring{\Gamma})$ the diffeomorphic covariant derivative, expressed in terms of the torsion-free Christoffel connection, which is the result of \cite{freidel}, as expected, because this term contains non derivative terms of the BI.

The action \eqref{fermionnonder} involves four-fermion interactions with {\it attractive} channels among the fermions. Such features may play a r\^ole in the physics of the early Universe, as we shall discuss in Subsection \ref{sec:cosmospin}.

We also observe from \eqref{fermionnonder} that the case $\alpha =0$ (minimal coupling), corresponds to a four-fermion axial-current \eqref{4fermi}, which however depends on the BI field. Thus, this limiting theory is not equivalent to our string-inspired model, in which the corresponding quantum-torsion-induced four-fermion axial-current-current interaction \eqref{4fermi} is independent from the KR axion field $b(x)$, although both cases agree with the sign of that interaction.

A different fermionic action than \eqref{fermionnonder}, using non-minimal coupling of fermions with $\gamma^5$, has been proposed in \cite{mercuri}  as a way to resolve an inconsistency of the Holst action, when coupled to fermions, in the case of constant $\gamma$. In that proposal, the $1 + \ii \alpha$ factor in \eqref{fermionnonder} below, is replaced by the Dirac-self-conjugate quantity $1 - \ii \, \alpha \, \gamma^5$. The decomposition of the torsion into its irreducible components in the presence of the Holst action with arbitrary (constant) BI parameter, leads to an inconsistency, implying that the vector component of the torsion is proportional to the axial fermion current, and hence this does not transform properly under improper Lorentz transformations. With the aforementioned modification of the fermion action the problem is solved, as demonstrated in \cite{mercuri}, upon choosing $\alpha = \overline \gamma$, which eliminates the vector component of the torsion. But this inconsistency is valid only if $\overline \gamma$ is considered as a constant. Promotion of the BI parameter $\overline \gamma$ to a {\it pseudoscalar} field, $\overline \gamma (x)$, resolves this issue, as discussed in \cite{holstferm}, given that one obtains in that case consistent results, in the sense that the vector component of the torsion transforms correctly under parity, as a vector, since it contains now, apart from terms proportional to the vector fermionic current \eqref{vector}, also terms proportional to the product of the BI pseudoscalar with the axial fermionic current \eqref{j5}, as well as terms of the form $\overline \gamma \partial_\mu \overline \gamma$, all transforming properly as vectors under improper Lorentz transformations.

%%%%%%%%%%%%%%%%%%%%%%%%%%%%%%%%%%%%%%%%%%%%%%%%%%%%%%%%%%%%%%%%%%%%%%%%%%
\section{Torsion on graphene}\label{SecGraphene}
%%%%%%%%%%%%%%%%%%%%%%%%%%%%%%%%%%%%%%%%%%%%%%%%%%%%%%%%%%%%%%%%%%%%%%%%%%

The use of graphene as a tabletop realization of some high-energy scenarios is now considerably well developed, see, e.g., \cite{iorio2011weyl}, the review \cite{iorio:2015} and the contribution \cite{GrapheneXonsUniverse2022} to this Special Issue. Let us here recall the main ideas and those features that make graphene a place where torsion is present.

\begin{figure}[H]
\begin{adjustwidth}{-\extralength}{0cm}
\centering
 \includegraphics[scale=0.30]{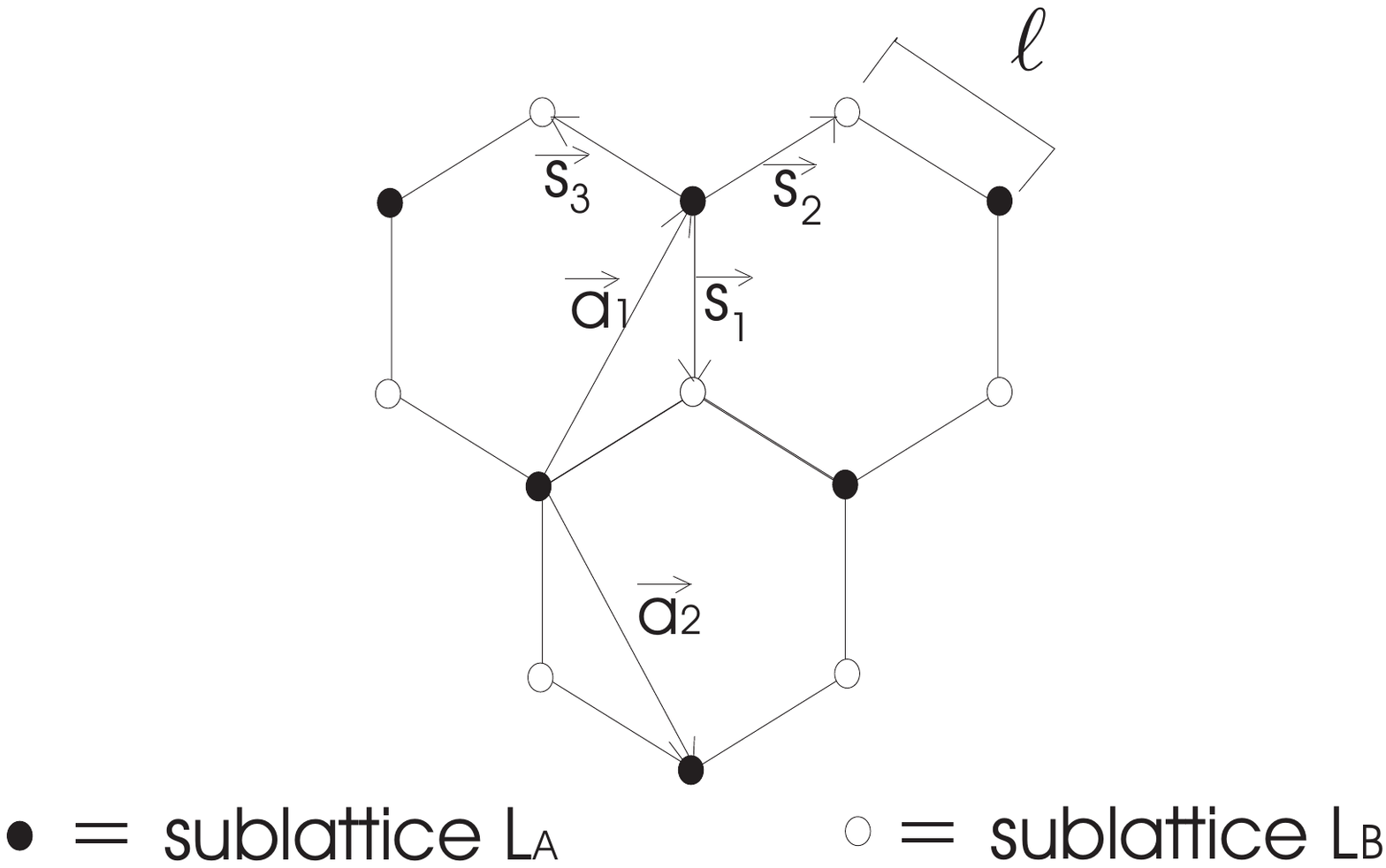}
\end{adjustwidth}
\caption{The honeycomb lattice of graphene, and its two triangular sublattices $L_A$ and $L_B$. The choice of the basis vectors, $(\vec{a}_1, \vec{a}_2)$ and   $(\vec{s}_1, \vec{s}_2, \vec{s}_3)$, is, of course, not unique. Figure taken from \cite{iorio2014}.}
\label{honeycombfig}
\end{figure}

Graphene is an allotrope of carbon and, being a one-atom-thick material, it is the closest to a two-dimensional object in nature. It is fair to say that was theoretically speculated \cite{wallace,semenoff} and, decades later, it was experimentally found \cite{geimnovoselovFIRST}. Its honeycomb lattice is made of two intertwined triangular sub-lattices $L_A$ and $L_B$, see Fig.~\ref{honeycombfig}. As is by now well known, this structure is behind a natural description of the electronic properties of graphene in terms of massless, $(2+1)$-dimensional, Dirac quasi-particles.
Indeed, starting from the tight-binding Hamiltonian for the conductivity electrons, and considering only near-neighbors contribution\footnote{It is possible to include in the description next-to-near neighbor contributions, while keeping a modified Dirac structure \cite{ip2}. In fact, such modifications
allow to reproduce scenarios related to generalized uncertainty principles both for commuting coordinates \cite{threelayers} and noncommuting coordinates \cite{ncgraphene}.}
\begin{equation}\label{hamiltonian-tb}
H = -t\,\sum\limits_{\vec{r}\in L_{A}}\,\sum\limits_{i=1}^{i=3}\,\left(a^{\dag}(\vec{r})b(\vec{r}+\vec{s}_{i})+b^{\dag}(\vec{r}+\vec{s}_{i})a(\vec{r})\right)\;,
\end{equation}
where $t$ is the nearest-neighbor hopping energy which is approximately $\SI{2.8}{\electronvolt}$, and $a,a^{\dag}(b,b^{\dag})$ are the anticommuting annihilation and creation operators for the planar electrons in the sub-lattice $L_{A}(L_{B})$.

\begin{figure}
\captionsetup[subfigure]{justification=centering}
\begin{subfigure}
[b]{0.45\textwidth}\includegraphics[width=\textwidth]{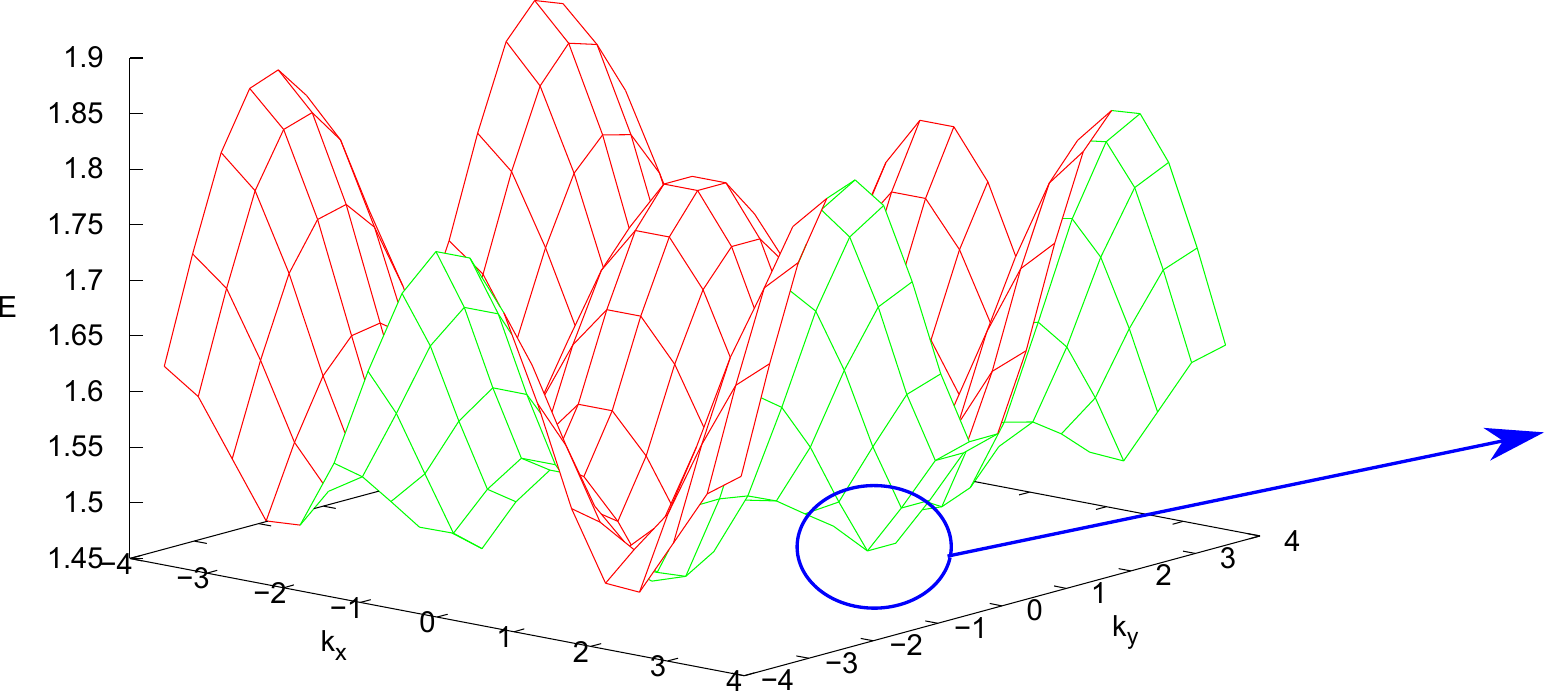}
\caption{ }
\end{subfigure}
\begin{subfigure}
{0.45\textwidth}\includegraphics[width=\textwidth]{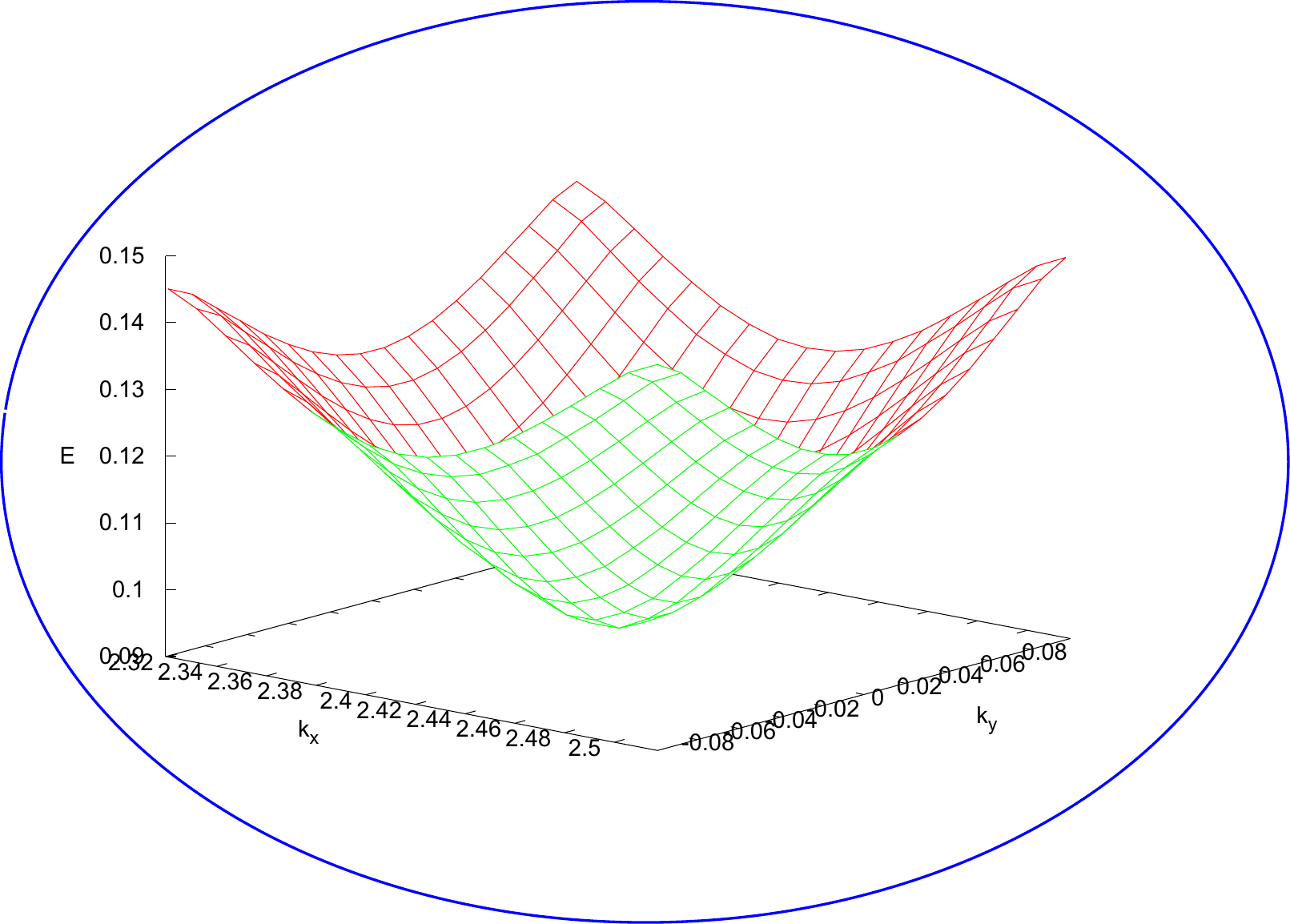}
\caption{ }
\end{subfigure}
\caption{ \textbf{(a)} The dispersion relation $E(\vec{k})$ for graphene, setting $t\ell=1$. We only take into account the near neighbors contribution in \eqref{hamiltonian-tb}. \textbf{(b)} A zoom near the Dirac point $K_{D+}$ showing the linear approximation works well in the low energies regime.}
\label{dispersion_unstrained_figure}
\end{figure}

If we Fourier-transform to momentum space, $\vec{k}=(k_{x},k_{y})$ annihilation and creation operators,
\begin{equation}\label{Fourier_trnasf}
a(\vec{r})=\sum_{\vec{k}}a_{\vec{k}}e^{i\vec{k}\cdot\vec{r}} \;,\; b(\vec{r})=\sum_{\vec{k}}b_{\vec{k}}e^{i\vec{k}\cdot\vec{r}} \;, {\rm etc} \;,
\end{equation}
then
\begin{equation*}
H=-t\,\sum_{\vec{k}}\,\sum\limits_{i=1}^{i=3}\,\left(a^{\dagger}_{\vec{k}}b_{\vec{k}}e^{i\vec{k}\cdot\vec{s}_{i}}+b^{\dagger}_{\vec{k}}a_{\vec{k}}e^{-i\vec{k}\cdot\vec{s}_{i}}\right)\;.
\end{equation*}
Using the conventions for $\vec{s}_{i}$  of Fig.~\ref{honeycombfig}, we find that
\begin{equation}\label{firstfunction}
  {\cal F} ({\vec{k}}) = -t\,\sum_{i= 1}^{3} e^{\ii{\vec{k}} \cdot {\vec{s}}_i} =
  -t \, e^{- \ii \ell k_y} \left[ 1 + 2 e^{\ii \frac{3}{2} \ell k_y} \cos\left(\frac{\sqrt{3}}{2} \ell k_x\right) \right] \;,
\end{equation}
leading to
\begin{equation*}
H=\sum_{\vec{k}} {\cal F} (\vec{k})a^{\dagger}_{\vec{k}}b_{\vec{k}}+ {\cal F}^{\ast} (\vec{k})b^{\dagger}_{\vec{k}}a_{\vec{k}}\;.
\end{equation*}
For graphene, the conduction and valence bands touch at two points\footnote{Actually, there are six such points, but the only two shown above are inequivalent under lattice discrete symmetry.} $K_{D\pm}=(\pm\frac{4\pi}{3\sqrt{3}\ell},0)$, as one can check by findig the zeroes of \eqref{firstfunction}. These points are called \emph{Dirac points}. The dispersion relation $E(\vec{k})=|f(\vec{k})|$, for $t\ell=1$, is shown in Figure \ref{dispersion_unstrained_figure} (a).
If we expand ${\cal F} (\vec{k})$ around the Dirac points, $\vec{k}_{\pm}=\vec{K}_{D\pm}+\vec{p}$, assuming $|p|\ll|K_{D}|$, we have
\begin{align*}
{\cal F}_{+} (\vec{p}) \ \equiv \ f(\vec{k}_{+}) \ &=\hspace{0.35cm} v_{F}\left(p_{x}+\ii \, p_{y}\right)\;,\\
{\cal F}_{-} (\vec{p}) \ \equiv \ f(\vec{k}_{-}) \ &=-\hspace{0.05cm} v_{F}\left(p_{x}-\ii \, p_{y}\right)\;,
\end{align*}
where $v_{F}\equiv\frac{3}{2}t\ell\sim c/300$ is the \emph{Fermi velocity}. We can see from this that the dispersion relations around the Fermi point is
\begin{equation}\label{dispersion_relation}
|E_{\pm}(\vec{p})|=v_{F}|\vec{p}|\;,
\end{equation}
which is the dispersion relation for a $v_F$-relativistic massless particle (see Figure \ref{dispersion_unstrained_figure} (b)).

Defining $a_{\pm}\equiv a(\vec{k}_{\pm})$ and $b_{\pm}\equiv b(\vec{k}_{\pm})$, and arranging the annihilation (creation) operators as a column (row) vector $\psi_{\pm}=\left(
                                                                                 \begin{array}{c}
                                                                                   b_{\pm} \\
                                                                                   a_{\pm} \\
                                                                                 \end{array}
                                                                               \right)$ ; $\psi^{\dagger}_{\pm}=\left(
                                                                                                                  \begin{array}{cc}
                                                                                                                    b^{\dagger}_{\pm} & a^{\dagger}_{\pm} \\
                                                                                                                  \end{array}
                                                                                                                \right)$, then
\begin{equation}\label{hamiltonian_massless}
H=v_{F}\sum_{\vec{p}}\left[\psi^{\dagger}_{+}\vec{\sigma}\cdot\vec{p}\psi_{+}-\psi^{\dagger}_{-}\vec{\sigma}^{\ast}\cdot\vec{p}\psi_{-}\right]\;,
\end{equation}
where $\vec{\sigma}=\left(\sigma_{1},\sigma_{2}\right)$ and $\vec{\sigma}^{\ast}=\left(\sigma_{1},-\sigma_{2}\right)$, being $\sigma_{i}$ the Pauli matrices.

Going back to the configuration space, which is equivalent to make the usual substitution $p^{\mu}\,\to-\ii\partial^{\mu}$,
\begin{equation}\label{hamiltonian_massless_configuration}
H=-\ii\,v_{F}\int d^{2}x\left[\psi^{\dagger}_{+}\,\sigma^{\mu}\,\partial_{\mu}\psi_{+}-\psi^{\dagger}_{-}\,\sigma^{\ast\mu}\,\partial_{\mu}\psi_{-}\right]\;,
\end{equation}
where sums turned into integrals because the continuum limit was assumed.

By including time to make the formalism fully relativistic, although with the speed of light $c$ traded for the Fermi velocity $v_{F}$, and making the Legendre transform of (\ref{hamiltonian_massless_configuration}), we obtain the action
\begin{equation}\label{flatactionWeyl2+1}
S = \ii \, v_{F} \, \int d^3 x \bar{\Psi} \gamma^a \partial_a \Psi \;,
\end{equation}
here $x^{a} = (t,x,y)$,   are the flat spacetime coordinates, $\Psi = (\psi_+, \psi_-)$ is a reducible representation for the Fermi field and the $\gamma^{a}$ are Dirac matrices in the same reducible representation in three dimensions.

%*** Here we show to possibility to spot torsion in Dirac materials through time loops \cite{ip4}. ***

\subsection{Torsion as continuous limit of dislocations}

Even if we will deal mainly with graphene, the considerations here apply to many other two dimensional crystals \cite{wehling}. For the purposes of this work, we can define  a  \emph{topological defect} as a lattice configuration that cannot be undone by continuous transformations. These are obtained by cutting and sewing the pristine material through what is customarily called a \emph{Volterra process} \cite{RuggieroTartaglia2003}. Probably, the easiest defects to visualize are the \emph{disclinations}. For this hexagonal lattice, a disclination defect is an $n$-sided polygon with $n \neq 6$, characterized by a \textit{disclination angle} $s$. When $n=3,4,5$, the defect has a \textit{positive} disclination angle $s=\SI{180}{\degree},\SI{120}{\degree},\SI{60}{\degree}$, respectively, whilst for $n=7,8,\ldots$, it has a \textit{negative} disclination angle $s=-\SI{60}{\degree},-\SI{120}{\degree},\ldots$, respectively. These defects carry intrinsic positive or negative curvature, according to the sign of the corresponding angle $s$, localized at the tip of a conical singularity. In a continuum description, obtained for large samples in the large wave-length regime, one can
associate\footnote{A deep study of how curvature and torsion emerge in a geometrical approach to quantum gravity, along the lines of how classical elastic-theory emerges from QED, can be found in \cite{SpaceFactoryIorioSmaldone2023}, see also \cite{SpaceFactoryDICE2022}. In those papers the authors elaborate on a model of quantum gravity inspired by graphene, but independent from it \cite{MartinAnnals2017,LucaGioPRD2020}, see also \cite{MartinPOS2017,LucaGioPOS2020}. A review can be found in \cite{GrapheneXonsUniverse2022}} \cite{Kleinert,Katanaev1992} to the disclination defect the spin-connection $\tensor{\omega}{^{ab}_{\mu}}$. Associated to $\omega$ is the curvature two-form tensor $R^{ab}$,
\begin{equation*}
\tensor{R}{^{ab}_{\mu\nu}} = \partial_{\mu}\tensor{\omega}{^{ab}_{\nu}} - \partial_{\nu}\tensor{\omega}{^{ab}_{\mu}} + \tensor{\omega}{^{a}_{c\mu}}\,\tensor{\omega}{^{cb}_{\nu}} - \tensor{\omega}{^{a}_{c\nu}}\,\tensor{\omega}{^{cb}_{\mu}} \;,
\end{equation*}
that we have already met in (\ref{curv2form}) and in (\ref{curvature_def}).

\begin{figure}
  \centering
  \includegraphics[width= .7\textwidth]{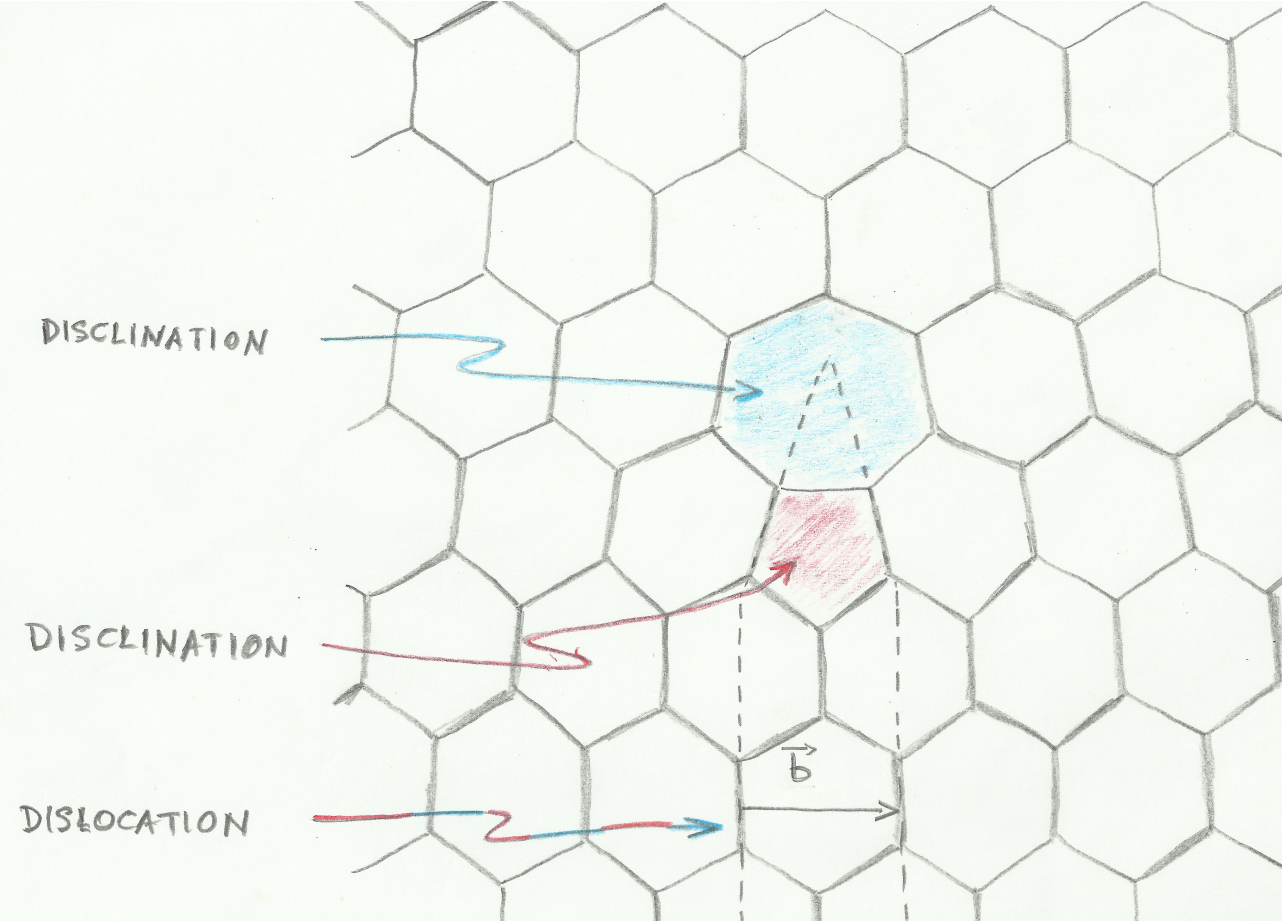}
 \caption{{\bf Edge dislocation from two disclinations}. Two disclinations, a heptagon and a pentagon, add-up to zero total intrinsic curvature, and make a dislocation with Burgers vector $\vec{b}$, as indicated. In the continuous long wave-length limit, this configuration carries nonzero torsion. Figure taken from \cite{ip3}.}
  \label{fig:edge-dislocation}
\end{figure}

A \emph{dislocation} can be produced as a dipole of disclinations with zero total curvature. In Fig. \ref{fig:edge-dislocation} it is shown a heptagon-pentagon dipole, which in Volterra process is equivalent to introducing a strip in the lower-half plane, whose width is the \emph{Burgers vector} $\vec{b}$, that characterizes this defect.
In the continuum limit one can associate the Burgers vector to the torsion tensor \cite{Kleinert,Katanaev1992}
\begin{equation}\label{Torsion}
\tensor{T}{^{a}_{\mu \nu}} = \partial_{\mu} \; \tensor{e}{^{a}_{\nu}} - \partial_{\nu} \; \tensor{e}{^{a}_{\mu}} + \tensor{\omega}{^{a}_{b}_{\mu}} \tensor{e}{^{b}_{\nu}} - \tensor{\omega}{^{a}_{b}_{\nu}} \tensor{e}{^{b}_{\mu}} \;,
\end{equation}
where $\tensor{T}{^{\rho}_{\mu \nu}} = \tensor{E}{_{a}^{\rho}}\tensor{T}{^{a}_{\mu \nu}}$. On this see our earlier discussion around (\ref{tordef}) and (\ref{cderviel}).

The explicit relation between Burgers vectors and torsion can be written as \cite{Katanaev2005}
\begin{equation}\label{torsion-Burgers}
b^{i}=\iint_{\Sigma}T^{i}_{\mu\nu}dx^{\mu}\wedge dx^{\nu}\;,
\end{equation}
where the surface $\Sigma$ has a boundary enclosing the defect. Roughly speaking, the torsion tensor is the surface density of the Burgers vector. Nonetheless, although the relation (\ref{torsion-Burgers}) looks simple, there are subtleties: given a distribution of Burgers vector, there is no simple procedure to assign a torsion tensor to it, even for the simple case of edge dislocations \cite{Lazar2003}.

The smooth way to introduce the effect of dislocations in the long wave-length regime, through torsion tensor, is to consider an action in a $(2+1)$-dimensional space with a spin-connection that carries torsion, i.e., a Riemann-Cartan space $U_{3}$ \cite{hehl}. Demanding only Hermiticity and local Lorentz invariance, starting by a simple action
\begin{equation} \label{pre-action_torsion}
S = \frac{\ii}{2} \, v_{F} \, \int d^{3}x \sqrt{|g|} \,
\left(\overline{\Psi}\gamma^{\mu}\overrightarrow{D}_{\mu} (\omega) \Psi-\overline{\Psi}\overleftarrow{D}_{\mu} (\omega) \gamma^{\mu}\Psi\right) \;,
\end{equation}
where
\begin{eqnarray}
\overrightarrow{D}_{\mu}(\omega)\Psi&=&\partial_{\mu}\Psi+\frac{1}{8}\omega^{ab}_{\mu} [\gamma_a, \gamma_b] \Psi\;,\\
\overline{\Psi}\overleftarrow{D}_{\mu}(\omega)&=&\partial_{\mu}\overline{\Psi}-\frac{1}{8} \overline{\Psi} [\gamma_a, \gamma_b] \omega^{ab}_{\mu}\;,
\end{eqnarray}
we obtain, besides possible boundary terms, (see details in Appendix A of \cite{ip4}),
\begin{equation}\label{action_torsion}
S = \ii \, v_{F} \, \int d^{3}x \; |e| \; \overline{\Psi}\left(E_{a}^{\mu}\,\gamma^{a} \overrightarrow{D}_{\mu}(\mathring{\omega}) - \frac{\ii}{4} \gamma^{5} \frac{\epsilon^{\mu\nu\rho}}{|e|} T_{\mu \nu\rho} \right)\Psi\;,
\end{equation}
where $|e|=\sqrt{|g|}$, the covariant derivative is based only on the torsion-free connection, $\gamma^{5} \equiv  i \gamma^{0} \gamma^{1} \gamma^{2} = \left(
                                                           \begin{array}{cc}
                                                             I_{2 \times 2} & 0 \\
                                                             0 & -I_{2 \times 2} \\
                                                           \end{array}
                                                         \right)$
(we used the conventions for $\gamma^{0},\gamma^{1},\gamma^{2}$ that give a $\gamma^{5}$ that {\it commutes} with the other three gamma matrices\footnote{This is due to the reducible, rather than irreducible, representation of the Lorentz group we use.}), and the contribution due to the torsion is all in the last term through its totally antisymmetric component \cite{ip4}.

We see that the last term couples torsion with the fermionic excitations describing the quasi-particles and is the three-dimensional version of (\ref{axial}), for $A_{\mu}=0$. It can be also seen that, to have a nonzero effect, we need $\epsilon^{\mu\nu\rho}T_{\mu \nu\rho}\neq0$, that requires at least three dimensions. This mathematical fact is behind the obstruction pointed out some time ago leading to the conclusion that, in two-dimensional Dirac materials, torsion can play no physical role \cite{deJuan2010,Vozmediano:2010zz,Amorim:2015bga}.

To overcome this obstruction, in \cite{ip4} the time dimension is included in the picture. With this in mind, we have two possibilities that a nonzero Burgers vector gives rise to $\epsilon^{\mu\nu\rho}T_{\mu \nu\rho}\neq0$:
\begin{enumerate}[label=(\roman*)]
\item a \textit{time-directed} screw dislocation (only possible if the crystal has a time direction)
\begin{equation}\label{timecrys}
  b_{t} \propto \int\int  T_{012} dx \wedge dy \;,
\end{equation}
or
\item an edge dislocation ``felt'' by an integration along a \textit{spacetime circuit} (only possible if we can actually go around a loop in time), e.g,
\begin{equation}\label{tloop}
b_{x} \propto \int\int  T_{102} dt \wedge dy \;.
\end{equation}
This last scenario is depicted in Fig. \ref{Fig3TIMELOOP}.
\end{enumerate}

\subsection{Time-loops in Graphene}

Scenario (i) could be explored in the context of the very intriguing time crystals introduced some time ago \cite{Wilczek2012, WilczekShapere2012}, and nowadays under intense experimental studies \cite{PhysRevLett.109.163001,PhysRevLett.121.185301}. Lattices that are discrete in all dimensions, including time, would be an interesting playground to probe quantum gravity ideas \cite{Loll1998}. In particular, it would have an impact to explore defect-based models of classical gravity/geometry, see for instance \cite{Kleinert} and \cite{Katanaev1992}. However, here we shall focus only on scenario (ii).

By assuming the Riemann curvature to be zero, $\tensor{\mathring{R}}{^{\mu}_{\nu\rho\sigma}}=0$, but nonzero torsion (or contorsion $\tensor{K}{^{\mu}_{\nu\rho}}\neq0$), and choosing a frame where $\mathring{\omega}_\mu^{ab}=0$ (see Appendix B of \cite{ip4}), the action (\ref{action_torsion}) is
\begin{equation}\label{action_pure_torsion}
S = \ii\, \,v_F \int d^{3}x |e| \,  \left(\overline{\Psi}\gamma^{\mu}\partial_{\mu}\Psi - \frac{\ii}{4} \overline{\psi}_{+} \phi \psi_{+} + \frac{\ii}{4} \overline{\psi}_{-} \phi \psi_{-} \right) \;,
\end{equation}
where $\Psi = (\psi_+,\psi_-)$ and $\phi \equiv \epsilon^{\mu \nu \rho} T_{\mu\nu\rho} / |e|$ is what we call \emph{torsion field}; it is a pseudo-scalar and the three-dimensional version of the $S_{\mu}$ we discussed earlier. Even in the presence of torsion, the two irreducible spinors, $\psi_{+}$ and $\psi_{-}$, are decoupled (however, with opposite signs).

\begin{comment}
To overcome the three-dimensional geometric obstruction through the ``time-loop'' in the $(y,t)$-plane of scenario (ii), see (\ref{tloop}), the proposal of \cite{ip4} is to make use of the particle-antiparticle description of the dynamics encoded in the action (\ref{action_pure_torsion}). By realizing that the regime of the materials we are describing is the ``half-filling'' \cite{pacoreview2009}, for which the energy states of the valence band ($E < 0$) have the vacancies completely filled (being the analog of the Dirac sea), while the vacancies of the conduction band ($E > 0$) stay empty, we think now of exciting a pair particle-hole out of this vacuum and making them oscillate, say, along the $y$-axis, as described in Fig.~\ref{Fig3TIMELOOP}. This amounts to a circuit of the particle-antiparticle pair in the $(y,t)$-plane. What is left to do is to fully exploit the emergent relativistic-like structure of the model and see the portion of the circuit described by the \textit{antiparticle} moving \textit{forwards} in time, as corresponding to the same \textit{particle} moving \textit{backwards} in time. This realizes what we may call a \textit{time-loop}.
\end{comment}

\begin{figure}
\begin{center}
\includegraphics[width=0.4\textwidth,angle=90]{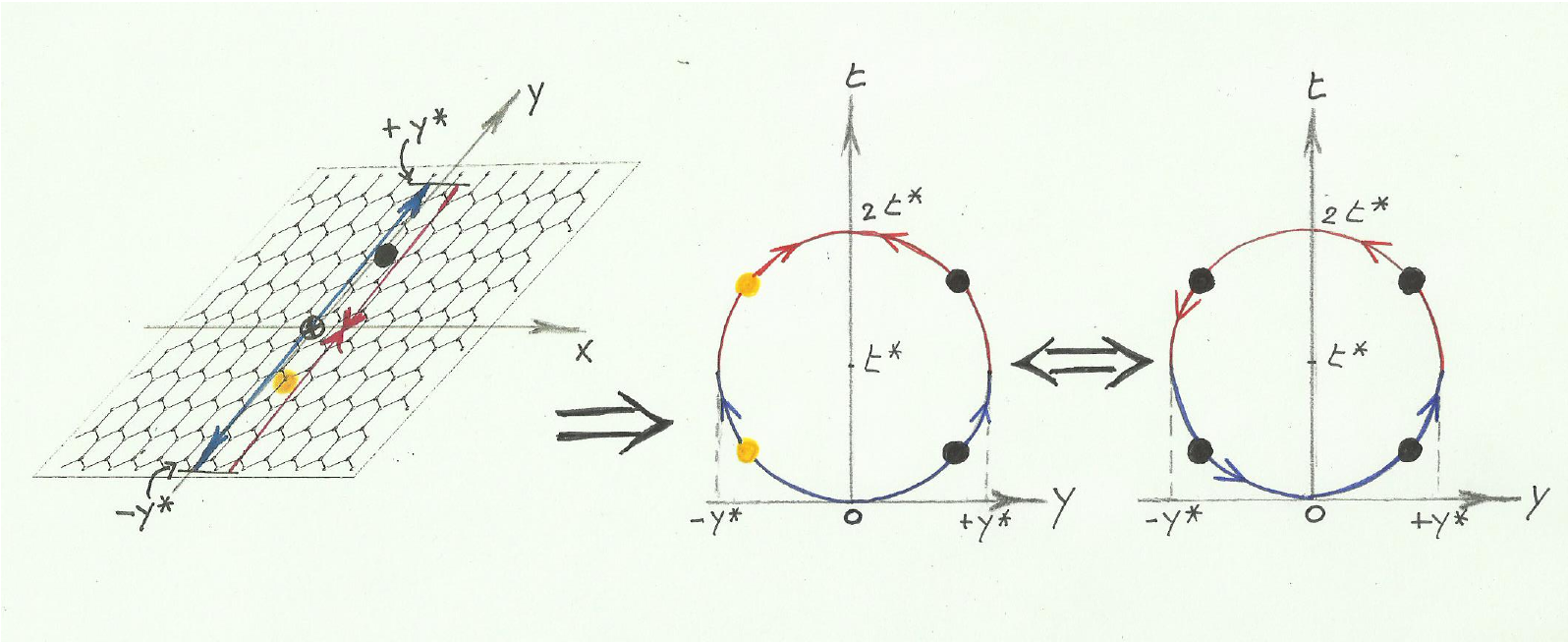}
\end{center}
\caption{Idealized \emph{time-loop}. At $t=0$, the hole (yellow) and the particle (black) start their movements from $y=0$, in opposite directions. At $t=t^*>0$, the hole is at position $-y^*$, while the particle is at position $+y^*$,  (the blue portion of the circuit). Then they come back to their original position, $y=0$, at $t=2t^*$ (red portion of the circuit). On the far right, is depicted an equivalent \textit{time-loop}, where the hole moving forward in time is replaced by a particle moving backward. Figure taken from \cite{ip4}.}%
\label{Fig3TIMELOOP}%
\end{figure}

The pictures in Fig.~\ref{Fig3TIMELOOP} refer to a defect-free honeycomb graphene-like sheet. The presence of a dislocation, with Burgers vector $\vec{b}$ directed along $x$, would result in a failure to close the loop proportional to $\vec{b}$ \cite{ip4}.

\begin{comment}
Therefore, going from a first to a second quantization approach, paying due attention to the subtleties involved in describing dislocation distributions by a suitable torsion tensor, the Dirac field theory emerging here can indeed include a nonzero coupling with torsion. This accounts for a field theoretical description of the effects of dislocations only when the third dimension is taken to be time, overcoming the geometrical obstruction discussed earlier \cite{deJuan2010,Vozmediano:2010zz,Amorim:2015bga}.
\end{comment}

The idea of time-looping is fascinating. The challenge is to bring this idealized picture close to experiments. We present below the first steps in that direction, as taken in \cite{ip4}.

\subsection{Towards spotting torsion in a lab}

The simplest way to realize the scenario just discussed is to have:
\begin{enumerate}[label=\roman*)]
\item the particle-hole pair required for the time-loop to be excited by an \emph{external electromagnetic field}, and
\item that what we shall call \emph{holonomy} — a proper disclination or torsion — provides the non-closure of the loop in the proper direction.
\end{enumerate}
Stated differently, we are searching for \textit{the quantifiable consequences of an holonomy, caused by disclination or torsion in a time-loop}. Only an appropriate combination of i) and ii) can yield the desired outcome.

With this in mind, the action governing such microscopic dynamics is
\begin{eqnarray}
S  & = & \ii  \int d^{3}x \,|e|\, \left(\overline{\Psi} \gamma^{\mu} (\partial_{\mu} - \ii g_{\mbox{em}}\, A_\mu) \Psi - \ii\, g_{\mbox{tor}}\, \overline{\psi}_{+} \phi \psi_{+} + \ii\,g_{\mbox{tor}}\,\overline{\psi}_{-}\,\phi\,\psi_{-} \right)  \label{action torsion external A} \\
& \rightarrow & \ii  \int d^{3}x \, \left(\overline{\psi} \gamma^{\mu} \partial_{\mu} \psi  - \ii\,g_{\mbox{em}}\,\,\hat{j}^{\mu}_{\mbox{em}}\,A_\mu - \ii\,g_{\mbox{tor}}\,\hat{j}_{\mbox{tor}}\,\phi \right) \equiv S_0 [\overline{\psi}, \psi] + S_I [A, \phi] \;. \label{action samples}
\end{eqnarray}
where $v_F$ is taken to be one, while $g_{\mbox{em}}$ and $g_{\mbox{tor}}$ are the electromagnetic and torsion coupling constants, respectively, the latter including the factor $1/4$. In (\ref{action samples}) we only have one Dirac point, say $\psi \equiv \psi_+$, as this simplifies calculations, and we focus on flat space, $|e|=1$. Finally, $\hat{j}^{\mu}_{\mbox{em}} \equiv  \overline{\psi} \gamma^{\mu} \psi$ and $\hat{j}_{\mbox{tor}} \equiv \overline{\psi} \psi$.

The electromagnetic field is \textit{external}, hence it is a four-vector $A_\mu \equiv (V, A_x, A_y,A_z)$. Nonetheless, the dynamics it induces on the electrons living on the membrane is two-dimensional. Therefore, the effective vector potential may be taken to be $A_\mu \equiv (V, A_x, A_y)$, see, e.g., \cite{lasergraphenePRL2018,natureLaserGraphene}. There are two alternatives to this approach. One is the \emph{reduced QED} of \cite{Marino:1992xi,Gorbar:2001qt}, where the gauge field propagates on a three-dimensional space and one direction is integrated out to obtain an effective interaction with the electrons, constrained to move on a two-dimensional plane. In this approach a Chern-Simons photon naturally appears (see, e.g.,  \cite{Dudal:2018mms,Dudal:2018pta}). Another approach is to engineer a $(2+1)-$dimensional $A_\mu$ by suitably straining the material, see, e.g., \cite{Vozmediano:2010zz,Amorim:2015bga}, and \cite{ipfirst}. In that case, one usually takes the temporal gauge and $A_x \sim u_{xx} - u_{yy}$, $A_x \sim 2 u_{xy}$, where $u_{i j}$ is the strain tensor.

As said, defects here are not dynamical, therefore the torsion field $\phi$ enters the action as an external field, just like the electromagnetic field. One could, as well, include the effects of the constant $\phi$ into the unperturbed action, as a mass term $S_0 \to S_m$, see, e.g., \cite{DauriaZanelli2019}, where $S_m = i \int d^{3}x  \; \overline{\psi}(\slashed{\partial} - m(\phi)) \psi$.

We are in the situation described by the microscopic perturbation
\begin{equation}\label{SIfx}
S_{I} [F_i] = \int d^{3}x \, \hat{X}_i (\vec{x},t) F_i (\vec{x},t) \;,
\end{equation}
with the system responding through $\hat{X}_i (\vec{x},t)$ to the external probes $F_i (\vec{x},t)$. The general goal is then to find
\begin{equation}\label{X[F]}
\hat{X}_i [F_i] \;,
\end{equation}
to the extent of predicting a measurable effect of the combined action of the two perturbations $F_i (\vec{x},t)$: $F^{\mbox{em}}_{1} (\vec{x},t) \propto A_\mu (\vec{x},t)$ that induces the response $\hat{j}^\mu_{\mbox{em}}$, and $F^{\mbox{tor}}_{2} (\vec{x},t) \propto \phi (\vec{x},t)$ inducing the response  $\hat{j}_{\mbox{tor}}$:
\begin{equation}\label{SIaphi}
S_{I} [A, \phi] = \int d^{3}x \, \left( \hat{j}^\mu_{\mbox{em}} A_\mu + \hat{j}_{\mbox{tor}} \phi \right) \;,
\end{equation}
where the couplings, $g_{\mbox{em}}$ and $g_{\mbox{tor}}$, are absorbed in the respective currents.

With no explicit calculations, simply based on the charge conjugation invariance of the action \eqref{action samples}, we can already predict that
\begin{equation}\label{firstFurry}
    \chi^{\mbox{torem}}_{\mu} (x,x') \sim \langle \hat{j}^{\mbox{em}}_{\mu} (x) \hat{j}^{\mbox{tor}} (x') \rangle \equiv 0 \;,
\end{equation}
that is just an instance of the Furry's theorem of quantum field theory \cite{Peskin}, that in QED reads
\begin{equation}\label{FurryEMgeneral}
    \chi^{\mbox{em}}_{\mu_{1} ... \mu_{2n+1}} (x_{1}, ..., x_{2n+1}) \sim \langle \hat{j}^{\mbox{em}}_{\mu_{1}} (x_{1}) \cdots \hat{j}^{\mbox{em}}_{\mu_{2n+1}} (x_{2n+1}) \rangle = 0 \;,
\end{equation}
and for us implies
\begin{equation}\label{Furry_ours}
    \chi^{\mbox{torem}}_{\mu_{1} ... \mu_{2n+1}} (x_{1}, ..., x_{2n+1}, y_{1}, ..., y_{m}) \sim \langle \hat{j}^{\mbox{em}}_{\mu_{1}} (x_{1}) \cdots \hat{j}^{\mbox{em}}_{\mu_{2n+1}} (x_{2n+1})
    \hat{j}^{\mbox{tor}} (y_1) \cdots \hat{j}^{\mbox{tor}} (y_{m})\rangle = 0 \;.
\end{equation}

This finding indicates that entering the nonlinear response domain is necessary to observe the desired consequences. High-order harmonic generation (HHG) is a well-established technique that has been used to analyze structural changes in atoms, molecules, and more recently, bulk materials (see, e.g.~\cite{StanislavRevModPhys} for a recent overview). Thus, the presence or absence of dislocations will significantly alter the intra-band harmonics in our system, which are controlled by the intra-band (electron-hole) current.

\subsection{On the continuum description of the two inequivalent Dirac points}

We have shown earlier that two Dirac points, associated to the reducible $\Psi=(\psi_{+},\psi{-})$ \cite{iorio:2015}, are important to treat torsion.  Two such points are actually relevant in a broader set of cases. From the material point of view \cite{pacoreview2009}, this generally has to do with an extra ``valley'' degree of freedom in a pristine material, also called color index \cite{Gusynin}. Things change more drastically when topological defects are present. For instance, to make a fullerene $C_{60}$ form pristine graphene we need twelve pentagons sitting at the vertices of an icosahedron, and this generates color mismatches, see a discussion of these effects in \cite{GONZALEZ1993771}. There, different magnetic flux are added for each vertex which contain a color line frustration, pointing out to a ``magnetic monopole'' at the center of the molecule structure \cite{GONZALEZ1993771}. Such ``monopole'' is associated to the SU(2) symmetry group stemming from the doublet structure of the valley degree of freedom (not to be confused with the doublet structure associated with each valley, that generates the irreducibles $\psi_\pm$).

\begin{figure}
  \centering
  \includegraphics[width= 1\textwidth]{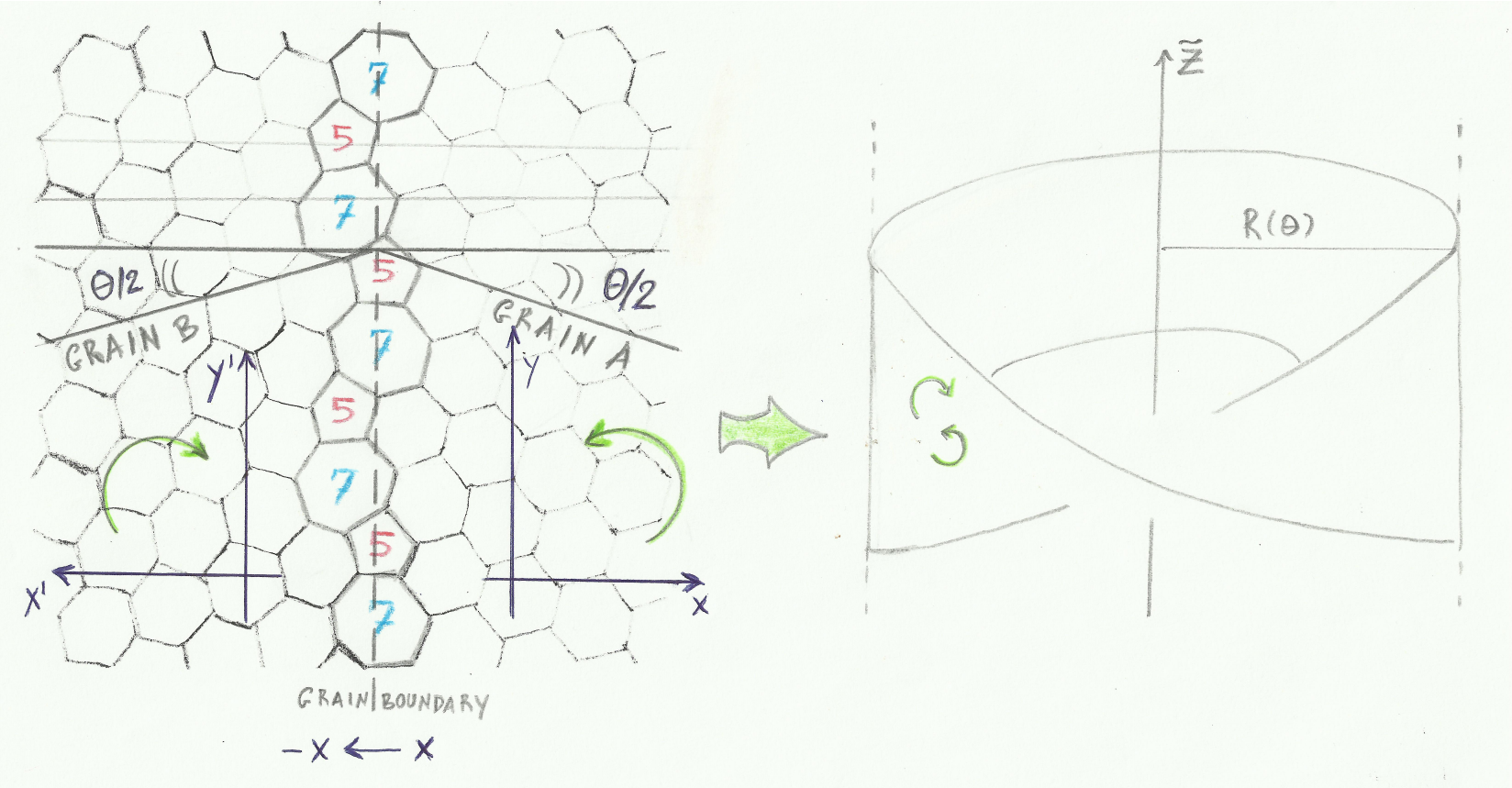}
 \caption{A grain boundary (left), and a possible modeling of its effects in a continuum (right). This is the prototypical GB, where grain A and grain B are related via a parity ($x \to - x$) transformation. With this, the right-handed frame in grain A is mapped to the left-handed frame in grain B, so that the net effect of a GB is that two orientations coexist on the membrane, and a discontinuous change happens at the boundary. If one wants to trade this discontinuous change for a continuous one, an equivalent coexistence is at work in the non-orientable M\"obius strip. One way to quantify the effects of different $\theta$s is to relate a varying $\theta$ to a varying radius $R(\theta)$ of the M\"obius strip. Notice that the third spatial axis is an abstract coordinate, $\tilde{z}$, whose relation with the real $z$ of the embedding space is not specified. Figure taken from \cite{ip3}.}
  \label{fig:GrainBoundaryTwoChiralities}
\end{figure}

Another instance where both Dirac points are needed for an effective description are \emph{grain boundaries} (GBs). A GB is a line of disclinations of opposite curvature, pentagonal and heptagonal here, arranged in such a way that the two regions (grains) of the membrane match. The two grains have lattice directions that make an angle $\theta/2$ with respect to the direction the lattice would have in the absence of the GB. Different arrangements of the disclinations, always carrying zero total curvature, correspond to different $\theta$s, the allowed number of which is of course finite, and related to the discrete symmetries of the lattice (hexagonal here). The most common (stable) being $\theta = \SI{21.8}{\degree}$, and $\theta = \SI{32.3}{\degree}$, see, e.g., \cite{Yazyev2014,Yazyev2010}. Other arrangements can be found in \cite{hirth1967theory}.  In general, one might expect that the angle of the left grain differs in magnitude from the angle of the right grain, $|\theta_L| \neq |\theta_R|$, nonetheless, high asymmetries are not common, and the symmetric situation depicted in Fig.\ref{fig:GrainBoundaryTwoChiralities} is the one the system tends to on annealing \cite{C5NR04960A}.

There exists \cite{Yazyev2010,hirth1967theory} a relation (the Frank formula) between $\theta$ and the resultant Burgers vector, obtained by adding all Burgers vectors $\vec{b}$s cut by rotating a vector, laying on the GB, of an angle $\theta$ with respect to the reference crystal. A possible modeling for this kind of defects was put forward in \cite{ip3}. That is a four-spinor living on a M\"obius strip, see Fig.\ref{fig:GrainBoundaryTwoChiralities}, and \cite{ip3}.

% Another theory that could take into account the two Dirac points, and includes torsion in a very natural way, is unconventional supersymmetry, and we shall discuss in the next Section.

%%%%%%%%%%%%%%%%%%%%%%%%%%%%%%%%%%%%%%%%%%%%%%%%%%%%%%%%%%%%%%%%%%%%%%%%%%

\section{Torsion in Standard Local Supersymmetry}\label{sec:sugra}

%%%%%%%%%%%%%%%%%%%%%%%%%%%%%%%%%%%%%%%%%%%%%%%%%%%%%%%%%%%%%%%%%%%%%%%%%%

As a prelude to the Section dedicated to cosmology, we should discuss fermionic (gravitino) torsion in SUGRA models, which can also lead to dynamical breaking of SUGRA. Such models can serve in inducing inflationary scenarios by providing sources for primordial gravitational waves which play a crucial role in inflation, to be discussed in detail in Section \ref{SecCosmology}.

SUGRA theories are Einstein-Cartan theories with fermionic torsion, provided by the gravitino field, $\psi_\mu(x)$, the spin-3/2 (local) supersymmetric fermionic partner of the graviton.

We commence our discussion with the first local SUSY constructed historically, the (3+1)-dimensional $N=1$ SUGRA~\cite{n=1sugra1,n=1sugra1b,n=1sugra2}, which in fact finds a plethora of (conjectural) applications to the phenomenology of particle physics~\cite{Nilles}.
In the remainder of this Section we shall work in units of the gravitational constant $\kappa=1$ for brevity.

The spectrum of the unbroken (3+1)-dimensional $N=1$ SUGRA is a massless spin 2 graviton field, described by the symmetric tensor field  $g_{\mu\nu}(x)=g_{\nu\mu}(x)$, $\mu, \nu = 0, \dots 3$ and
a massless gravitino spin 3/2 Rarita-Schwinger Majorana fermion $\psi_\mu(x)$.

The standard action is given by~\cite{n=1sugra2}
\begin{align}\label{sugract}
 S_{\rm SG1} = \frac{1}{2}  \int d^4x \, \sqrt{-g} \, \Big(\Sigma_{ab}^{\mu\nu} \, R_{\mu\nu}^{\quad ab}(\omega) - \epsilon^{\mu\nu\rho\sigma} \, \overline{\psi}_\mu  \, \gamma^5 \, \gamma_\nu \, D_\rho(\omega)\, \psi_\sigma\Big)\,,
\end{align}
where $\Sigma_{ab}^{\mu\nu} = \frac{1}{2} E^\mu_{\,[a} \, E^\nu_{\,b]}$ and $D_\mu (\omega) = \partial_\mu + \frac{1}{8} \, \omega_{a b\,\mu} \, [\gamma^a\,,\, \gamma^b]$
is the diffeomorphic covariant derivative, with respect to a spin connection $\tensor{\omega}{^{a}_{b}_{\mu}}$ which, as we shall discuss below, necessarily contains fermionic (gravitino-induced) torsion.

As shown in \cite{kaul,tsuda}, the action \eqref{sugract} can be augmented by adding to it a total derivative Holst type action, which preserves the on-shell $N=1$ SUSY for an arbitrary coefficient $t$:
\begin{align}
 S_{\rm Holst1} = \ii\, \frac{\eta}{2} \, \int d^4x \, \sqrt{-g}\, \Big( \Sigma_{ab}^{\mu\nu} \, \widetilde R_{\mu\nu}^{\quad ab}(\omega)
 - \epsilon^{\mu\nu\rho\sigma} \, \overline{\psi}_\mu  \, \gamma_\nu \, D_\rho(\omega)\, \psi_\sigma\Big)\,,
\end{align}
with $\widetilde R_{\mu\nu}^{\quad ab}(\omega)$ the dual Lorentz curvature tensor.

Indeed, as demonstrated in \cite{kaul,tsuda}, the combined action
\begin{align}\label{totact}
&S_{\rm total~SG} = S_{\rm SG1} + S_{\rm Holst1}
= \nonumber \\
&\frac{1}{2}\,\int d^4x \, \Big(\sqrt{-g} \, \Big[E^\mu_{\,a} \, E^\nu_{\,b}\, R^{ab}_{\,\,\,\,\mu\nu} - \frac{t}{2}\,
\epsilon^{ab}_{\,\,\,\,cd} \, R^{cd}_{\,\,\,\,\mu\nu}\Big]
+ \epsilon^{\mu\nu\rho\sigma} \, \overline{\psi}_\mu \, \gamma^5\, \gamma_\rho \, \frac{1 - \ii\, \eta\, \gamma^5}{2}\, D_\sigma (\omega) \, \psi_\nu \Big) \;,
\end{align}
is invariant under the {\it local} SUSY transformation with infinitesimal Grassmann parameter $\alpha(x)$:
\begin{align}
 &\delta \psi_\mu = D_\mu (\omega) \, \alpha, \,\, \,
 \delta e^a_{\,\mu} = \frac{\ii}{2} \, \overline{\alpha} \, \gamma^a\, \psi_\mu, \,
 \,\, \delta B_{ab\mu} = \frac{1}{2} \Big(C_{\mu ab} - e_{\mu[a}\, C^c_{\,cb]}\Big)\,, \nonumber \\
\end{align}
where, by definition,
\begin{align}
 & C^{\lambda\mu\nu} \equiv \frac{1}{\sqrt{-g}} \, \epsilon^{\mu\nu\rho\sigma} \, \overline{\alpha}  \, \gamma^5 \, \gamma^\lambda \, \frac{1- i \eta\,\gamma^5}{2} \, D_\rho (\omega) \, \psi_\sigma \;.
\end{align}
We remark for completion that in the special case where $\eta= \pm \ii$ we obtain Ashtekar's chiral SUGRA extension, while for $\eta=0$ one recovers the standard $N=1$ SUGRA transformations.

We next remark that variation of the action \eqref{totact} with respect to the spin connection, leads to the well-known
gravitational equation of motion in first order formalism~\cite{n=1sugra1,n=1sugra1b,n=1sugra2}, which leads to  an expression of the gravitino-induced torsion  $T^{\quad \mu}_{\rho\sigma}(\psi)$ in terms of the gravitino fields:
\begin{align}\label{torsN=1}
 D_{[\mu}(\omega)\, e^a_{\,\nu]} \equiv  2T^{\quad a}_{\mu\nu}(\psi) = \frac{1}{2} \, \overline{\psi}_\mu \, \gamma^a \,
 \psi_\nu \;,
\end{align}
with the contorted spin connection being given by:
\begin{align}
\omega_\mu^{\,\,ab} (e, \psi) = \mathring{\omega}_{\mu}^{\,\, ab}(e) + K_\mu^{\,\,ab}(\psi)\, ,
\end{align}
where $\mathring{\omega}_\mu^{\,\,ab}(e)$ is the torsion-free spin connection (expressible, as in standard GR, in terms of the vielbeins $e_{\,\mu}^a$), and $K_\mu^{\,\,ab}(\psi)$ is the contorsion, given in terms of the gravitino field as:
\begin{align}
K_{\mu \rho \sigma}(\psi) = \frac{1}{4} \Big(\overline \psi_\rho \, \gamma_\mu \, \psi_\sigma +  \overline \psi_\mu \, \gamma_\rho \, \psi_\sigma  - \overline \psi_\mu \, \gamma_\sigma \, \psi_\rho \Big)\,.
\end{align}
The parameter $\eta$ does not enter the expression for the contorsion, which thus assumes the standard form of $N=1$ SUGRA without the Holst terms.

Substitution of the solution of the torsion equations of motion into the first-order lagrangian density, corresponding to the action
\eqref{totact}, leads to a second-order Lagrangian density that
can be written as the sum of the standard $N=1$ SUGRA Lagrangian density~\cite{n=1sugra2} and a total derivative, depending on the gravitino fields only:
\begin{align}\label{totderN=1sugra}
\mathcal L({\rm second~order}) = \mathcal L_{{\rm usual~N=1~SUGRA}}({\rm second~order}) + \frac{i}{4} \, \eta \, \partial_\mu (\epsilon^{\mu\mu\rho\sigma} \, \overline \psi_\nu\, \gamma_\rho\, \psi_\sigma) \,,
\end{align}
where the standard N=1 SUGRA in the second-order formalism includes four-gravitino terms,
\begin{align}\label{stansugra} \mathcal L_{\rm usual~N=1~SUGRA} &=  \sqrt{-g}\, \frac{1}{2} R(e) +
 \frac{1}{4} \partial_\mu [E_a^\mu \, E^\nu_b \sqrt{-g}] \, \Big(\overline{\psi}_\nu \, \gamma^a \, \psi^b - \overline{\psi}_\nu \, \gamma^b\, \psi^a + \overline{\psi}^a\, \gamma_nu \, \psi^b\Big) \nonumber \\
 &-\frac{1}{2} \, \epsilon^{\mu\nu\lambda\rho} \overline{\psi}_\mu \, \gamma^5 \, \gamma_\nu \Big[\partial_\lambda  + \frac{1}{2} \omega_\lambda^{\,ab}(e) \sigma_{ab}\Big]\, \psi_\rho
 \nonumber \\
& - \frac{11}{16} \sqrt{-g} \Big[ (\overline{\psi}_a \, \psi^a)^2 - (\overline \psi_b \, \gamma^5\, \psi^b)^2\Big] + \frac{33}{64}\, \sqrt{-g}\, (\overline{\psi}_b \, \gamma^5\, \gamma_c \, \gamma^b)^2 \nonumber \\
&+ {\rm appropriate~auxilliary-field~terms}\,, \qquad \sigma_{ab} = \frac{i}{4} [\gamma_a\,,\, \gamma_b]\,,
\end{align}
and as standard~\cite{n=1sugra2} the Lagrangian density is computed by requiring the irreducibility condition:
\begin{align}\label{ginofix}\gamma^\mu \, \psi_\mu =0~,
\end{align}
which ensures that the spin is exactly  $3/2$ and not a mixture of this and lower spins.
We note that the four-gravitino terms of \eqref{stansugra} have been used in \cite{Alexandre1,Alexandre2} in order to discuss, upon appropriate inclusion of Goldstino terms~\cite{deser},\footnote{The Goldstino $\lambda$ is a Majorana spin 1/2 fermion which plays the r\^ole of the Goldstone-type fermionic mode arising from the spontaneous breaking of global SUSY. To incorporate the relevant dynamics into the
dynamically-broken SUGRA scenario, one adds to the SUGRA Lagrangian \eqref{stansugra} the
terms
\begin{align}\label{VAlag}
\mathcal{L}_{\rm golds} = -f^2 \, {\rm det}\Big(\delta^\mu_\nu  + i\frac{1}{2\, f^2} \overline{\lambda} \, \gamma^\mu\, \partial_\nu \, \lambda\Big) = - f^2 -\frac{1}{2}\,i\, \overline{\lambda} \, \gamma^\mu \, \partial_\mu \lambda + \dots \,
\end{align}
where $f \in \mathbb R$ is the energy scale of SUSY breaking, and the $\dots$ denote higher order self-interaction terms of $\lambda$. Such a term
realises SUSY non linearly in the sense of Volkov and Akulov~\cite{volkov}. After an appropriate gauge fixing \eqref{ginofix}
the derivative  $\partial_\mu \lambda$ can then be absorbed, by a suitable redefinition of the gravitino field $\psi_\mu$ in the schematic combination $\psi_\mu^\prime = \psi_\mu + \partial_\mu \lambda$, so that the gravitino field acquires a non zero mass, proportional to the gravitino condensate $\sigma$. Then, all that is left from the lagrangian density \eqref{VAlag}
is a negative cosmological constant term $-f^2 < 0$, and thus the final, gauge fixed, SUGRA  lagrangian encoding dynamical breaking of local SUSY, is given by:
\begin{align}\label{totstansugra}
\mathcal{L}_{\rm total} = -f^2 +
\mathcal{L}_{\rm{N=1~SUGRA}}~.
\end{align}
We shall not give further details here on this dynamical mechanism for SUGRA breaking, referring the interested reader to the literature (see ref.~\cite{Alexandre1,Alexandre2} and references therein).} the possibility of dynamical breaking of SUGRA, via the formation of condensates of gravitino fields $\sigma_c = \langle \overline \psi_\mu \, \psi^\mu \rangle \ne 0$. The gravitino field becomes massive, with mass which can be close to Planck mass, which implies its eventual decoupling from the low-energy (non supersymmetric) theory.

\begin{figure}[ht]
		\centering
		\includegraphics[width=0.4\textwidth]{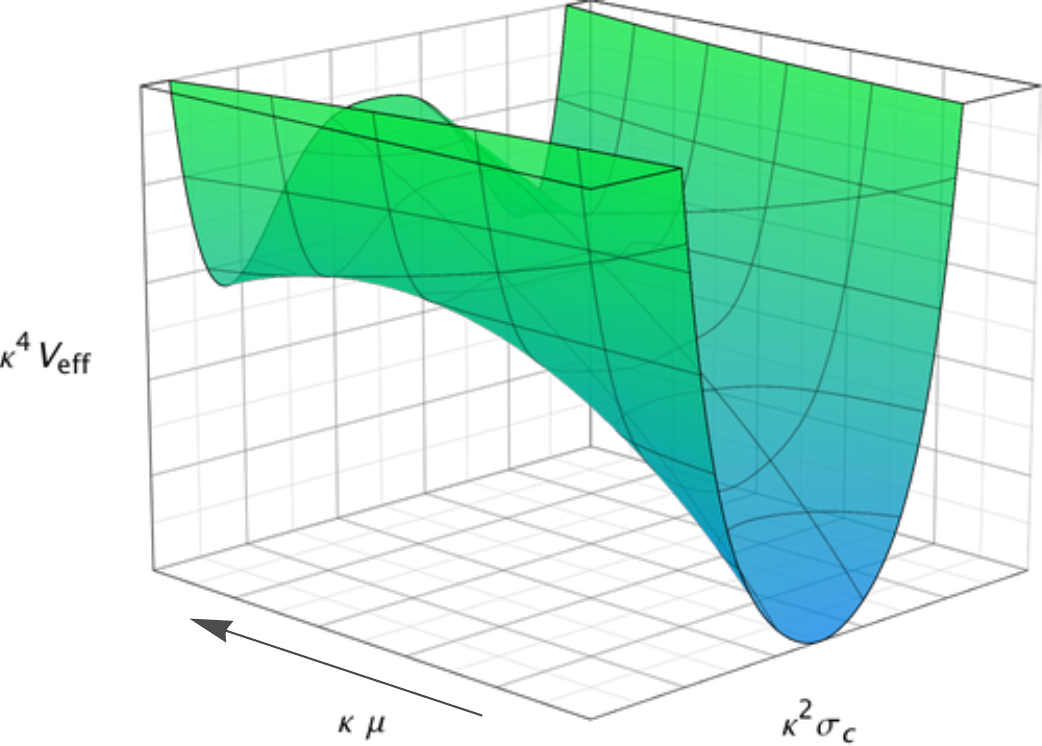}
		\includegraphics[width=0.4\textwidth]{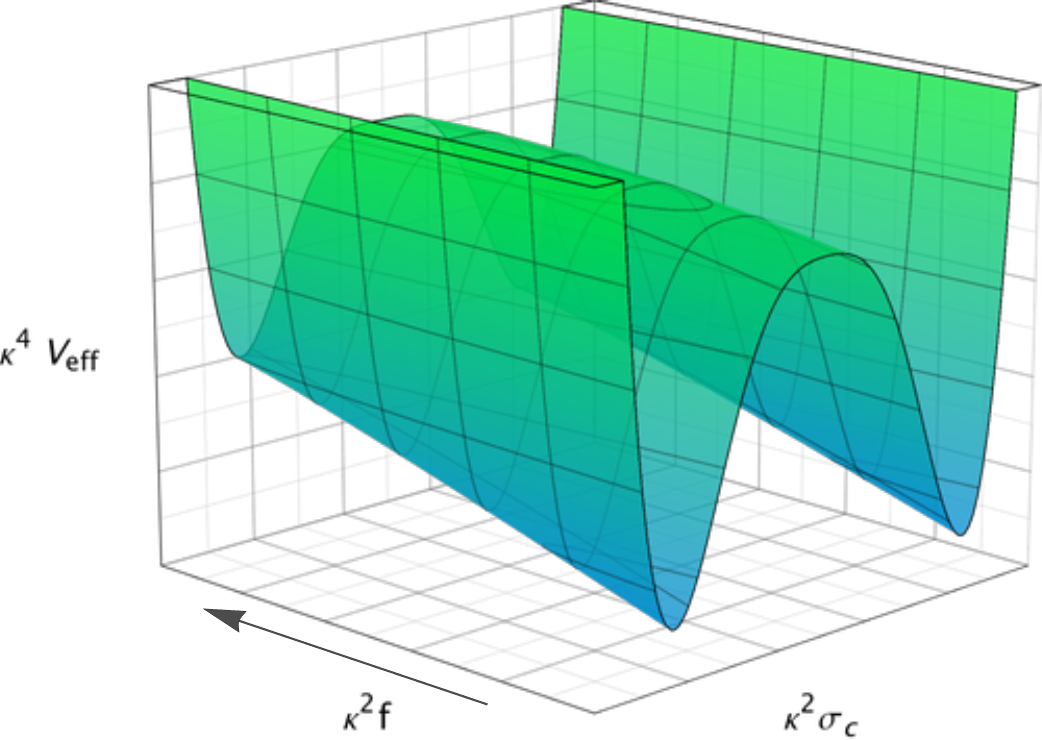}
		\caption{The effective potential of the torsion-induced gravitino condensate $\sigma_c=\langle \overline{\psi}_\mu\, \psi^\mu\rangle$ in the dynamical breaking of $N=1$ SUGRA scenario of \cite{Alexandre1}, in which, for simplicity, the one-loop-corrected cosmological constant $\Lambda \to 0^+$ (for an analysis with $\Lambda > 0$ see \cite{Alexandre2} and references therein). The figures show
  schematically the effect of tuning the inverse-proper-time (renormalization-group like) scale $\mu$ and the scale of SUSY breaking $f$, whilst holding, respectively, $f$ and $\mu$ fixed. The arrows in the respective axes correspond to the direction of increasing $\mu$ and $f$. The reader should note (see left panel) that the double-wall shape of the potential, characteristic of the super-Higgs effect (dynamical SUGRA breaking), appears for values of $\mu $ larger than a critical value, in the direction of increasing $\mu$, that is as we flow from Ultraviolet (UV) to infrared (IR) regions. Moreover, as one observes from the right panel of the figure,
  tuning $f$ allows us to shift the value of the effective potential $V_{\text{eff}}$
  appropriately so as to attain the correct vacuum structure, that is,  non-trivial minima $\sigma_c$ such that $V_{\text{eff}}\left(\sigma_c\right)=\Lambda \to 0^+$. Picture taken from \cite{Alexandre1}.}
\label{fig:brsugra}	
	\end{figure}

Such scenarios have been used to discuss hill-top inflation, as a consequence of the double-well shape of the effective gravitino potential. Indeed, for small condensates $\kappa^6\, \sigma_c(x) \ll 1 $,
one may obtain an inflationary epoch, not necessarily slow roll, as the gravitino rolls down towards one of the local minima of its double well potential~\cite{hilltopEllis} ({\it cf.} Fig.~\ref{fig:brsugra}). Such scenarios will be exploited further in  Section \ref{sec:string}, from the point of view of the generation of gravitational waves in the very early Universe, which can lead to a second inflationary era in such models, that could provide interesting, and compatible with the data, phenomenology/cosmology.

We complete the discussion on $N=1$ SUGRA as an Einstein-Cartan theory, by noticing that, on using  \eqref{totderN=1sugra}, \eqref{holsttotalterm}, \eqref{niehinvterm}, we may write for the super Holst term in this case~\cite{kaul,tsuda}:
\begin{align}\label{superholstN=1sugra}
&S_{{\rm Super~Holst~N=1~SUGRA}}(e,\psi) = -\frac{\ii \, \eta}{2} \, \int d^4 x \Big[T_{\rm NY} + \partial_\mu J^\mu(\psi) \Big]\,,
\end{align}
with
\begin{align}
 J^\mu (\psi) =  \epsilon^{\mu\nu\rho\sigma}\, \overline{\psi}_\nu\, \gamma_\rho\, \psi_\sigma   \,,
\end{align}
the axial gravitino current, and the Nieh-Yan invariant is given by \eqref{niehinvterm}.

Finally, combining the Fierz identity $\epsilon^{\mu\nu\rho\sigma} \, (\overline{\psi}_\mu \, \gamma_a \, \psi_\nu )\, \gamma^a \, \psi_\rho = 0$, with the expression for the N=1 SUGRA torsion $T_{\mu\nu}^{\quad a}(\psi)$ \eqref{torsN=1}, we arrive at
$\epsilon^{\mu\nu\rho\sigma} \, T_{\mu\nu a}(\psi) \, T_{\rho\sigma}^{\quad a}(\psi) =0$, we may write for the
on-shell-local-SUSY preserving Holst term \eqref{superholstN=1sugra}:
\begin{align}\label{superholstN=1sugraB}
&S_{{\rm Super~Holst~N=1~SUGRA}}(e,\psi) = -\frac{\ii \, \eta}{4} \, \int d^4 x \partial_\mu J^\mu(\psi) ] =
\frac{i\, \eta}{2} \, \int d^4 x \, \epsilon^{\mu\nu\rho\sigma} \,  \partial_\mu T_{\nu \rho\sigma} (\psi)  \,.
\end{align}

The Section is concluded by mentioning that super Holst modifications have been constructed~\cite{kaul} for extended SUGRAs, such as $N=2,4$, following and extending appropriately the $N=1$ case.
The spectrum of the $N=2$ SUGRA consists of a massless spin-2 graviton, two massless chiral  spin-3/2 gravitinos, $\gamma^5 \, \psi^I_\mu = + \psi^I_\mu$,
$\gamma^5 \, \psi_{I\mu} = - \psi_{I\mu}$, $I=1,2$, and an Abelian gauge field $A_\mu$. This is also an Einstein-Cartan theory, with torsion
\begin{align}
2T_{\mu\nu}^{\quad a} = \frac{1}{2}\Big(\overline{\psi}^I_\mu\, \gamma^a \, \psi_{I\nu} + \overline{\psi}_{I\mu}\, \gamma^a \, \psi^I_{\nu}\Big)\,,
\end{align}
and contorsion
\begin{align}\label{contN=2}
 K_{\mu\rho\sigma} =    \frac{1}{4}\Big[\overline{\psi}^I_\rho \, \gamma_\mu \, \psi_{I\sigma} +  \overline{\psi}^I_\mu \, \gamma_\rho \, \psi_{I\sigma} - \overline{\psi}^I_\mu \, \gamma_\sigma \, \psi_{I\rho} + {\rm c.c.} \Big] \;,
\end{align}
where c.c. denotes complex conjugate, whilst the super Holst term has the form~\cite{kaul}:
\begin{align}\label{superholstN=2sugra}
&S_{{\rm Super~Holst~N=2~SUGRA}}(e,\psi) = -\frac{\ii \, \eta}{4} \, \int d^4 x \partial_\mu J^\mu(\psi) ] =
\frac{i\, \eta}{2} \, \int d^4 x \, \epsilon^{\mu\nu\rho\sigma} \,  \partial_\mu T_{\nu \rho\sigma} (\psi)  \,,
\end{align}
with $J^\mu (\psi) = \epsilon^{\mu\nu\rho\sigma}\, \overline{\psi}^I_\nu\, \gamma_\rho\, \psi_{I \sigma}$ the axial gravitino current in this case. We observe from \eqref{contN=2} that then contorsion is again independent, as in the $N=1$ case, from the super Holst action parameter $\eta$.

Finally, we complete the discussion with the $N=2$ gauged SU(4) SUGRA. For our discussion, we restrict our attention only to the relevant part of its spectrum,  consisting of massless spin-2 gravitons, four chiral Majorana spin-3/2 gravitinos $\psi^I_\mu$, $I=1, \dots 4$, in the $4$ and $4^\star$ representations of SU(4), and 4 Majorana chiral gauginos $\Lambda^I$, $I=1, \dots 4$. The torsion of this theory depends on both the gravitino and gaugino fields~\cite{kaul},
\begin{align}\label{torN=4}
2T_{\mu\nu}^{\quad a} = 2T_{\mu\nu}^{\quad a}(\psi) +
2T_{\mu\nu}^{\quad a}(\psi) =
\frac{1}{2} \overline{\psi}^I_{[\mu} \, \gamma^a \, \psi_{\nu]I} + \frac{1}{2\, \sqrt{-g}} \, e^{a\rho} \, \epsilon_{\mu\nu\rho\sigma} \,
\overline{\Lambda}_I \, \gamma^\sigma \, \lambda^I\,,
\end{align}
and the contorsion reads
\begin{align}
 K_{\mu\nu\rho} = \frac{1}{4} \Big( \overline{\psi}^I_\nu \, \gamma_\mu \, \psi_{\rho\, I} + \overline{\psi}^I_\mu \, \gamma_\nu \, \psi_{\rho\, I}
 - \overline{\psi}^I_\mu \, \gamma_\rho \, \psi_{\nu\, I} + {\rm c.c.} \Big) -
 \frac{1}{4\, \sqrt{-g}} \, \epsilon_{\mu\nu\rho\sigma}\, \overline{\Lambda}_I \, \gamma^\sigma \, \Lambda^I \;,
\end{align}
which again is independent of the  parameter $\eta$ of the super Holst term, which has the form~\cite{kaul}:
\begin{align}\label{superholstN=4sugra}
S_{{\rm Super~Holst~N=2~SUGRA}}(e,\psi) &= -\frac{\ii \, \eta}{4} \, \int d^4 x \partial_\mu [J^\mu(\psi) - J^\mu (\Lambda) ] \nonumber\\
&=
\frac{i\, \eta}{2} \, \int d^4 x \, \epsilon^{\mu\nu\rho\sigma} \,  \partial_\mu \Big(T_{\nu \rho\sigma} (\psi) - \frac{1}{3}T_{\nu \rho\sigma} (\Lambda) \Big) \;,
\end{align}
where $J^\mu(\Lambda) = \sqrt{-g} \, \overline{\Lambda_I}\, \gamma^\mu \, \Lambda^I$, and the torsion quantities have been defined in \eqref{torN=4}.

%%%%%%%%%%%%%%%%%%%%%%%%%%%%%%%%%%%%%%%%%%%%%%%%%%%%%%%%%%%%%%%%%%%%%%%%%%
\section{Torsion in Unconventional Supersymmetry}\label{SecUSUSY}
%%%%%%%%%%%%%%%%%%%%%%%%%%%%%%%%%%%%%%%%%%%%%%%%%%%%%%%%%%%%%%%%%%%%%%%%%%

USUSY is an appealing theory where all the fields belong to a one-form connection $\mathbb{A}$, in $(2+1)$ dimensions, and the vielbein is realized in a different way than in standard SUGRA models \cite{AVZ}. It has nontrivial dynamics, and leads to a scenario where local SUSY is absent (although there is still diffeomorphism invariance), but rigid SUSY can survive for certain background geometries. Because there is no local SUSY, there are no SUSY pairings. Likewise, no gauginos are present. The only propagating degrees of freedom are fermionic \cite{GPZ}, and the parameters that appear in the model are either dictated by gauge invariance, or arise as integration constants. We take the one-form connection spanned by the Lorentz generators $\mathbb{J}_{a}$, the SU(2) generators corresponding to the internal gauge symmetry $\mathbb{T}_{I}$ (or a other internal group generator, including the Abelian U(1)), the supercharges $\overline{\mathbb{Q}}^{i}$ and $\mathbb{Q}_{i}$ (note that these last generators contain the index corresponding to the fundamental group of SU(2) as well as the spinors)\footnote{It is possible to add a central extension generator $\mathbb{Z}$ and its corresponding one-form coefficient $b$ \cite{USUSYSU2}. However, we shall not consider this extension in the present work.} \cite{USUSYSU2}
\begin{equation}\label{conncetion_su(2)}
\mathbb{A}=A^{I}\mathbb{T}_{I}+\overline{\psi}^{i}\slashed{e}\mathbb{Q}_{i}+\overline{\mathbb{Q}}^{i}\slashed{e}\psi_{i}+\omega^{a}\mathbb{J}_{a}\;,
\end{equation}
where $A^{I}=A^{I}_{\mu}dx^{\mu}$ is the one-form SU(2) connection, $\omega^{a}=\tensor{\omega}{^{a}_{\mu}}dx^{\mu}$ is the one-form Lorentz connection in $(2+1)$ dimensions, and we defined the one-form $\slashed{e}\equiv\tensor{e}{^{a}_{\mu}}\gamma_{a}dx^{\mu}$.

We can construct a three-form Chern-Simons Lagrangian from \eqref{conncetion_su(2)}, namely\footnote{Here, we omitted the wedge notation for the exterior product. For instance, $\mathbb{A}^{3}$ stands for the three-form $\mathbb{A}\wedge\mathbb{A}\wedge\mathbb{A}$.}
\begin{equation}\label{action_su(2)}
L=\frac{\kappa}{2}\langle\mathbb{A}d\mathbb{A}+\frac{2}{3}\mathbb{A}^{3}\rangle\;,
\end{equation}
where $\langle\ldots\rangle$ is the invariant supertrace of $\mathfrak{usp}(2,1|2)$ graded Lie algebra (for the case of internal SU(2) group) and $\kappa$ is a dimensionless constant. This way, the Lagrangian can be written simply as
\begin{equation}\label{action_su(2)_expanded}
L=\frac{\kappa}{4}\left(A^{I}dA_{I}+\frac{1}{3}\epsilon_{IJK}A^{I}A^{J}A^{K}\right)+\frac{\kappa}{4}\left(\omega^{a}d\omega_{a}+\frac{1}{3}\epsilon_{abc}\omega^{a}\omega^{b}\omega^{c}\right)+L_{\psi}\;,
\end{equation}
where the fermionic part is
\begin{equation*}
L_{\psi}=\kappa\overline{\psi}\left(\gamma^{\mu}\overrightarrow{D}_{\mu}-\overleftarrow{D}_{\mu}\gamma^{\mu}-\frac{i}{2}\tensor{\epsilon}{_{a}^{bc}}\tensor{T}{^{a}_{bc}}\right)\psi|e|d^{3}x\;.
\end{equation*}
We can see the action (\ref{action_su(2)_expanded}) possesses also a local scale (Weyl) symmetry. Indeed, by scaling the dreibein and the fermions as
\begin{equation*}
\tensor{e}{^{a}_{\mu}} \to \tensor{e}{^{a}_{\mu}}'=\lambda\tensor{e}{^{a}_{\mu}} \; , \; \psi \to \psi'=\lambda^{-1}\psi \;,
\end{equation*}
where $\lambda=\lambda(x)$ is a non-singular function on the spacetime manifold, the action (\ref{action_su(2)_expanded}) is invariant. This is a consequence of the particular construction of the connection (\ref{conncetion_su(2)}), where the fermions always appear along with the dreibein field, forming a composite field.

For the case of the internal group SU(2) the internal index can be interpreted as valley index, making USUSY another good scenario to describe the continuous limit of both Dirac points (see details in \cite{ip3}).

\begin{comment}
If the geometric background is fixed and the non-Abelian gauge field is external (there is no dynamics for the phonons and gauge fields), then the Lagrangian \eqref{action_su(2)_expanded} leads to the following action
\begin{equation}\label{action_su(2)_fermions}
S_{\psi}=\kappa\int\overline{\psi}\left(\gamma^{\mu}\overrightarrow{\widetilde{D}}_{\mu}-\overleftarrow{\widetilde{D}}_{\mu}\gamma^{\mu}-\frac{i}{4}\tensor{\epsilon}{_{a}^{bc}}\tensor{T}{^{a}_{bc}}\right)\psi|e|d^{3}x\;.
\end{equation}
Therefore, we can see that the only difference of $S_{\psi}$ with respect to (\ref{action_pure_torsion}) is the coefficient in front of the torsion term. The role of torsion in USUSY is to give an effective mass to the electrons (without mixing the Dirac points).

A peculiarity of USUSY is the absence of gravitini, although it includes gravity and SUSY. All the parameters involved in the system are either protected by gauge invariance or emerge as integration constants. Interesting enough, the vacuum sector is defined by configurations with locally flat Lorentz and $SU(2)$ connections carrying nontrivial global charges, as is the case of BTZ black holes \cite{BTZ1992}. Moreover, the only propagating degrees of freedom are the fermionic ones \cite{GPZ}.
\end{comment}

The action of USUSY in $(2+1)$ dimensions, for fixed background bosonic fields, apart for possible boundary terms, is obtained from the Chern-Simons three-form for $\mathbb{A}$ with an SU(2) internal gauge group \cite{USUSYSU2}%and it is very similar to the action \eqref{action_torsion}
\begin{equation}\label{action_ususy}
S_{USUSY}=\kappa\int\overline{\psi}^{i}\left(\gamma^{\mu}\mathring{\overrightarrow{D}}_{\mu}-\frac{\ii}{8}\tensor{\epsilon}{_{a}^{bc}}\tensor{T}{^{a}_{bc}}\right)\psi_{i}|e|d^{3}x\;,
\end{equation}
where lower case Latin letters, $a,b,\ldots$, represent tangent space Lorentz indices, and $\tensor{T}{^{a}_{bc}}=\tensor{T}{^{a}_{\mu\nu}}\,E^{\mu}_{b}\,E^{\nu}_{c}$.

This action immediately points to (\ref{action_torsion}), that is the action with torsion we have seen emerging in graphene, where we only need to fix the dimensionfull $\kappa$ to include $v_F$ rather than $c$. Notice that, as discussed at length in \cite{ip3} the two Dirac points are both necessary, so that the emergent action is made of two parts, one per Dirac point. This makes possible to have both the internal SU(2) symmetry and torsion, that is necessary for the USUSY description. So far as for similarities between (\ref{action_ususy}) and (\ref{action_torsion}). There are differences, though. The first is the coefficient of the torsion term, which appears in USUSY as an integration constant \cite{AVZ}. The second difference is the index $i$ (here taken as an internal colour index, considering both Dirac points in the model). Both differences are due to the starting point to get (\ref{action_torsion}), which is an Hermitian action with local Lorentz invariance in a Riemann-Cartan space. In contrast, the starting point of USUSY is an action with a supergroup $USP(2,1|2)$ invariance, which is allowed by using another representation for $\psi$ and the Dirac matrices (see details in Appendix B of \cite{ip3}). In addition, it is also possible to take into account the two Dirac points by using other internal supergroups, such as $OSp(p|2)\times OSp(q|2)$ in this USUSY context \cite{DauriaZanelli2019}. In any case, (\ref{action_ususy}), (\ref{action_torsion}) and the model proposed in \cite{DauriaZanelli2019} are top-down approaches to describe the $\psi$ electrons in graphene-like systems. Therefore, we should keep in mind these (and others) models to compare them with the results of a real experiment in the lab.

\begin{comment}
Apart from a global factor $\kappa$ that can be adjusted to be $\ii\,\hbar\, v_{F}$, let us comment the differences of (\ref{action_ususy}) with respect to (\ref{action_torsion}). The first one is the coefficient in front of the torsion term, which appears in U-SUSY as an integration constant \cite{AVZ}. The second difference is the index $i$ (here taken as a colour internal index, considering both Dirac points in the model). Both differences are due the starting point to get (\ref{action_torsion}), which is an Hermitian action with local Lorentz invariance in a Riemann-Cartan space, while the starting point of USUSY is an action with a supergroup $USP(2,1|2)$ invariance, which is allowed by using another representation for $\psi$ and the Dirac matrices (see details in Appendix B of \cite{ip3}). In addition, it is also possible to take into account the two Dirac points by using other internal supergroups, as $OSp(p|2)\times OSp(q|2)$ in this USUSY context \cite{DauriaZanelli2019}. In any case, (\ref{action_ususy}), (\ref{action_torsion}) and the model proposed in \cite{DauriaZanelli2019} are top-down approaches to describe the $\pi$ electrons in graphene-like systems, therefore we should keep in mind these (and others) models to compare with the results of a real experiment in the lab.
\end{comment}

Finally, let us comment that the Bañados-Zanelli-Teitelboim (BTZ) black hole \cite{BTZ1992}, in a pure bosonic vacuum state ($\psi=0$), is a solution of USUSY \cite{AVZ}. This follows from the fact that the BTZ black hole, whose metric in cylindrical coordinates ($-\infty<t<+\infty$, $0<r<+\infty$, and $0\leq\phi\leq2\pi$) is
\begin{equation}\label{BTZ_metric}
ds^{2} = -N^{2}(r)\,dt^{2} + N^{2}(r)\,dr^{2} + r^{2}\left(N^{\phi}(r)\,dt + d\phi\right)^{2} \;,
\end{equation}
\begin{equation}
N^{2}(r) = -M + \frac{r^{2}}{\ell^{2}} + \frac{J^{2}}{4r^{2}}\;, N^{\phi}(r) =  \frac{J^{2}}{2r^{2}} \;,
\end{equation}
can be obtained from a Lorentz-flat connection with torsion \cite{BTZLorentzflat}. The spectrum of these black holes is given in terms of their mass, $M$, and angular momentum, $J$, including the extremal, $M\ell=|J|$ and $M=0$ cases\footnote{The case $M=-1$ is the globally anti-De Sitter space, while the other cases are conical singularities \cite{Miskovic2009}.}. We also mention here that the $M=0$ case could play a very important role in the Generalized Uncertainty Principle induced by gravity \cite{GUPBTZ,ip5}, and in Hawking-Unruh phenomenon on graphene and graphene-like materials \cite{Iorio:2011yz}.

\section{Torsion in Cosmology}\label{SecCosmology}
%%%%%%%%%%%%%%%%%%%%%%%%%%%%%%%%%%%%%%%%%%%%%%%%%%%%%%%%%%%%%%%%%%%%%%%%%%

A Plethora of precision cosmological data~\cite{Planck} in the past twenty five years, have indicated that the energy budget of the current cosmological epoch of our (observable) Universe is dominated (by $\sim 95\%)$ by a dark sector of unknown, at present, microscopic origin. If one fits the available data at large scales, corresponding to the modern era of the Universe, within the so-called $\Lambda$CDM framework, which consists of a de Sitter Universe (dominated by a positive cosmological constant $\Lambda$) and a Cold Dark Matter (CDM) component, then one obtains excellent agreement. On the other hand, there appear to be tensions to such data at smaller scales~\cite{tensions1,tensions2,tensions3}, arising either from discrepancies between the value of the Hubble parameter in the modern era obtained from direct observations of nearby galaxies and that inferred by $\Lambda$CDM fits (``$H_0$ tension''), or from discrepancies in the value of the parameter $\sigma_8$ characterising galactic growth data between direct observations and $\Lambda$CDM fits (``$\sigma_8$ tension'').

To these tensions, provided of course the latter do not admit more mundane astrophysical explanations or are mere artefacts of relatively low statistics~\cite{Freedman}, and thus will be absent from future data, one should add theoretical obstacles to the self consistency of the $\Lambda$CDM framework, when viewed as a viable gravity model embeddable in microscopic models of quantum gravity, such as string theory~\cite{string1,string2} and its brane extensions~\cite{Polch}. Indeed, the existence of eternal de Sitter horizons, in spacetimes with a constant $\Lambda > 0$, prohibits the definition of asymptotic states, and thus a perturbative scattering S-matrix, which is the cornerstone of perturbative string theories, appears not to be well defined, thus posing problems with the compatibility of a de Sitter spacetime as a consistent background of perturbative strings~\cite{smatrix1,smatrix2}. Such problems extend to fully quantum gravity considerations, when one attempts to embed de Sitter spacetimes in microscopic ultraviolet complete models such as strings or branes, due to the so-called swampland conjectures~\cite{Palti1,Palti2,swamp1,swamp2,swamp3,swamp4}, which are violated by the $\Lambda$CDM framework.

Barring the (important) possibility of misinterpretation of the Planck data as far as dark energy is concerned, by, e.g.,  {\it relaxing} the assumption of homogeneity and isotropy of the Universe at cosmological scales~\cite{subir1,subir2},  one is therefore tempted to seek for theoretical alternatives to $\Lambda$CDM, which will not be characterised by a positive constant $\Lambda$, but rather having the de Sitter vacuum as a {\it metastable one}, in such a way that there are no asymptotic in future time de Sitter horizons. The current literature has a plethora of potential theoretical resolutions to the de Sitter $\Lambda$ problem~\cite{Tensol1}, which simultaneously alleviate the aforementioned tensions in small-scale cosmological data.
What we would like to discuss below, in the context of our review, is the potential r\^ole of a purely geometric origin of such a metastable dark sector, including both Dark Energy (DE) and Dark Matter (DM), which is associated with the existence of torsion in the geometry of the early-universe~\cite{philtrans,torsionmav,anommav}.

To this end, we consider as a first example, in the next Subsection, string-inspired cosmologies with chiral anomalies. Our generic discussion in Section \ref{SecTorsion} on the r\^ole of (quantum) torsion in Einstein-Cartan QED~\cite{olivetorsion}, will find interesting application in this case. There we argued that, as a generic feature, the torsion degrees of freedom implied the existence of pseudoscalar (axion-like) massless dynamical fields in the spectrum, coupled to chiral anomalies.

\subsection{Quantum Torsion in string-inspired Cosmologies and the Universe Dark Sector}\label{sec:string}

We have seen that in Einstein-Cartan theories, which have been exemplified here by massless contorted QED, torsion conservation \eqref{torsioncons} introduces an axionic degree of freedom to the system, associated with the totally antisymmetric part of the torsion which is the only part that couples to matter (fermions). The axion-like field becomes a dynamical part of the theory as a result of (chiral) anomalies, otherwise it would decouple from the quantum path integral. A similar situation characterises string-inspired theories in which anomalies are not supposed to be cancelled in the (3+1)-dimensional spacetime after string compactification, which, as we shall review below, provide interesting cosmological models~\cite{bms1,bms2,ms1,ms2} in which the dark sector of the Universe, including the origin of its inflationary epoch, admits a geometric interpretation.

The starting point of such an approach to cosmology is that the early Universe is described by the (bosonic) gravitational theory of
the degrees of freedom that constitute the massless gravitational multiplet of
the string (which in the case of superstring is also their ground state). The latter consists of spin-0 dilatons, $\Phi$, spin-2 gravitons $g_{\mu\nu}$, and the spin-1 antisymmetric KR tensor field~\cite{string1,string2} $B_{\mu\nu}=-B_{\nu\mu}$.

Due to an Abelian gauge symmetry that characterises the closed string sector
$B_{\mu\nu} \to B_{\mu\nu} + \partial_{[\mu}\, \theta_{\nu]}$, the $(3+1)$-dimensional effective target spacetime action arising in the low-energy limit of strings (compared to the string mass scale $M_s$) depends only on the totally antisymmetric
field strength of the KR field $B_{\mu\nu}$,
\begin{equation}\label{hfield}
{H}_{\mu\nu\rho} = \partial_{[\mu}\, B_{\nu\rho]}\,.
\end{equation}
As explained in \cite{bms2}, one can assume self consistently a constant dilaton, so that the low-energy particle phenomenology is not affected. In this case, to lowest non-trivial order in a derivative expansion, or equivalent to $\mathcal O((\alpha^\prime)^0)$, with
$\alpha^\prime = M_s^2$ the Regge slope, the effective gravitational action reads~\cite{gross,MT}:
\begin{align}\label{sea2}
S_B =&\; \int d^{4}x\sqrt{-g}\left( \dfrac{1}{2\kappa^{2}}\, R - \frac{1}{6}\, {\mathcal H}_{\lambda\mu\nu}\, {\mathcal H}^{\lambda\mu\nu} + \dots \right)\,,
\end{align}
where  ${\mathcal H}_{\mu\nu\rho} \equiv \kappa^{-1} H_{\mu\nu\rho}$ has dimension [mass]$^2$,
and the $\dots$ represent higher derivative terms.

Comparing \eqref{sea2} with \eqref{gravact} one observes that
the quadratic in the $H$-field terms
can be viewed as a contorsion, in such a way that the effective action
\eqref{sea2} can be expressed in terms of a generalised scalar curvature in a contorted geometry, with a generalised Christoffel symbols:
\begin{align}\label{torcon}
{\overline \Gamma}_{\mu\nu}^{\rho} = \mathring{\Gamma}_{\mu\nu}^\rho + \frac{\kappa}{\sqrt{3}}\, {\mathcal H}_{\mu\nu}^\rho  \ne {\overline \Gamma}_{\nu\mu}^{\rho}~,
\end{align}
where $\mathring{\Gamma}_{\mu\nu}^\rho = \mathring{\Gamma}_{\nu\mu}^\rho$ is the torsion-free Christoffel symbols.\footnote{We note for completeness that, by exploiting local field redefinition ambiguities~\cite{gross,MT,bento,olivetorsion}, which do not affect the perturbative scattering amplitudes, one may extend the above conclusion to the fourth order in derivatives, that is, to the ${\mathcal O}(\alpha^{\prime\,2})$ effective low-energy action, which includes quadratic curvature terms.}

The requirement of cancellation of gauge versus gravitational anomalies lead Green and Schwarz~\cite{gs} to add appropriate counterterms in the effective target space action of strings, expressed by the modification of the
field strength of the KR field \eqref{hfield} by the Lorentz (L) and Yang-Mills (Y) gauge Chern-Simons (CS) terms~\cite{string2}:
\begin{align}\label{csterms}
{\mathcal H} &= d\,B + \frac{\alpha^\prime}{8\, \kappa} \, \Big(\Omega_{\rm 3L} - \Omega_{\rm 3Y}\Big),  \nonumber \\
\Omega_{\rm 3L} &= \omega^a_{\,\,c} \wedge d\,\omega^c_{\,\,a}
+ \frac{2}{3}  \omega^a_{\,\,c} \wedge  \omega^c_{\,\,d} \wedge \omega^d_{\,\,a},
\quad \Omega_{\rm 3Y} = A \wedge  d\,A + A \wedge A \wedge A,
\end{align}
where $\omega$ is the standard torsion-free spin connection, and $  A$ the non-Abelian gauge fields that characterise strings.

The modification (\ref{csterms}) of the KR field strength \eqref{hfield} leads to the following Bianchi identity~\cite{string2}
\begin{equation}\label{modbianchi}
d\,\mathcal{H} = \frac{\alpha^\prime}{8 \, \kappa} {\rm Tr} \Big(R \wedge R - F \wedge F\Big) \;,
\end{equation}
with $F = d\,A + A \wedge  A$ the Yang-Mills field strength two form  and $R^a_{\,\,b} = d\,\omega^a_{\,\,b} + \omega^a_{\,\,c} \wedge \omega^c_{\,\,b}$, the curvature two form and the trace ($\mbox{Tr}$) is over gauge and Lorentz group indices.
The non zero quantity on the right hand side of \eqref{modbianchi} is the ``mixed (gauge and gravitational) quantum anomaly'' we have seen previously in the non-conservation of the axial fermion current \eqref{anomcurr}.\footnote{We stress once again that the modifications (\ref{csterms}) and the right-hand-side of the Bianchi (\ref{modbianchi}) contain the {\it torsion-free} spin connection, given that, as explained previously, any $H$-torsion contribution can be removed by an appropriate addition of counterterms~\cite{toranom2,toranom3}.}

In \cite{bms1} the crucial assumption made was the (3+1)-dimensional gravitational anomalies are not cancelled in the very early Universe. This was the
consequence of the assumption that only fields from the massless gravitational string multiplets characterised the early universe gravitational theory, appearing as external fields. Chiral fermionic matter, radiation and in general gauge fields, which constitute the physical content of the low-energy particle physics models derived from strings,
appear as the result of the decay of the false vacuum at the end of inflation in the scenario of \cite{bms1,bms2,ms1,ms2}.

In this sense, the gauge fields $  A$
in \eqref{csterms} can be set to zero, $A=0$. In such a case, the Bianchi identity \eqref{modbianchi} becomes (in component form):
\begin{equation}\label{modbianchi2}
 \varepsilon_{abc}^{\;\;\;\;\;\mu}\, {\mathcal H}^{abc}_{\;\;\;\;\;\; ;\mu}
 %=   \frac{ \alpha^\prime}{64\, \kappa}\, \varepsilon_{\mu\nu\rho\sigma}  \Big(R_{ab}^{\,\,\,\,\,\,\,\mu\nu}\, R_{\rho\sigma ab} - F^{\mu\nu}\, F^{\rho\sigma} \Big)
 =  \frac{\alpha^\prime}{32\, \kappa} \, \sqrt{-g}\, R_{\mu\nu\rho\sigma}\, \widetilde R^{\mu\nu\rho\sigma}  \equiv - \sqrt{-g}\, {\mathcal G}(\omega),
\end{equation}
where the semicolon denotes covariant derivative with respect to the standard
Christoffel connection, and
\begin{equation}\label{leviC}
\varepsilon_{\mu\nu\rho\sigma} = \sqrt{-g}\,  \epsilon_{\mu\nu\rho\sigma}, \quad \varepsilon^{\mu\nu\rho\sigma} =\frac{{\rm sgn}(g)}{\sqrt{-g}}\,  \epsilon^{\mu\nu\rho\sigma},
\end{equation}
with $\epsilon^{0123} = +1$, {\emph etc.}, are the gravitationally covariant Levi-Civita tensor densities, totally antisymmetric in their indices, and the dual is defined as
\begin{align}\label{duals}
\widetilde R_{\mu\nu\rho\sigma} = \frac{1}{2} \varepsilon_{\mu\nu\lambda\pi} R_{\,\,\,\,\,\,\,\rho\sigma}^{\lambda\pi}\,.
\end{align}
The alert reader should have observed similarities between the contorted QED model, examined in the previous Section \ref{sec:TorsQED}, and the string inspired gravitational theory, insofar as the constraints imposed by the torsion conservation \eqref{torsioncons} in the QED case, and the Bianchi
constraint \eqref{modbianchi2}. They are both exact results that are valid in the quantum theory (the Bianchi \eqref{modbianchi2} is an exact one-loop result due to the nature of the chiral anomalies).
In fact the dual of $H^{\mu\nu\rho}$, $\varepsilon_{\mu\nu\rho\sigma} \, H^{\nu\rho\sigma}$ plays a r\^ole analogous with the pseudovector $S_\mu$ of the contorted QED case, associated with the
totally antisymmetric
component of the torsion.
In the string theory example, this is all there is from torsion, as we infer from \eqref{torcon}.

Following the contorted QED case, one may implement the Bianchi constraint \eqref{modbianchi2} via a $\delta$-functional in the corresponding path integral, represented by means of an appropriate Lagrange multiplier pseudoscalar field $b(x)$, canonically normalized:
\begin{align}\label{deltastring}
&\Pi_{x}\, \delta\Big(\varepsilon^{\mu\nu\rho\sigma} \, {{\mathcal H}_{\nu\rho\sigma}(x)}_{; \mu} + {\mathcal G}(\omega) \Big)
\Rightarrow  \nonumber \\ &\int {\mathcal D}b \, \exp\Big[i \, \,\int d^4x \sqrt{-g}\, \frac{1}{\sqrt{3}}\, b(x) \Big(\varepsilon^{\mu\nu\rho\sigma }\, {{\mathcal H}_{\nu\rho\sigma}(x)}_{; \mu} - {\mathcal G}(\omega) \Big) \Big] \nonumber \\
&= \int {\mathcal D}b \, \exp\Big[-i \,\int d^4x \sqrt{-g}\, \Big( \partial ^\mu b(x) \, \frac{1}{\sqrt{3}} \, \epsilon_{\mu\nu\rho\sigma} \,{\mathcal H}^{\nu\rho\sigma}  + \frac{b(x)}{\sqrt{3}}\, {\mathcal G}(\omega) \Big)\Big] \;,
\end{align}
where to arrive at the second equality we performed partial integration, upon assuming that fields die out properly at spatial infinity, so that no boundary terms arise.
We remark at this point that the similarity~\cite{torsionmav} of the exponent in the right-hand side of the last equality in
\eqref{deltastring}, upon performing a partial integration of the first term, and identifying the anomaly with $\partial_\mu j^{5\mu}$,
with the total Holst action (including the Nieh-Yan invariant) \eqref{holsttot},
in the case where the BI parameter is promoted to a pseudoscalar field~\cite{calcagni}.

Inserting the identity \eqref{deltastring} in the path
integral over $H$ of the theory
\eqref{sea2}, we observe that the equations of motion of the (non-derivative) field $H$ yield
$\epsilon_{\mu\nu\rho\sigma} \, H^{\nu\rho\sigma} \propto \partial_\mu b$, implying an analogy of the pseudovector field $S_\mu$ with $\partial_\mu b$. After path-integrating out the $H$-torsion, one obtains an effective target space action with a dynamical torsion-induced axion $b$:
\begin{align}\label{sea3}
S^{\rm eff}_B =&\; \int d^{4}x\sqrt{-g}\Big[ \dfrac{1}{2\kappa^{2}}\, R + \frac{1}{2}\, \partial_\mu b \, \partial^\mu b +  \sqrt{\frac{2}{3}} \, \frac{\alpha^\prime}{96\, \kappa} \, b(x) \, R_{\mu\nu\rho\sigma}\, \widetilde R^{\mu\nu\rho\sigma}  + \dots \Big],
\end{align}
where the dots $\dots$ denote higher derivative terms appearing in the target-space string effective action~\cite{gross,MT,olivetorsion}.

With the exception of the four-fermion interactions, which are absent here, as the theory is bosonic, the action \eqref{sea3} has the same form as the effective action
\eqref{axionanom}, with the pseudoscalar field $b$ having similar origin related to torsion as its contorted QED counterpart. But the action \eqref{sea3} is purely bosonic, and the anomalies here arise from the Green-Schwarz counterterms \eqref{csterms}. In the model of \cite{bms1} these are primordial anomalies, unrelated to chiral matter fermions as in the QED case, but because of the presence of such anomalies, the torsion (through its dual axion field  $b(x)$) maintains its non trivial r\^ole via its coupling to the gravitational anomaly CS term.
The gravitational model \eqref{sea3} is a Chern-Simons modified gravity model~\cite{JackiwPiChernSimons,yunes}.

The massless axion field $b(x)$ is the so-called string-model independent axion~\cite{svrcek}, and is one of the many axion fields that string models have. The other axions are due to compactification. The string axions lead to a rich
phenomenology and cosmology~\cite{arvanitaki,marsh}.

\begin{figure}[ht]
\begin{center}
\includegraphics[width=\textwidth]{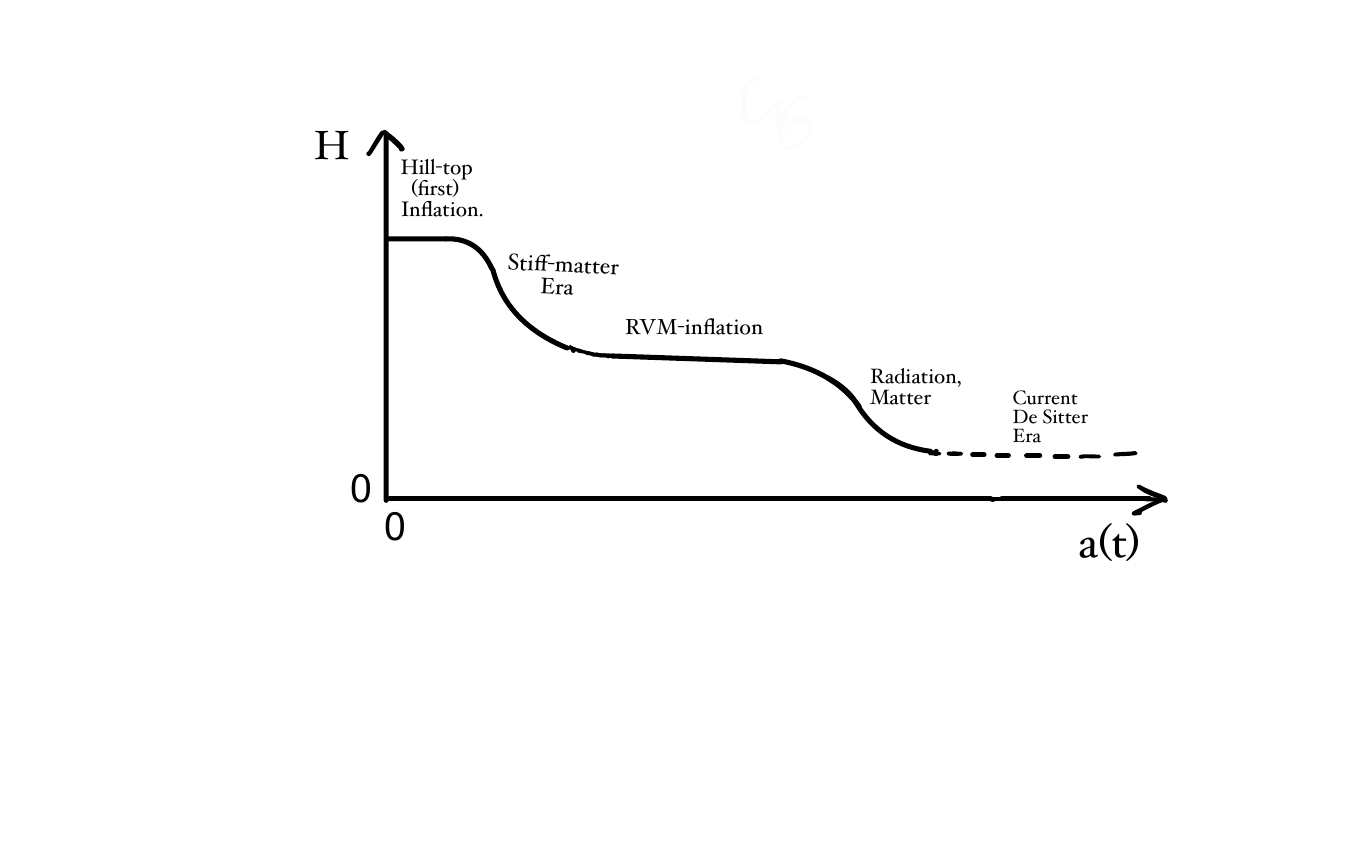}
\end{center}
\vspace{-2.5cm}
\caption{Schematic representation of the RVM cosmological evolution of the contorted cosmological model
of \cite{bms1,bms2,ms1,ms2}. The figure depicts the evolution of the Hubble parameter with the scale factor of an expanding stringy-RVM Universe, involving two torsion-induced inflationary eras, interpolated by a stiff KR-axion ``matter'' epoch: a first hill-top first inflation, which exists immediately
after the Big-Bang, and is due to dynamical breaking of SUGRA, as a result of gravitino-torsion-induced condensates of the gravitino field, and second an RVM inflation, due to gravitational anomaly condensates, that are coupled to the torsion-induced KR axion field $b(x)$. The latter can also play the role of a dark-matter component during post-RVM inflationary eras. Picture taken from \cite{ms2}.}
\label{fig:Hubbleevol}
\end{figure}

From our point of view we restrict ourselves to the r\^ole of the KR axion in implying a geometric origin of the dark sector of the Universe, including non conventional inflation.
Indeed, in \cite{bms1,bms2,ms1,ms2} it was argued that condensation of primordial gravitational waves (GW) leads to a non-vanishing contribution of the gravitational Chern-Simons term $\langle R_{\mu\nu\rho\sigma}\, \widetilde R^{\mu\nu\rho\sigma}\rangle$, where $\langle \dots \rangle$ denote weak graviton condensates associated with primordial chiral GW~\cite{stephon,lyth}. If one assumes a density of sources for primordial GW, which have been formed in he very early Universe, before the inflationary stage in the model of \cite{ms1,ms2}, then, the weak quantum graviton  calculation of \cite{lyth}, adopted to include densities of GW sources, leads~\cite{mavrosourcegw}:
 \begin{align}\label{condensateN2}
\langle R_{\mu\nu\rho\sigma} \, \widetilde R^{\mu\nu\rho\sigma} \rangle_{\rm condensate\, \mathcal N}
=\frac{\mathcal N(t)}{\sqrt{-g}}  \, \frac{1.1}{\pi^2} \,
\Big(\frac{H}{M_{\rm Pl}}\Big)^3 \, \mu^4\, \frac{\dot b(t)}{M_s^{2}} \equiv n_\star \, \frac{1.1}{\pi^2} \,
\Big(\frac{H}{M_{\rm Pl}}\Big)^3 \, \mu^4\, \frac{\dot b(t)}{M_s^{2}}~.
\end{align}
In the above expression, $\mu$ is an Ultraviolet (UV) cutoff for the graviton modes entering the chiral GW, and $n_\star \equiv \frac{\mathcal N(t)}{\sqrt{-g}} $ denotes the number density (over the proper de Sitter volume)  of the sources of GW. Without loss of generality, we may take this density to be (approximately) time independent during the very early universe. The parameter $H(t)$ is the Hubble parameter of a FLRW Universe, which is assumed slowly varying with the cosmic time.\footnote{To ensure homogeneity and isotropy conditions, the authors of \cite{ms1} assumed the existence of a stiff-axion-$b$-dominated era ({\it i.e.} with equation of state $w_b=+1$) that succeeds a first hill-top inflation~\cite{hilltopEllis} ({\it cf.} Fig.~\ref{fig:Hubbleevol}), which is the result of dynamical breaking of local SUSY (SUGRA) right after the Big Bang, that is assumed to characterise the superstring inspired theories. This breaking is achieved by a condensation of the gravitino (supersymmetric partner of gravitons) as a result of the existence of attractive channels in the four-gravitino interactions that characterise the SUGRA Lagrangian due to fermionic torsion~\cite{Alexandre1,Alexandre2}, as discussed in Section \ref{sec:sugra}. As argued in \cite{ms1,ms2}, unstable domain walls (DW) are formed as a result of the gravitino condensate double well potential (Fig.~\ref{fig:brsugra}), whose degeneracy can be lifted by percolation effects~\cite{ross}. The non-spherical collapse of such DW leads to primordial GW, which then condense leading to \eqref{condensateN2}.} The analysis of \cite{bms1,ms1} then, shows that there is a metastable de Sitter spacetime emerging, given that the condensate \eqref{condensateN2} is only mildly depending on cosmic time through $H(t)$ mainly, and thus can be considered approximately constant.
It can be shown~\cite{bms1}, that as a consequence of the axion $b$ equations of motion, the existence of a condensate leads to approximately constant $\dot b$ during the inflationary period (for which $H \simeq  $ constant) 
\begin{align}\label{constbdot}
\dot b \simeq \epsilon H \, M_{\rm Pl} \;,
\end{align}
where the overdot denotes derivative with respect to the cosmic time $t$. The parameter $\epsilon$ is phenomenological and to satisfy the Planck data~\cite{Planck} on slow-roll inflation, one should set it to $\epsilon = \mathcal O(10^{-2})$~\cite{ms1}.
Then conditions for an approximately constant
\begin{align}\label{LbCS}
\langle b(t)\, R_{\mu\nu\rho\sigma} \, \widetilde R^{\mu\nu\rho\sigma} \rangle_{\rm condensate\, \mathcal N} \simeq {\rm constant} \;,
\end{align}
for some period $\Delta t $ can be ensured,
which then leads to a {\it metastable} de Sitter spacetime (inflation), with $\Delta t$ the duration of inflation. Taking into account that the scale of inflation, set by the current Planck data~\cite{Planck} is
\begin{align}
 H_I \lesssim 10^{-5}\, M_{\rm Pl} \;,
\end{align}
and that the
the number of e-foldings is estimated to be (in single-field models of inflation) $\mathcal N=\mathcal O(60-70)$, these conditions can be stated as:
\begin{align}
 |\overline b(t_0)| \gtrsim N_e \, \sqrt{2\epsilon} \, M_{\rm Pl}\, = \mathcal O(10^2)\,\sqrt{\epsilon}\, M_{\rm Pl} \;,
\end{align}
with $b(t_0)$ the initial value of the axion field at the onset ($t=t_0$) of inflation.

In view of the $H$-dependence of the condensate the inflation is of the so-called Running-Vacuum-Model (RVM) type~\cite{sola1,sola2,sola3,lima1,lima2,sola4}, which involves a time-dependent, rather than a constant de Sitter parameter $\Lambda(t) \propto H^2(t)$, but with a de Sitter equation of state for the vacuum:
\begin{align}\label{rvmeos}
  p_{\rm rvm} = - \rho_{\rm rvm} \;,
\end{align}
where $p$ ($\rho$) denotes pressure (energy) density.
In the model of \cite{ms1}, detailed calculations have shown that in the phase of the GW-induced condensate \eqref{condensateN2}, \eqref{LbCS}, the de Sitter-RVM equation of state \eqref{rvmeos} is satisfied. The corresponding energy density, comprising of contributions from $b$ field (superscript $b$), the gravitational CS terms (superscript gCS) and the condensate term (superscript) $\Lambda$), acquires~\cite{bms1,bms1,ms1,ms2,mavrosourcegw} the familiar RVM form~\cite{lima1,lima2,sola4}
\begin{align}\label{totalenerden}
\rho^{\rm total} = \rho^b + \rho^{\rm gCS} + \rho_{\rm condensate}^\Lambda = -\frac{1}{2}\, \epsilon \, M_{\rm Pl}^2\, H^2 + 4.3 \times 10^{10} \, \sqrt{\epsilon}\, \frac{|\overline b(0)|}{M_{\rm Pl}} \, H^4\,.
\end{align}

The important point to notice is that the RVM inflation does not require a fundamental inflaton scalar field, but is due to the non-linear $H^4$ terms in the respective vacuum energy density \eqref{totalenerden}~\cite{lima1,lima2,sola4}, arising in our case by the form of the condensate~\eqref{condensateN2}. Such terms are dominant in the early Universe and drive inflation. During the RVM inflation in our string-inspired CS gravity the $H^2$ term is negative in
contrast to standard RVM formalisms with a smooth evolution from inflation to the current era~\cite{lima1,lima2}. In our case, it is the CS quadratic curvature corrections to GR that leads to such negative contributions tom the stress-energy tensor, in full analogy to the dilaton-Gauss-Bonnet string-inspired theories~\cite{kanti}. Nevertheless, the dominance of the condensate (i.e. $\mathcal O(H^4))$ terms in \eqref{totalenerden} ensures the positivity of the vacuum energy density during the RVM inflationary era. We stress that the $H^4$ term in the vacuum energy density \eqref{totalenerden} arises exclusively from the gravitational anomaly condensate in our string-inspired cosmology. In standard quantum field theories in curved spacetime, RVM energy densities arise after appropriate renormalization of the quantum matter fields in the FLRW spacetime background, but in such cases an $H^4$ term is {\it not} generated in the vacuum energy density. Instead one has the generation of order $H^6$ terms and higher~\cite{sola4,pul1,pul2,pul3,pul4}. Such non linear terms, which will be dominant in the early Universe, can still, of course, drive RVM inflation.

During the final stages of RVM inflation, the decay of the RVM metastable vacuum~\cite{lima1,lima2} results in the generation of chiral matter fermions in the cosmology model of \cite{bms1,bms2,ms1,ms2} we are analysing here. The chiral fermions would generate their own mixed (gauge and gravitational) chiral anomaly terms through the non conservation of the chiral current \eqref{totcurr} over the various chiral fermion species (\ref{anomcurr}). The effective action during such an era will, therefore, contain fermions, which will couple universally to the torsion $H_{\mu\nu\rho}$ via the diffeomorphic covariant derivative.
After integrating out the $H$-field, we arrive at the following effective action including fermions~\cite{bms1}:
 \begin{align}\label{sea6}
&S^{\rm eff} =\; \int d^{4}x\sqrt{-g}\Big[ \dfrac{1}{2\kappa^{2}}\, R + \frac{1}{2}\, \partial_\mu b \, \partial^\mu b  + \sqrt{\frac{2}{3}}\,
\frac{\alpha^\prime}{96\, \kappa} \, b(x) \, R_{\mu\nu\rho\sigma}\, \widetilde R^{\mu\nu\rho\sigma}\Big]  \nonumber \\
&+ S_{\rm Dirac~or~Majorana}^{Free} + \int d^{4}x\sqrt{-g}\, \Big[\Big( {\mathcal F}_\mu - \frac{\alpha^\prime}{2\, \kappa} \, \sqrt{\frac{3}{2}} \, b \,J_{\quad ;\mu}^{5\mu} \Big)\,   - \dfrac{3\alpha^{\prime\, 2}}{16 \, \kappa^2}\,J^{5}_{\mu}J^{5\mu} \Big] + \dots \;,
\end{align}
where the $S_{\rm Dirac~or~Majorana}^{Free}$ fermionic terms denote the standard Dirac or Majorana fermion kinetic terms in a curved spacetime without torsion,
and ${\mathcal F}^a  =   \varepsilon^{abcd} \, e_{b\lambda} \,  \partial_d \, e^\lambda_{\,\,c}$.

The gravitational part of the anomaly is assumed in \cite{bms1} to {\it cancel} the primordial gravitational anomalies, but the chiral gauge anomalies remain in general. Thus in \cite{bms1} we assumed that at the exit phase from RVM inflation one has the condition:
\begin{align}
   \label{anom2}
   \partial_\mu \left[\sqrt{-g}\, \left(  \sqrt{\frac{3}{8}} \kappa\, J^{5\mu}  -  \sqrt{\frac{2}{3}}\,
\frac{\kappa}{96} \, {\mathcal K}^\mu  \right) \right] \! & =\!   \sqrt{\frac{3}{8}}\, \frac{\alpha^\prime}{\kappa}\,  \frac{e^2}{8\pi^2}  \, \sqrt{-g}\,  {F}^{\mu\nu}\,  \widetilde{F}_{\mu\nu}  \nonumber \\ &+  \sqrt{\frac{3}{8}} \,\frac{\alpha^\prime}{\kappa}\, \frac{\alpha_s}{8\pi}\, \sqrt{-g} \, G_{\mu\nu}^a \, \widetilde G^{a\mu\nu} \,,
\end{align}
where we used the fact that the gravitational CS anomaly is a total derivative of an appropriate topological current $\mathcal K^\mu$~\cite{anomalies1,anomalies2,anomalies3},
\begin{align}\label{totderK}
R_{\mu\nu\rho\sigma}\,\widetilde R^{\mu\nu\rho\sigma} = \mathcal K^\mu_{\,\,\,;\mu}\,,
\end{align}
$F_{\mu\nu}$ denotes the electromagnetic U(1) Maxwell tensor,
which corresponds to radiation fields in the post inflationary epoch,
and $G_{\mu\nu}^a$, $a=1, \dots 8$ is the gluon tensor associated with the SU(3) (of colour) strong interactions with (squared) coupling $\alpha_s = g_s^2/(4\pi)$, which dominate the Universe during the QCD epoch,
and the $\widetilde{(\dots)}$ denotes the corresponding duals, as usual ({\it cf.} \eqref{duals}), with $\widetilde F^{\mu\nu} = \frac{1}{2} \varepsilon^{\mu\nu\rho\sigma}\, F_{\rho\sigma}$.

At the exit from RVM inflation, it was assumed in \cite{bms1,bms2,ms1,ms2} that no chiral gauge anomalies are dominant. Such dominance comes much later in the post inflationary Universe evolution. In such a case, it can be shown~\cite{bms1} that the $b$-field equation of motion implies a scaling of $\dot b$ with the temperature as
\begin{align}\label{bT3}
\dot b \propto T^3\,.
\end{align}
In this case one may obtain an unconventional leptogenesis of the type discussed in \cite{lepto1,lepto2} in theories involving massive sterile right handed neutrinos, as a result of the decay of the latter to standard-model particles in the presence of the Lorentz-violating background \eqref{bT3}. Hence, in such scenarios the torsion is also linked to matter-antimatter asymmetry, given that the so-generated lepton asymmetry can be communicated to the baryon sector vial Baryon (B) and Lepton number (L) violating, but B-L conserving sphaleron processes in the standard-model sector~\cite{sarkarbaryo}.

Connection of torsion to DM might be obtained by noting that the QCD dominance era (which in the models of \cite{bms1,bms2} comes much after the leptogenesis epoch)
might be characterised by SU(3) instanton effects, which in turn break the axionic shift symmetry by inducing appropriate potential, and mass terms, ({\it cf.} \eqref{inst} ) for the torsion-induced axion field $b$, which could play a r\^ole as a DM component. The electromagnetic U(1) chiral anomalies may be dominant in the modern eras, and their effects have been discussed in detail in \cite{bms1}.

 We also mention for completion that, as a result of the (anomalous) coupling $\dot b J^{5\, 0}$ ({\it cf.} \eqref{sea6}), one obtains a Standard-Model-Extension (SME) situation, with the Lorentz and CPT Violating SME background being provided by $\dot b$. It is the latter that is constrained by a plethora of precision experiments, which provide stringent bounds for Lorentz and CPT violation~\cite{smebounds}.
Using the chiral gauge anomalies at late eras of the Universe,
as appearing in \eqref{anom2},
the thermal evolution of the Lorentz- and CPT- symmetry-Violating torsion-induced background $\dot b (T)$ at late eras of the Universe, including the current epoch, has also been estimated in \cite{bms1}, and found to be comfortably consistent with the aforementioned existing bounds of  Lorentz and CPT Violation, as well as torsion today~\cite{smebounds}.

In the above cosmological scenarios, the entire dark sector of the Universe and its cosmological evolution are one way or another linked to some sort of torsion in the geometry. During the very early epochs after the Big bang, it is the gravitino torsion of a SUGRA theory, which the effective string cosmology model  of \cite{bms1,ms1} is embedded to, that leads to a first inflationary epoch~\cite{hilltopEllis}, whilst it is the stringy torsion associated with the field strength of the antisymmetric spin-one KR field, which in turn gives rise to the KR axion $b(x)$, that is responsible for the second RVM type inflation, and the eventual cosmological evolution until the present era, during which the field $b(x)$ can also develop a mass, thus becoming a dark-matter candidate. Schematically, such a cosmological evolution is depicted in Fig.~\ref{fig:Hubbleevol}~\cite{ms2}.

Before closing this section, we would like to mention the very recent related work of \cite{Capanelli:2023uwv}, which explores further the Cosmology of Kalb-Ramond-like particles (KRLP), which one encounters in string models,  and which contain also massive pseudovector excitations, in addition to the massless pseudoscalar ones, discussed in this review in connection with the totally antisymmetric part of torsion. Although the non-interacting KRLP
are related to either pseudoscalar or pseudovector excitations, the interacting massive KRLP can be distinguished from its scalar and vector counterparts, and can have important phenomenological implications for the dark sector of the Universe, which are described in detail in \cite{Capanelli:2023uwv}.

\subsection{Comments on other contorted cosmological models with a spin}\label{sec:cosmospin}

In the previous Subsection we discussed cosmological models corresponding to the standard generic type of Einstein-Cartan theories with fermionic torsion, involving in their Lagrangian densities repulsive four fermion interactions, of axial-current-current terms $j^5-j^5$, with fixed coefficient depending on the theory, proportional to the gravitational coupling $\kappa^2$. Condensates of such repulsive terms, when formed, have been interpreted  as providers of dark energy components in both the early~\cite{poplinfl} and the late~\cite{poplcosm} Universe, thus leading to a current-era acceleration of the Universe.

In this Subsection we shall discuss briefly generalizations, involving more general four-fermi structures among {\it chiral} (Weyl) spinors~\cite{kibble} , which include vector fermionic currents in addition to the axial ones, in similar spirit to the models \eqref{fermionnonder}, but with more general coefficients (on the other, hand, unlike the situation encountered in \eqref{fermionnonder}, the BI parameter in \cite{kibble} is assumed constant, which, as we have discussed in Section \ref{SecTorsion}, and mention below as well, is a problematic feature). Depending on the couplings considered, such fermion self-interactions may conserve or break  parity invariance, while they may contribute positively or negatively to the energy density, thus having the feature that they could also be attractive. Thus, such  ``cosmologies with a spin''~\cite{kibble} exhibit a broad spectrum of possibilities, ranging from cases for which no significant cosmological novelties arise, to cases in which the fermion self-interaction can turn a mass potential into an upside-down Mexican hat potential, leading to cosmologies with a bounce~\cite{kibble,poplbounc}, without a cosmic singularity.

However, as we shall discuss below, there are some subtleties in the treatment of \cite{kibble}, which,
in view of what we discussed in Section \ref{SecTorsion}, require some discussion.
Let us first describe the approach of \cite{kibble}.
On defining Dirac spinors $\Psi(x)$ from the chiral ones $\xi, \chi$ as
\begin{align}
  \Psi(x) = \begin{pmatrix} \xi(x) \\\chi(x)\end{pmatrix}\,,
\end{align}
the authors of \cite{kibble} constructed fermionic field theories in a contorted curved spacetime, with action given by:
\begin{align}\label{cfact}
 \mathcal S_\Psi [e, \omega^{ab}, \Psi] &= \frac{1}{3}\, \int d^4x \,
 \epsilon_{abcd} \, e^b\, e^c \Big[ \frac{1}{2} \, e^a \, \overline{\Psi} \, \gamma^d D (\omega) \, \Psi - \overline{D(\omega) \Psi} \, \gamma^d \, \Psi) + \frac{3}{2} \, T^a (\alpha \, V^d + \beta A^d ) \Big] \nonumber \\
 &-\frac{1}{4} \int d^4x \, U\, \epsilon_{abcd} \, e^a\, e^b\, e^c\, e^d + S_{\rm int}[\xi,\chi, A]\,,
\end{align}
where $T^a$ is the torsion two-form, \eqref{tordef}, $U$ is a fermion-self-interaction potential which is assumed to be a function of scalars constructed from $\overline{\Psi}\, \Psi$ and $\overline{\Psi}\, \gamma^5\, \Psi$, while $S_{\rm int}$ denotes an interaction term of the chiral spinors $\xi, \chi$ with (in general, non-Abelian) gauge fields $A$. We also defined
$V^d = \overline{\Psi} \gamma^d \Psi$ as the vector chiral current, and $A^d = \overline{\Psi} \gamma^5 \, \gamma^d\, \Psi$ its axial counterpart. Finally, the quantities $\alpha, \beta \in \mathbb  R$ are real couplings that characterise the model.

The gravitational dynamics, on the other hand, is described by the standard Einstein-Hilbert term plus the Holst action, this is the combination \eqref{torsiondecompaction} and \eqref{holst}, which in the parametrization and normalization of \cite{kibble} is written as:
\begin{align}
\mathcal S_{\rm grav+Holst} = \frac{1}{2\kappa^2}\, \int d^4 x \Big(\epsilon_{abcd} + \frac{1}{\gamma} \eta_{ac}\, \eta_{bd}\Big)\, e^a\, e^b \, R^{cd}\,,
\end{align}
with $R^{ab}$ the Riemann curvature two-form, and $\gamma \in \mathbb R$ is related to the BI parameter $\beta=-1/\gamma$ (\ref{holst}).

This is not a minimal torsion model, as the generic Einstein-Cartan theories examined before, given that it includes several postulated interaction potentials.
Because of this, this model leads to more general four-fermion interactions than the standard Einstein-Cartan theory. The effective four fermion interaction is found by using, as in the standard Einstein-Cartan theories, the Euler-Lagrange equations of motion for the fermions, torsion and gravity fields. By varying the action with respect to the contorted spin connection, we determine the torsion $T^a$ and contorsion $K_{abc}$ for this model~\cite{kibble}:
\begin{align}
&\frac{1}{\kappa^2} \, \Big(\epsilon_{abcd} + \frac{2}{\gamma}\, \eta_{a[c}\, \eta_{d]b}\Big) \, T^a\, e^b = \frac{1}{4\!}
\, \epsilon_{amnp} \, e^a\, e^m\, e^n\, \epsilon^{dp}_{\quad cd}\, A_d  - \frac{1}{4\!} \epsilon_{[c|mnq}\, e^m\, e^n e_{|d]}\Big(\alpha \, V^q + \beta A^q\Big)\,,
\nonumber \\
&K_{abc} = \kappa^2 \frac{\gamma^2}{4 \, (\gamma^2 +1)}\, \Big[\epsilon^d_{\,\,\,abc}\frac{1}{2}\Big(A_d + \frac{1}{\gamma} (\alpha V_d + \beta A_d) \Big) - \frac{1}{\gamma}\, A_{[b}\, \eta_{a]c} +
\alpha \, V_{[b}\,\eta_{a]c}  + \beta\,
A_{[b}\,\eta_{a]c}\Big]\,.
\end{align}
From the graviton (vielbein) and fermion equations of motion, on the other hand, we obtain, respectively:
\begin{align}
&\frac{2}{\kappa^2} \, \mathring{G}_{\mu\nu} = -\frac{\ii}{2} e_{d\mu} (\overline{\Psi}\, \gamma^d \, \mathring{D}_\nu \, \Psi - \overline{\mathring{D}_\nu \Psi}) \, \gamma^d \, \Psi ) + \frac{\ii}{2} e^\sigma_{\, d} (\overline{\Psi}\, \gamma^d \, \mathring{D}_\sigma \, \Psi - \overline{(\mathring{D}_\sigma \Psi} \, \gamma^d \, \Psi ) - g_{\mu\nu} \, W\,, \nonumber \\
& i\gamma^d \, e^\mu_{\, d} \, \mathring{D}_\mu \, \Psi =\frac{\delta W}{\delta \overline \Psi}\,,
\end{align}
where $\mathring{G}_{\mu\nu}$ is the standard Einstein tensor of GR, to ease the notation we used $\mathring{D}_\mu \equiv D(\mathring{\omega})_\mu$ and $W$ is the effective four-fermion interaction potential, which depends on the contorsion:
\begin{align}\label{Wpot}
 W = U + \frac{3\, \kappa^2}{16} \, \frac{\gamma^2}{\gamma^2 + 1} \, \Big[ (1-\beta^2 + \frac{2}{\gamma}\, \beta) A_a\, A^a  - \alpha^2 \, V_a\, V^a - 2\alpha (\beta - \frac{1}{\gamma})\, A_a\, V^a \Big]\,.
\end{align}
The mixed axial-vector current term in \eqref{Wpot} breaks parity.
One should compare these four-fermion interactions with the ones in the models \eqref{fermionnonder}, discussed in Section \ref{sec:immiraxion}.

However, the analysis of \cite{kibble} leading to \eqref{Wpot} is not entirely formally correct, as we have explained in Section \ref{SecTorsion}, following the careful analysis of \cite{mercuri}. The presence of the (constant) BI parameter in the effective potential \eqref{Wpot} would imply that a parameter that appears in a total derivative term does affect physics at the end. As explained above, this paradox leads also to another inconsistency, that  of equation \eqref{false}, in which, for non-zero $1/\gamma$, one obtains the inconsistent result that the vector component of torsion is proportional to the pseudovector of the axial current. As we discussed in Section \ref{SecTorsion}, the resolution of this paradox is achieved by considering the addition of the Nieh-Yan topological invariant~\cite{Nieh} \eqref{holsttotalterm}.

We do mention at this stage that, naively, the independence of the potential $W$ on the (constant) BI parameter $\gamma$ can be achieved in the specific cases
\begin{align}
\beta= \frac{1}{\gamma} \quad {\rm and} \quad
\alpha^2 = c_0^2 \frac{\gamma^2 + 1}{\gamma^2}\,,
\end{align}
where $c_0 \in \mathbb  R$ is an arbitrary real constant.  This case preserves parity, since the mixed term $A_a\, V^a$ in the potential $W$ \eqref{Wpot} is absent. In such a case the effective four-fermion interactions become
\begin{align}\label{Wpotfinal}
 W = U + \frac{3\kappa^2}{16} \Big(A_a\, A^a - c_0^2 \, V_a\, V^a\Big)\,.
\end{align}
This model, contains, in addition to the potential term $U$, the standard repulsive axial-current-current four-fermion interactions of the Einstein-Cartan theory, augmented by vector-current-current four fermion interactions.

Superficially looking at \eqref{Wpotfinal}, one may think that
the contributions to the vacuum energy density due to such interactions could be positive or negative, depending on the relative magnitude of the parameter $c_0^2$, and in general the terms in \eqref{Wpotfinal}.
However, this is not the case. Indeed, as discussed in \cite{kibble}, for classical spinors, as appropriate for solutions of Euler-Lagrange equations of motion, one may argue that
\begin{align}\label{timelike}
 \langle A_a \, A^a \rangle = - \langle V_a \, V^a \rangle\,,
\end{align}
given that the axial term is always space-like, while the vector time-like. From \eqref{timelike} and \eqref{Wpotfinal}
we obtain that
in this case
$W = U +
\frac{\kappa^2}{16} \, (1 + c_0^2)\,A_a\, A^a $
and due to the space-like nature of the classical axial-current-current term $\langle A_a \, A^a\rangle$, the four-fermion interaction is always repulsive, as in the standard Einstein-Cartan theory,
but with a coefficient whose magnitude is unconstrained, given the phenomenological nature of the parameter $c_0$.
In that case, one can show that there are no bouncing cosmologies or other effects, such as for instance turning a positive mass potential into a Higgs one, which arose in the treatment of \cite{kibble}. Nonetheless, doubt is cast on the mathematical consistency of such solutions in view of \eqref{false}, which is still valid in such special cases, even if the potential \eqref{Wpotfinal} is independent from the BI parameter.

The above criticisms, however, may be bypassed in the case one promotes the BI parameter to a pseudoscalar (axion-like) field $1/\gamma \to a(x)$, as discussed previously in Section \ref{sec:immiraxion}.
Indeed in such a case, the corresponding effective four-fermion interactions \eqref{Wpot} have to be reworked in accordance with the fact that the BI parameter is now a fully fledged pseudoscalar field, as in the case of the action \eqref{fermionnonder}.
Thus, cosmologies based on such models, with four-fermion interactions that may include {\it attractive} fermion channels,
may justify (some of) the expectations of \cite{neubert} on the r\^ole of torsion-induced {\it fermion condensates} in the early universe cosmology, which cannot characterise the  repulsive terms \eqref{4fermi}. In this latter respect, in the context of SUGRA theories ({\it cf.} Section \ref{sec:sugra}), the torsion-connected four gravitino interactions can also lead, due to the existence of attractive channels, to the formation of appropriate condensates~\cite{Alexandre1,Alexandre2}, which, as we have discussed in Section \ref{sec:string},  may play an important role in the early eras of string-inspired cosmologies.

%%%%%%%%%%%%%%%%%%%%%%%%%%%%%%%%%%%%%%%%%%%%%%%%%%%%%%%%%%%%%%%%%%%%%%%%%%
\section{Concluding remarks: other observational effects of torsion}\label{SecConclusions}
%%%%%%%%%%%%%%%%%%%%%%%%%%%%%%%%%%%%%%%%%%%%%%%%%%%%%%%%%%%%%%%%%%%%%%%%%%

We reviewed various aspects of torsion, both in emergent geometric descriptions of  graphene, or other Dirac materials,  and in fundamental theories of spacetime, especially cosmology. These two scenarios have enormously different scales, yet the physical properties of torsion appear to be universal, and can in principle be appreciated in experiments in both frameworks.

On the cosmological side, we focused on specific string-inspired models in which the totally antisymmetric component of torsion is represented as an axion-like field. Condensates associated with torsion can lead, as we have discussed, to inflationary physics of RVM type, characterised under some conditions, by torsion-induced-axion background that spontaneously violates  Lorentz symmetry. Such a situation may leave imprints in the early Universe cosmic microwave background.

In general, however, in generic Einstein-Cartan theories,  torsion has more components. In \cite{Kostelecky:2007kx}, a plethora of tests involving coupling of the various torsion components to fermions
in combination with Lorentz violation, in the context of the Standard Model Extension~\cite{smebounds},
have been discussed which exhibit sensitivity for some of the pertinent Lorentz-violating parameters down to $10^{-1}$~GeV.

The presence of torsion may also have important consequences for cosmological observations independent of Lorentz violation. For instance, as discussed in \cite{Bolejko:2020nbw}, non-zero torsion affects the relation between the angular-diameter ($D_A$) and luminosity ($D_L$) distances used in astrophysical/cosmological measurements, such that the quantity $\eta= \frac{D_L}{D_A (1+z)^2}-1$ is linked to various types of torsion. This may affect low-redshift measurements, and thus contribute to the observed Hubble-parameter ($H_0$) tensions~\cite{Aluri:2022hzs}. Of course, contributions to such tensions, including the growth of structure ones ($\sigma_8$)~\cite{tensions1,tensions2,tensions3}, can also come, as we discussed in Section \ref{sec:string}, from the late-Universe RVM cosmology, which the contorted string-inspired models lead to, but the combination of the plethora of late-time cosmological measurements, and details of structure formation~\cite{Gomez-Valent:2023hov} can provide information that can distinguish between the quantum string-inspired RVM cosmology and generic torsion models.

Other constraints on late-Universe torsion of relevance to our discussion here, namely of associating axions to torsion, come from CP (rather than Lorentz) violation effects in axion-photon cosmic plasma through dynamo primordial-magnetic-field amplification~\cite{GarciadeAndrade:2011zzc} (see also \cite{Campanelli:2005ye} on the role of axion fields), which torsion is a specific species of for cosmic magnetic helicity generation).

An alternative way to probe experimentally the role of torsion is to realize in graphene, or other Dirac materials, the scenarios described in this review. At this time, there is still nothing going on in that direction. There are two  steps that will make this enterprise possible. On the theory side, we should identify the best experimental setting to have a precise correspondence between the specific dislocation defects (the nonzero Burgers vectors) and the torsion term in the Dirac action. On the experimental side, we should be able to realize, with the help of suitable external electromagnetic fields, the time-loop that will spot the nonzero torsion in the time direction.

We mention for completeness that we have not covered here certain interesting aspects of torsion, such as those characterising teleparallel theories~\cite{telep}, in which torsion replaces the metric, or the so-called $f(Q)$ gravity theories~\cite{qgrav}, which involve the non-metricity tensor $Q_{\alpha \mu\nu} = D_{\alpha} (\omega) \, g_{\mu\nu} \ne 0$. The interested reader is referred to the rich relevant literature for more details on the formal, phenomenological and cosmological aspects of such models.

We would like also to mention here that, in the current literature, there are several works which deal with topics  partially overlapping with those of our review, but from a different perspective than ours.

In a revisited Einstein-Cartan approach to graphene dislocations, in particular wedge disclination in a planar graphene sheet, the authors of \cite{Fernandez:2023kwc},
studied the properties of its electronic degrees of freedom in a novel approach which relates to elasticity theory, given that the aforementioned disclination is found there. An important novel result, as these authors claim, is the demonstrated explicit dependence of the energy on the elasticity (Poisson's) constant.
The works \cite{volovik1,volovik2} examine effects of the thermal Nieh-Yan anomaly terms of the axial fermion current, of the form
$\partial_\mu J^{5\mu} \propto T^2 \, T^a \wedge T_a $, where $T$ is the temperature, and  $T^a$ the effective/emergent torsion, where the proportionality constant is determined by the geometry and topology of the material, and the number of chiral quantum fields. In the case of Weyl superfluids, the authors show that such anomalous terms characterise the hydrodynamics of a chiral p-wave superfluid, such as $^3$He-A, or a chiral superconductor.

The role of torsion, when induced by the BI field within Holst and Nieh-Yan formulations, in modified general relativity and bounce cosmology has been studied in a series of works \cite{Bombacigno,Bombacigno1,Bombacigno2,Bombacigno3,Bombacigno4}, which complement our treatment of the $H$-torsion in this review and related references.
In this context, the  role of
spacetime torsion, sourced from antisymmetric tensor (Kalb-Ramond ) fields in various modified gravity theories is discussed in \cite{tan1,tan2,tan3,tan4,tan5}, including phenomenological aspects, providing potential explanation for the invisibility of torsion in late eras of the Universe, due to dynamical suppression in its couplings with standard model fields.

Claims on potential connection between Lorentz symmetry breaking
and torsion are provided in \cite{lvt}, where the one-loop fermionic effective action in Einstein-Cartan theories, computed by the proper time method, results in
a contact interaction term between the two topological terms of the Nieh-Yan topological current (axial vector torsion $S_\mu$ in our review) and the Chern-Simons topological current, which is thoroughly determined by the metric. Such terms may lead to spontaneous breaking of Lorentz symmetry, through appropriate vacuum expectation values of $S_\mu$. We note that similar mechanisms for spontaneous breaking of Lorentz symmetry arise in our stringy RVM model~\cite{bms1,ms1}, where the time derivative of the Kalb-Ramond axion acquires a constant vacuum expectation value.

The effects of torsion
on gravitational waves in extended theories of gravity, in particular in Einstein-Cartan gravity using the post-Newtonian
formalism devised by Blanchet-Damour, that goes beyond the linearized gravitational theory is discussed in the works \cite{falco1,falco2,falco3,falco4,falco5,falco6}.

In \cite{chak1},  the Lorentzian gravitational path integral has been evaluated in the presence of non-vanishing torsion (with the application of the Picard-Lefschetz theory for minisuperspaces  corresponding to a number of phenomenological bouncing cosmological models as well as for the inflationary paradigm). In addition, in \cite{chak2}, it was demonstrated that, unlike any other non-trivial modifications of the Einstein gravity,  the presence of spacetime torsion does not affect the entropy of a black hole. In \cite{chak3}, a shift-symmetric Galileon model in the presence of spacetime torsion has been constructed, with applications to the study of the evolution of the universe at a cosmological scale. For a wide class of torsional structures, the model leads to late time cosmic acceleration, while the standard results are obtained in the limit of vanishing torsion which is a smooth one.

The r\^ole of metric-scalar-torsion couplings and their impact on the growth of matter perturbations in the Universe has been discussed in \cite{Sharma:2021fou} within the context of an interacting dark-energy scenario in which the matter density of a scalar field that sources a torsion mode ceases to
be self-conserved, thereby affecting not only the background cosmological evolution but also
the perturbative spectrum of the local inhomogeneities, thus leading to cosmic growth. As argued in \cite{Sharma:2021fou}, the model can become phenomenologically viable.

A rather surprising  feature of spacetimes with torsion was pointed out in \cite{length1}, where, the authors, on considering the coupling of fermions in the presence of torsion have demonstrated
the emergence of a possibly new length scale (in analogy to the electroweak theory, as we shall explain below), which turns out to be transplanckian, and actually much larger than the Planck length.
The new scale arises as a result of
the non-renormalizable, gravitational four-fermion contact interaction, which characterises generic Einstein-Cartan theories, as we discussed repeatedly in this review.
The authors of \cite{length1}, argued that, by augmenting the Einstein-Cartan Lagrangian with suitable kinetic terms quadratic in the torsion and curvature, gives rise to new, massive propagating gravitational degrees of freedom. The whole situation is to be viewed in close analogy to the Fermi's effective four-fermion weal interaction, which is the effective low energy theory of the standard model and arises from virtual exchange of the (emergent)  W and Z weak bosons of the electroweak theory.

In an interesting recent work \cite{Gialamas:2023emn}, torsion was associated with potentially measurable properties of the electroweak vacuum, in the sense that the latter can be stabilized provided one assumes the metric-affine framework instead of the usual metric
formulation of gravity. In this framework the Holst invariant is present since in general the torsion does not vanish and this leads to important physical consequences, according to the claims of the authors of that work. Specifically, by using measured quantities such as the Higgs and
top quark masses, the authors claim that, in principle, the Einstein-Cartan theory can be differentiated from the standard General Relativity.

Last but not least, we mention the work of ref.~\cite{Pal:2022szb}, where the authors, with the help of appropriate conformal transformations, explored the use of non-symmetric contorted connections in ``Fisher information geometry''. As well known, the latter corresponds to a probability distribution function ubiquitous in the study of the effective ``geometry'' entering information theory. They introduced the idea of both metric and torsion playing equal roles in such a context, and studied the corresponding scalar curvature for a few statistical systems, which served as concrete examples for pointing out the relevant properties. As the authors claimed, this study helps to solve some long-standing problems in the field of information geometry, concerning the uniqueness of the Fisher information metric.

Our report would not be complete if we did not mention the role of torsion in the hydrodynamics of a fluid system with spin currents, as discussed in \cite{Gallegos:2021bzp}. This could be of interest in the case of, say, heavy-ion collisions in particle physics, where there is
experimental evidence for correlations between the spin polarization of $\Lambda$-hyperons and the angular momentum of the quark-gluon plasma in off-center collisions~\cite{STAR:2017ckg,STAR:2018gyt} or in the case of liquid metals, where an experimental realization of spin currents has been demonstrated~\cite{takahashi}.

A fully consistent theory of spin-current hydrodynamics is currently lacking.
In constructing such a theory, the first open issue to be addressed is identifying a canonical spin current. At this stage, we remind the reader that, in a relativistic theory, on a flat background without torsion, Lorentz invariance dictates that energy and momentum are conserved, which, as a result, implies also the conservation of angular momentum. In the absence of torsion, it is always possible to add an improvement term to the energy momentum tensor, $T_{\mu\nu}$ such that the symmetry property $T_{\mu\nu}=T_{\nu\mu}$ follows from an additional equation of motion, implying angular momentum conservation from energy-momentum conservation.
Recalling that the angular momentum  tensor $\mathcal J^{\mu\nu\rho}$ is related to the
spin current $S^{\mu\nu\rho}$ via
$\mathcal J^{\mu\nu\rho} = x^\nu \, T^{\mu\rho} - x^\rho \, T^{\mu\nu} - S^{\mu\nu\rho}$,
there follows that the spin current
suffers from ambiguities due to the possibility of adding improvement terms to the stress tensor. Specifically, by a judicious choice of
such terms, it can be set to zero. In the work of \cite{Gallegos:2021bzp}, it has been argued that one way of dealing with such ambiguity is to couple the theory to an external spin connection with torsion (which is thus  independent of the vielbeins). As discussed in that work, the presence of such a background torsion leads to a uniqueness of the spin current by precluding the addition of improvement terms to the stress tensor. After the computation of the spin current, one can set the background torsion to zero, going back to Minkowski spacetime. The presence of torsion plays an important role in ensuring uniqueness, i.e. the absence of ambiguities, in the so-called entropy current that enters the local version of the second law of thermodynamics in the pertinent fluid. The formalism of turning on the background torsion, and eventually turning it off, ensures that the total entropy current is independent of the choice of improvement terms, which in turn resolves some issues regarding the effect of the improvement terms (called {\it pseudo-gauge transformations}) on the entropy production in the system.

%%%%%%%%%%%%%%%%%%%%%%%%%%%%%%%%%%%%%%%%%%

\funding{N.E.M. is supported in part by the UK Science and Technology Facilities research Council (STFC)  under the research grants ST/T000759/1 and ST/X000753/1, and UK Engineering and Physical Sciences Research Council (EPSRC) under the research grant
EP/V002821/1; he also acknowledges participation in the COST Association Action CA18108 {\it Quantum Gravity Phenomenology in the Multimessenger Approach (QG-MM)}. P.~P. thanks Fondo Nacional de Desarrollo Cient\'{i}fico y Tecnol\'{o}gico--Chile (Fondecyt Grant No.~3200725). A.~I. and P.~P. gladly acknowledge support from Charles University Research Center (UNCE/SCI/013).}

\institutionalreview{Not applicable.}

\informedconsent{Not applicable.}

\dataavailability{Not applicable.}

\acknowledgments{A.I. and P.P. are indebted to Jorge Zanelli, for explaining to them the special role of torsion in general, and in USUSY and graphene, in particular.}

\conflictsofinterest{The authors declare no conflict of interest.}

%\section[\appendixname~\thesection]{}
%All appendix sections must be cited in the main text. In the appendices, Figures, Tables, etc. should be labeled, starting with ``A''---e.g., Figure A1, Figure A2, etc.

%%%%%%%%%%%%%%%%%%%%%%%%%%%%%%%%%%%%%%%%%%
\begin{adjustwidth}{-\extralength}{0cm}
%\printendnotes[custom] % Un-comment to print a list of endnotes

\reftitle{References}

% Please provide either the correct journal abbreviation (e.g. according to the “List of Title Word Abbreviations” http://www.issn.org/services/online-services/access-to-the-ltwa/) or the full name of the journal.
% Citations and References in Supplementary files are permitted provided that they also appear in the reference list here.

%=====================================
% References, variant A: external bibliography
%=====================================
%\bibliography{your_external_BibTeX_file}

\bibliography{Bibliografia_Universe}

\begin{thebibliography}{999}

\bibitem[Cartan(2001)]{cartan}
Cartan, E.
\newblock {\em {Riemannian geometry in an orthogonal frame}}; World Scientific,
   2001.

\bibitem[Hehl \em{et~al.}(1976)Hehl, Von Der~Heyde, Kerlick, and Nester]{hehl}
Hehl, F.W.; Von Der~Heyde, P.; Kerlick, G.D.; Nester, J.M.
\newblock {General Relativity with Spin and Torsion: Foundations and
  Prospects}.
\newblock {\em Rev. Mod. Phys.} {\bf 1976}, {\em 48},~393--416.
\newblock {\url{https://doi.org/10.1103/RevModPhys.48.393}}.

\bibitem[Shapiro(2002)]{shapiro}
Shapiro, I.L.
\newblock {Physical aspects of the space-time torsion}.
\newblock {\em Phys. Rept.} {\bf 2002}, {\em 357},~113,
  \href{http://xxx.lanl.gov/abs/hep-th/0103093}{{\normalfont
  [hep-th/0103093]}}.
\newblock {\url{https://doi.org/10.1016/S0370-1573(01)00030-8}}.

\bibitem[Eguchi \em{et~al.}(1980)Eguchi, Gilkey, and Hanson]{Eguchi}
Eguchi, T.; Gilkey, P.B.; Hanson, A.J.
\newblock {Gravitation, Gauge Theories and Differential Geometry}.
\newblock {\em Phys. Rept.} {\bf 1980}, {\em 66},~213.
\newblock {\url{https://doi.org/10.1016/0370-1573(80)90130-1}}.

\bibitem[Nakahara(2003)]{Nakahara}
Nakahara, M.
\newblock {\em Geometry, topology and physics}; CRC press,  2003.

\bibitem[Duncan \em{et~al.}(1992)Duncan, Kaloper, and Olive]{olivetorsion}
Duncan, M.J.; Kaloper, N.; Olive, K.A.
\newblock {Axion hair and dynamical torsion from anomalies}.
\newblock {\em Nucl. Phys. B} {\bf 1992}, {\em 387},~215--235.
\newblock {\url{https://doi.org/10.1016/0550-3213(92)90052-D}}.

\bibitem[Santaló(1973)]{Santalo}
Santaló, L.A.
\newblock {\em Vectores y Tensores con sus Aplicaciones}; Editorial
  Universitaria de Buenos Aires,  1973.

\bibitem[Hehl and Obukhov(2007)]{HehlObukhov2007}
Hehl, F.W.; Obukhov, Y.N.
\newblock {Elie Cartan's torsion in geometry and in field theory, an essay}.
\newblock {\em Annales Fond. Broglie} {\bf 2007}, {\em 32},~157,
  \href{http://xxx.lanl.gov/abs/0711.1535}{{\normalfont
  [arXiv:gr-qc/0711.1535]}}.

\bibitem[Iorio and Pais(2023)]{DICEtorsion}
Iorio, A.; Pais, P.
\newblock {Time-loops to spot torsion on bidimensional Dirac materials with
  dislocations}.
\newblock In Proceedings of the {Spacetime, Matter, Quantum Mechanics},  2023,
  \href{http://xxx.lanl.gov/abs/2302.02491}{{\normalfont
  [arXiv:hep-th/2302.02491]}}.

\bibitem[Capozziello \em{et~al.}(2001)Capozziello, Lambiase, and
  Stornaiolo]{Capozziello2001}
Capozziello, S.; Lambiase, G.; Stornaiolo, C.
\newblock {Geometric classification of the torsion tensor in space-time}.
\newblock {\em Annalen Phys.} {\bf 2001}, {\em 10},~713--727,
  \href{http://xxx.lanl.gov/abs/gr-qc/0101038}{{\normalfont [gr-qc/0101038]}}.
\newblock
  {\url{https://doi.org/10.1002/1521-3889(200108)10:8<713::AID-ANDP713>3.0.CO;2-2}}.

\bibitem[Capolupo \em{et~al.}(2023)Capolupo, Maria, Monda, Quaranta, and
  Serao]{capolupo2023quantum}
Capolupo, A.; Maria, G.D.; Monda, S.; Quaranta, A.; Serao, R.
\newblock Quantum Field Theory of neutrino mixing in spacetimes with torsion,
  2023,  \href{http://xxx.lanl.gov/abs/2310.09309}{{\normalfont
  [arXiv:hep-ph/2310.09309]}}.

\bibitem[Adler(1969)]{adler}
Adler, S.L.
\newblock Axial-Vector Vertex in Spinor Electrodynamics.
\newblock {\em Phys. Rev.} {\bf 1969}, {\em 177},~2426--2438.
\newblock {\url{https://doi.org/10.1103/PhysRev.177.2426}}.

\bibitem[Bell and Jackiw(1969)]{belljackiw}
Bell, J.S.; Jackiw, R.
\newblock A PCAC puzzle: $\pi$0{\textrightarrow}$\gamma$$\gamma$ in the
  $\sigma$-model.
\newblock {\em Il Nuovo Cimento A (1965-1970)} {\bf 1969}, {\em 60},~47--61.
\newblock {\url{https://doi.org/10.1007/BF02823296}}.

\bibitem[Bardeen and Zumino(1984)]{toranom1}
Bardeen, W.A.; Zumino, B.
\newblock {Consistent and Covariant Anomalies in Gauge and Gravitational
  Theories}.
\newblock {\em Nucl. Phys. B} {\bf 1984}, {\em 244},~421--453.
\newblock {\url{https://doi.org/10.1016/0550-3213(84)90322-5}}.

\bibitem[Zumino \em{et~al.}(1984)Zumino, Wu, and Zee]{anomalies1}
Zumino, B.; Wu, Y.S.; Zee, A.
\newblock {Chiral Anomalies, Higher Dimensions, and Differential Geometry}.
\newblock {\em Nucl. Phys. B} {\bf 1984}, {\em 239},~477--507.
\newblock {\url{https://doi.org/10.1016/0550-3213(84)90259-1}}.

\bibitem[Fujikawa(1980)]{anomalies2}
Fujikawa, K.
\newblock {Comment on Chiral and Conformal Anomalies}.
\newblock {\em Phys. Rev. Lett.} {\bf 1980}, {\em 44},~1733.
\newblock {\url{https://doi.org/10.1103/PhysRevLett.44.1733}}.

\bibitem[Alvarez-Gaume and Witten(1984)]{anomalies3}
Alvarez-Gaume, L.; Witten, E.
\newblock {Gravitational Anomalies}.
\newblock {\em Nucl. Phys. B} {\bf 1984}, {\em 234},~269.
\newblock {\url{https://doi.org/10.1016/0550-3213(84)90066-X}}.

\bibitem[Hull(1986)]{toranom2}
Hull, C.M.
\newblock {Anomalies, Ambiguities and Superstrings}.
\newblock {\em Phys. Lett. B} {\bf 1986}, {\em 167},~51--55.
\newblock {\url{https://doi.org/10.1016/0370-2693(86)90544-7}}.

\bibitem[Mavromatos(1988)]{toranom3}
Mavromatos, N.E.
\newblock {A Note on the Atiyah-singer Index Theorem for Manifolds With Totally
  Antisymmetric $H$ Torsion}.
\newblock {\em J. Phys. A} {\bf 1988}, {\em 21},~2279.
\newblock {\url{https://doi.org/10.1088/0305-4470/21/10/008}}.

\bibitem[Kim and Carosi(2010)]{Kim}
Kim, J.E.; Carosi, G.
\newblock {Axions and the Strong CP Problem}.
\newblock {\em Rev. Mod. Phys.} {\bf 2010}, {\em 82},~557--602,
  \href{http://xxx.lanl.gov/abs/0807.3125}{{\normalfont
  [arXiv:hep-ph/0807.3125]}}.
\newblock [Erratum: Rev.Mod.Phys. 91, 049902 (2019)],
  {\url{https://doi.org/10.1103/RevModPhys.82.557}}.

\bibitem[Jackiw and Pi(2003)]{JackiwPiChernSimons}
Jackiw, R.; Pi, S.Y.
\newblock Chern-Simons modification of general relativity.
\newblock {\em Phys. Rev. D} {\bf 2003}, {\em 68},~104012.
\newblock {\url{https://doi.org/10.1103/PhysRevD.68.104012}}.

\bibitem[Guralnik \em{et~al.}(2003)Guralnik, Iorio, Jackiw, and
  Pi]{GURALNIK2003222}
Guralnik, G.; Iorio, A.; Jackiw, R.; Pi, S.Y.
\newblock Dimensionally reduced gravitational Chern–Simons term and its kink.
\newblock {\em Annals of Physics} {\bf 2003}, {\em 308},~222--236.
\newblock
  {\url{https://doi.org/https://doi.org/10.1016/S0003-4916(03)00142-8}}.

\bibitem[Alexander and Yunes(2009)]{yunes}
Alexander, S.; Yunes, N.
\newblock {Chern-Simons Modified General Relativity}.
\newblock {\em Phys. Rept.} {\bf 2009}, {\em 480},~1--55,
  \href{http://xxx.lanl.gov/abs/0907.2562}{{\normalfont
  [arXiv:hep-th/0907.2562]}}.
\newblock {\url{https://doi.org/10.1016/j.physrep.2009.07.002}}.

\bibitem[Mavromatos(2022)]{philtrans}
Mavromatos, N.E.
\newblock {Geometrical origins of the universe dark sector: string-inspired
  torsion and anomalies as seeds for inflation and dark matter}.
\newblock {\em Phil. Trans. A. Math. Phys. Eng. Sci.} {\bf 2022}, {\em
  380},~20210188,  \href{http://xxx.lanl.gov/abs/2108.02152}{{\normalfont
  [arXiv:gr-qc/2108.02152]}}.
\newblock {\url{https://doi.org/10.1098/rsta.2021.0188}}.

\bibitem[Ashtekar and Lewandowski(2004)]{loop1}
Ashtekar, A.; Lewandowski, J.
\newblock {Background independent quantum gravity: A Status report}.
\newblock {\em Class. Quant. Grav.} {\bf 2004}, {\em 21},~R53,
  \href{http://xxx.lanl.gov/abs/gr-qc/0404018}{{\normalfont [gr-qc/0404018]}}.
\newblock {\url{https://doi.org/10.1088/0264-9381/21/15/R01}}.

\bibitem[Rovelli(2004)]{loop2}
Rovelli, C.
\newblock {\em {Quantum gravity}}; Cambridge Monographs on Mathematical
  Physics, Univ. Pr.: Cambridge, UK,  2004.
\newblock {\url{https://doi.org/10.1017/CBO9780511755804}}.

\bibitem[Immirzi(1997{\natexlab{a}})]{immirzi1}
Immirzi, G.
\newblock {Real and complex connections for canonical gravity}.
\newblock {\em Class. Quant. Grav.} {\bf 1997}, {\em 14},~L177--L181,
  \href{http://xxx.lanl.gov/abs/gr-qc/9612030}{{\normalfont [gr-qc/9612030]}}.
\newblock {\url{https://doi.org/10.1088/0264-9381/14/10/002}}.

\bibitem[Immirzi(1997{\natexlab{b}})]{immirzi2}
Immirzi, G.
\newblock {Quantum gravity and Regge calculus}.
\newblock {\em Nucl. Phys. B Proc. Suppl.} {\bf 1997}, {\em 57},~65--72,
  \href{http://xxx.lanl.gov/abs/gr-qc/9701052}{{\normalfont [gr-qc/9701052]}}.
\newblock {\url{https://doi.org/10.1016/S0920-5632(97)00354-X}}.

\bibitem[Holst(1996)]{holst}
Holst, S.
\newblock {Barbero's Hamiltonian derived from a generalized Hilbert-Palatini
  action}.
\newblock {\em Phys. Rev. D} {\bf 1996}, {\em 53},~5966--5969,
  \href{http://xxx.lanl.gov/abs/gr-qc/9511026}{{\normalfont [gr-qc/9511026]}}.
\newblock {\url{https://doi.org/10.1103/PhysRevD.53.5966}}.

\bibitem[Barbero~G.(1995{\natexlab{a}})]{barbero1}
Barbero~G., J.F.
\newblock {Real Ashtekar variables for Lorentzian signature space times}.
\newblock {\em Phys. Rev. D} {\bf 1995}, {\em 51},~5507--5510,
  \href{http://xxx.lanl.gov/abs/gr-qc/9410014}{{\normalfont [gr-qc/9410014]}}.
\newblock {\url{https://doi.org/10.1103/PhysRevD.51.5507}}.

\bibitem[Barbero~G.(1995{\natexlab{b}})]{barbero2}
Barbero~G., J.F.
\newblock {Reality conditions and Ashtekar variables: A Different perspective}.
\newblock {\em Phys. Rev. D} {\bf 1995}, {\em 51},~5498--5506,
  \href{http://xxx.lanl.gov/abs/gr-qc/9410013}{{\normalfont [gr-qc/9410013]}}.
\newblock {\url{https://doi.org/10.1103/PhysRevD.51.5498}}.

\bibitem[Ashtekar(1986)]{ashtekar1}
Ashtekar, A.
\newblock {New Variables for Classical and Quantum Gravity}.
\newblock {\em Phys. Rev. Lett.} {\bf 1986}, {\em 57},~2244--2247.
\newblock {\url{https://doi.org/10.1103/PhysRevLett.57.2244}}.

\bibitem[Ashtekar(1987)]{ashtekar2}
Ashtekar, A.
\newblock {New Hamiltonian Formulation of General Relativity}.
\newblock {\em Phys. Rev. D} {\bf 1987}, {\em 36},~1587--1602.
\newblock {\url{https://doi.org/10.1103/PhysRevD.36.1587}}.

\bibitem[Ashtekar \em{et~al.}(1989)Ashtekar, Romano, and Tate]{art}
Ashtekar, A.; Romano, J.D.; Tate, R.S.
\newblock {New Variables for Gravity: Inclusion of Matter}.
\newblock {\em Phys. Rev. D} {\bf 1989}, {\em 40},~2572.
\newblock {\url{https://doi.org/10.1103/PhysRevD.40.2572}}.

\bibitem[Perez and Rovelli(2006)]{rovelli}
Perez, A.; Rovelli, C.
\newblock {Physical effects of the Immirzi parameter}.
\newblock {\em Phys. Rev. D} {\bf 2006}, {\em 73},~044013,
  \href{http://xxx.lanl.gov/abs/gr-qc/0505081}{{\normalfont [gr-qc/0505081]}}.
\newblock {\url{https://doi.org/10.1103/PhysRevD.73.044013}}.

\bibitem[Freidel \em{et~al.}(2005)Freidel, Minic, and Takeuchi]{freidel}
Freidel, L.; Minic, D.; Takeuchi, T.
\newblock {Quantum gravity, torsion, parity violation and all that}.
\newblock {\em Phys. Rev. D} {\bf 2005}, {\em 72},~104002,
  \href{http://xxx.lanl.gov/abs/hep-th/0507253}{{\normalfont
  [hep-th/0507253]}}.
\newblock {\url{https://doi.org/10.1103/PhysRevD.72.104002}}.

\bibitem[Mercuri(2006)]{mercuri}
Mercuri, S.
\newblock {Fermions in Ashtekar-Barbero connections formalism for arbitrary
  values of the Immirzi parameter}.
\newblock {\em Phys. Rev. D} {\bf 2006}, {\em 73},~084016,
  \href{http://xxx.lanl.gov/abs/gr-qc/0601013}{{\normalfont [gr-qc/0601013]}}.
\newblock {\url{https://doi.org/10.1103/PhysRevD.73.084016}}.

\bibitem[Calcagni and Mercuri(2009)]{calcagni}
Calcagni, G.; Mercuri, S.
\newblock {The Barbero-Immirzi field in canonical formalism of pure gravity}.
\newblock {\em Phys. Rev. D} {\bf 2009}, {\em 79},~084004,
  \href{http://xxx.lanl.gov/abs/0902.0957}{{\normalfont
  [arXiv:gr-qc/0902.0957]}}.
\newblock {\url{https://doi.org/10.1103/PhysRevD.79.084004}}.

\bibitem[Nieh and Yan(1982)]{Nieh}
Nieh, H.T.; Yan, M.L.
\newblock {An Identity in Riemann-cartan Geometry}.
\newblock {\em J. Math. Phys.} {\bf 1982}, {\em 23},~373.
\newblock {\url{https://doi.org/10.1063/1.525379}}.

\bibitem[Kaul(2008)]{kaul}
Kaul, R.K.
\newblock {Holst Actions for Supergravity Theories}.
\newblock {\em Phys. Rev. D} {\bf 2008}, {\em 77},~045030,
  \href{http://xxx.lanl.gov/abs/0711.4674}{{\normalfont
  [arXiv:gr-qc/0711.4674]}}.
\newblock {\url{https://doi.org/10.1103/PhysRevD.77.045030}}.

\bibitem[Mavromatos(2021)]{torsionmav}
Mavromatos, N.E.
\newblock {Torsion in String-Inspired Cosmologies and the Universe Dark
  Sector}.
\newblock {\em Universe} {\bf 2021}, {\em 7},~480,
  \href{http://xxx.lanl.gov/abs/2111.05675}{{\normalfont
  [arXiv:hep-th/2111.05675]}}.
\newblock {\url{https://doi.org/10.3390/universe7120480}}.

\bibitem[Castellani \em{et~al.}(1991)Castellani, D'Auria, and Fre]{castel}
Castellani, L.; D'Auria, R.; Fre, P.
\newblock {\em {Supergravity and superstrings: A Geometric perspective. Vol. 1:
  Mathematical foundations}}; World Scientific,  1991.

\bibitem[Tsuda(2000)]{tsuda}
Tsuda, M.
\newblock {Generalized Lagrangian of N=1 supergravity and its canonical
  constraints with the real Ashtekar variable}.
\newblock {\em Phys. Rev. D} {\bf 2000}, {\em 61},~024025,
  \href{http://xxx.lanl.gov/abs/gr-qc/9906057}{{\normalfont [gr-qc/9906057]}}.
\newblock {\url{https://doi.org/10.1103/PhysRevD.61.024025}}.

\bibitem[Taveras and Yunes(2008)]{taveras}
Taveras, V.; Yunes, N.
\newblock {The Barbero-Immirzi Parameter as a Scalar Field: K-Inflation from
  Loop Quantum Gravity?}
\newblock {\em Phys. Rev. D} {\bf 2008}, {\em 78},~064070,
  \href{http://xxx.lanl.gov/abs/0807.2652}{{\normalfont
  [arXiv:gr-qc/0807.2652]}}.
\newblock {\url{https://doi.org/10.1103/PhysRevD.78.064070}}.

\bibitem[Torres-Gomez and Krasnov(2009)]{holstferm}
Torres-Gomez, A.; Krasnov, K.
\newblock {Remarks on Barbero-Immirzi parameter as a field}.
\newblock {\em Phys. Rev. D} {\bf 2009}, {\em 79},~104014,
  \href{http://xxx.lanl.gov/abs/0811.1998}{{\normalfont
  [arXiv:gr-qc/0811.1998]}}.
\newblock {\url{https://doi.org/10.1103/PhysRevD.79.104014}}.

\bibitem[Iorio(2011)]{iorio2011weyl}
Iorio, A.
\newblock Weyl-gauge symmetry of graphene.
\newblock {\em Annals of Physics} {\bf 2011}, {\em 326},~1334--1353.

\bibitem[Iorio(2015)]{iorio:2015}
Iorio, A.
\newblock {Curved Spacetimes and Curved Graphene: a status report of the
  Weyl-symmetry approach}.
\newblock {\em Int. J. Mod. Phys. D} {\bf 2015}, {\em 24},~1530013,
  \href{http://xxx.lanl.gov/abs/1412.4554}{{\normalfont
  [arXiv:hep-th/1412.4554]}}.
\newblock {\url{https://doi.org/10.1142/S021827181530013X}}.

\bibitem[Acquaviva \em{et~al.}(2022)Acquaviva, Iorio, Pais, and
  Smaldone]{GrapheneXonsUniverse2022}
Acquaviva, G.; Iorio, A.; Pais, P.; Smaldone, L.
\newblock Hunting Quantum Gravity with Analogs: the case of graphene.
\newblock {\em Universe} {\bf 2022}, {\em 8},~455.

\bibitem[Wallace(1947)]{wallace}
Wallace, P.R.
\newblock The Band Theory of Graphite.
\newblock {\em Phys. Rev.} {\bf 1947}, {\em 71},~622--634.
\newblock {\url{https://doi.org/10.1103/PhysRev.71.622}}.

\bibitem[Semenoff(1984)]{semenoff}
Semenoff, G.W.
\newblock Condensed-Matter Simulation of a Three-Dimensional Anomaly.
\newblock {\em Phys. Rev. Lett.} {\bf 1984}, {\em 53},~2449--2452.
\newblock {\url{https://doi.org/10.1103/PhysRevLett.53.2449}}.

\bibitem[Novoselov \em{et~al.}(2004)Novoselov, Geim, Morozov, Jiang, Zhang,
  Dubonos, Grigorieva, and Firsov]{geimnovoselovFIRST}
Novoselov, K.S.; Geim, A.K.; Morozov, S.V.; Jiang, D.; Zhang, Y.; Dubonos,
  S.V.; Grigorieva, I.V.; Firsov, A.A.
\newblock Electric Field Effect in Atomically Thin Carbon Films.
\newblock {\em Science} {\bf 2004}, {\em 306},~666--669,
  \href{http://xxx.lanl.gov/abs/https://www.science.org/doi/pdf/10.1126/science.1102896}{{\normalfont
  [https://www.science.org/doi/pdf/10.1126/science.1102896]}}.
\newblock {\url{https://doi.org/10.1126/science.1102896}}.

\bibitem[Iorio and Lambiase(2014)]{iorio2014}
Iorio, A.; Lambiase, G.
\newblock Quantum field theory in curved graphene spacetimes, Lobachevsky
  geometry, Weyl symmetry, Hawking effect, and all that.
\newblock {\em Phys. Rev. D} {\bf 2014}, {\em 90},~025006.
\newblock {\url{https://doi.org/10.1103/PhysRevD.90.025006}}.

\bibitem[Iorio \em{et~al.}(2018)Iorio, Pais, Elmashad, Ali, Faizal, and
  Abou-Salem]{ip2}
Iorio, A.; Pais, P.; Elmashad, I.A.; Ali, A.F.; Faizal, M.; Abou-Salem, L.I.
\newblock {Generalized Dirac structure beyond the linear regime in graphene}.
\newblock {\em Int. J. Mod. Phys.} {\bf 2018}, {\em D27},~1850080,
  \href{http://xxx.lanl.gov/abs/1706.01332}{{\normalfont
  [arXiv:physics.gen-ph/1706.01332]}}.
\newblock {\url{https://doi.org/10.1142/S0218271818500803}}.

\bibitem[Iorio \em{et~al.}(2022)Iorio, Iveti\'c, Mignemi, and
  Pais]{threelayers}
Iorio, A.; Iveti\'c, B.; Mignemi, S.; Pais, P.
\newblock {Three \textquotedblleft{}layers\textquotedblright{} of graphene
  monolayer and their analog generalized uncertainty principles}.
\newblock {\em Phys. Rev. D} {\bf 2022}, {\em 106},~116011,
  \href{http://xxx.lanl.gov/abs/2208.02237}{{\normalfont
  [arXiv:gr-qc/2208.02237]}}.
\newblock {\url{https://doi.org/10.1103/PhysRevD.106.116011}}.

\bibitem[Iorio \em{et~al.}(2023)Iorio, Iveti\'c, and Pais]{ncgraphene}
Iorio, A.; Iveti\'c, B.; Pais, P.
\newblock {Turning graphene into a lab for noncommutativity} {\bf 2023}.
\newblock  \href{http://xxx.lanl.gov/abs/2306.17196}{{\normalfont
  [arXiv:physics.gen-ph/2306.17196]}}.

\bibitem[Wehling \em{et~al.}(2014)Wehling, Black-Schaffer, and
  Balatsky]{wehling}
Wehling, T.; Black-Schaffer, A.; Balatsky, A.
\newblock Dirac materials.
\newblock {\em Advances in Physics} {\bf 2014}, {\em 63},~1--76.
\newblock {\url{https://doi.org/10.1080/00018732.2014.927109}}.

\bibitem[Ruggiero and Tartaglia(2003)]{RuggieroTartaglia2003}
Ruggiero, M.L.; Tartaglia, A.
\newblock {Einstein-Cartan theory as a theory of defects in space-time}.
\newblock {\em Am. J. Phys.} {\bf 2003}, {\em 71},~1303--1313,
  \href{http://xxx.lanl.gov/abs/gr-qc/0306029}{{\normalfont [gr-qc/0306029]}}.
\newblock {\url{https://doi.org/10.1119/1.1596176}}.

\bibitem[Iorio and Smaldone(2023{\natexlab{a}})]{SpaceFactoryIorioSmaldone2023}
Iorio, A.; Smaldone, L.
\newblock Quantum black holes as classical space factories.
\newblock {\em International Journal of Modern Physics D} {\bf 2023}, {\em
  32},~2350063,
  \href{http://xxx.lanl.gov/abs/https://doi.org/10.1142/S0218271823500633}{{\normalfont
  [https://doi.org/10.1142/S0218271823500633]}}.
\newblock {\url{https://doi.org/10.1142/S0218271823500633}}.

\bibitem[Iorio and Smaldone(2023{\natexlab{b}})]{SpaceFactoryDICE2022}
Iorio, A.; Smaldone, L.
\newblock {Classical space from quantum condensates}.
\newblock {\em J. Phys. Conf. Ser.} {\bf 2023}, {\em 2533},~012030,
  \href{http://xxx.lanl.gov/abs/2302.04847}{{\normalfont
  [arXiv:hep-th/2302.04847]}}.
\newblock {\url{https://doi.org/10.1088/1742-6596/2533/1/012030}}.

\bibitem[Acquaviva \em{et~al.}(2017)Acquaviva, Iorio, and
  Scholtz]{MartinAnnals2017}
Acquaviva, G.; Iorio, A.; Scholtz, M.
\newblock On the implications of the Bekenstein bound for black hole
  evaporation.
\newblock {\em Annals of Physics} {\bf 2017}, {\em 387},~317--333.
\newblock {\url{https://doi.org/https://doi.org/10.1016/j.aop.2017.10.018}}.

\bibitem[Acquaviva \em{et~al.}(2020)Acquaviva, Iorio, and
  Smaldone]{LucaGioPRD2020}
Acquaviva, G.; Iorio, A.; Smaldone, L.
\newblock Bekenstein bound from the Pauli principle.
\newblock {\em Phys. Rev. D} {\bf 2020}, {\em 102},~106002.
\newblock {\url{https://doi.org/10.1103/PhysRevD.102.106002}}.

\bibitem[Acquaviva \em{et~al.}(2017)Acquaviva, Iorio, and
  Scholtz]{MartinPOS2017}
Acquaviva, G.; Iorio, A.; Scholtz, M.
\newblock {Quasiparticle picture from the Bekenstein bound}.
\newblock {\em PoS} {\bf 2017}, {\em CORFU2017},~206,
  \href{http://xxx.lanl.gov/abs/1712.05275}{{\normalfont
  [arXiv:hep-th/1712.05275]}}.
\newblock {\url{https://doi.org/10.22323/1.318.0206}}.

\bibitem[Acquaviva \em{et~al.}(2021)Acquaviva, Iorio, and
  Smaldone]{LucaGioPOS2020}
Acquaviva, G.; Iorio, A.; Smaldone, L.
\newblock {Bekenstein bound from the Pauli principle: a brief introduction}.
\newblock {\em PoS} {\bf 2021}, {\em ICHEP2020},~681,
  \href{http://xxx.lanl.gov/abs/2011.05176}{{\normalfont
  [arXiv:hep-th/2011.05176]}}.
\newblock {\url{https://doi.org/10.22323/1.390.0681}}.

\bibitem[Kleinert(1989)]{Kleinert}
Kleinert, H.
\newblock {\em Gauge Fields in Condensed Matter}; WORLD SCIENTIFIC,  1989;
  \href{http://xxx.lanl.gov/abs/https://www.worldscientific.com/doi/pdf/10.1142/0356}{{\normalfont
  [https://www.worldscientific.com/doi/pdf/10.1142/0356]}}.
\newblock {\url{https://doi.org/10.1142/0356}}.

\bibitem[Katanaev and Volovich(1992)]{Katanaev1992}
Katanaev, M.; Volovich, I.
\newblock Theory of defects in solids and three-dimensional gravity.
\newblock {\em Annals of Physics} {\bf 1992}, {\em 216},~1--28.
\newblock {\url{https://doi.org/https://doi.org/10.1016/0003-4916(52)90040-7}}.

\bibitem[Iorio and Pais(2018)]{ip3}
Iorio, A.; Pais, P.
\newblock (Anti-)de Sitter, Poincar{\'e}, Super symmetries, and the two Dirac
  points of graphene.
\newblock {\em Annals of Physics} {\bf 2018}, {\em 398},~265 -- 286.
\newblock {\url{https://doi.org/https://doi.org/10.1016/j.aop.2018.09.011}}.

\bibitem[Katanaev(2005)]{Katanaev2005}
Katanaev, M.O.
\newblock {Geometric theory of defects}.
\newblock {\em Phys. Usp.} {\bf 2005}, {\em 48},~675--701,
  \href{http://xxx.lanl.gov/abs/cond-mat/0407469}{{\normalfont
  [arXiv:cond-mat.mtrl-sci/cond-mat/0407469]}}.
\newblock [Usp. Fiz. Nauk175,705(2005)],
  {\url{https://doi.org/10.1070/PU2005v048n07ABEH002027}}.

\bibitem[Lazar(2003)]{Lazar2003}
Lazar, M.
\newblock {A Nonsingular solution of the edge dislocation in the gauge theory
  of dislocations}.
\newblock {\em J. Phys. A} {\bf 2003}, {\em 36},~1415,
  \href{http://xxx.lanl.gov/abs/cond-mat/0208360}{{\normalfont
  [cond-mat/0208360]}}.
\newblock {\url{https://doi.org/10.1088/0305-4470/36/5/316}}.

\bibitem[Ciappina \em{et~al.}(2020)Ciappina, Iorio, Pais, and Zampeli]{ip4}
Ciappina, M.F.; Iorio, A.; Pais, P.; Zampeli, A.
\newblock Torsion in quantum field theory through time-loops on Dirac
  materials.
\newblock {\em Phys. Rev. D} {\bf 2020}, {\em 101},~036021.
\newblock {\url{https://doi.org/10.1103/PhysRevD.101.036021}}.

\bibitem[de~Juan \em{et~al.}(2010)de~Juan, Cortijo, and Vozmediano]{deJuan2010}
de~Juan, F.; Cortijo, A.; Vozmediano, M.A.H.
\newblock {Dislocations and torsion in graphene and related systems}.
\newblock {\em Nucl. Phys. B} {\bf 2010}, {\em 828},~625,
  \href{http://xxx.lanl.gov/abs/0909.4068}{{\normalfont
  [arXiv:cond-mat.mes-hall/0909.4068]}}.
\newblock {\url{https://doi.org/10.1016/j.nuclphysb.2009.11.012}}.

\bibitem[Vozmediano \em{et~al.}(2010)Vozmediano, Katsnelson, and
  Guinea]{Vozmediano:2010zz}
Vozmediano, M.A.H.; Katsnelson, M.I.; Guinea, F.
\newblock {Gauge fields in graphene}.
\newblock {\em Phys. Rept.} {\bf 2010}, {\em 496},~109,
  \href{http://xxx.lanl.gov/abs/1003.5179}{{\normalfont
  [arXiv:cond-mat.mes-hall/1003.5179]}}.
\newblock {\url{https://doi.org/10.1016/j.physrep.2010.07.003}}.

\bibitem[Amorim \em{et~al.}(2016)Amorim et~al.]{Amorim:2015bga}
Amorim, B.;  et~al.
\newblock {Novel effects of strains in graphene and other two dimensional
  materials}.
\newblock {\em Phys. Rept.} {\bf 2016}, {\em 617},~1,
  \href{http://xxx.lanl.gov/abs/1503.00747}{{\normalfont
  [arXiv:cond-mat.mes-hall/1503.00747]}}.
\newblock {\url{https://doi.org/10.1016/j.physrep.2015.12.006}}.

\bibitem[Wilczek(2012)]{Wilczek2012}
Wilczek, F.
\newblock {Quantum Time Crystals}.
\newblock {\em Phys. Rev. Lett.} {\bf 2012}, {\em 109},~160401,
  \href{http://xxx.lanl.gov/abs/1202.2539}{{\normalfont
  [arXiv:quant-ph/1202.2539]}}.
\newblock {\url{https://doi.org/10.1103/PhysRevLett.109.160401}}.

\bibitem[Shapere and Wilczek(2012)]{WilczekShapere2012}
Shapere, A.; Wilczek, F.
\newblock {Classical Time Crystals}.
\newblock {\em Phys. Rev. Lett.} {\bf 2012}, {\em 109},~160402,
  \href{http://xxx.lanl.gov/abs/1202.2537}{{\normalfont
  [arXiv:cond-mat.other/1202.2537]}}.
\newblock {\url{https://doi.org/10.1103/PhysRevLett.109.160402}}.

\bibitem[{Li} \em{et~al.}(2012){Li}, {Gong}, {Yin}, {Quan}, {Yin}, {Zhang},
  {Duan}, and {Zhang}]{PhysRevLett.109.163001}
{Li}, T.; {Gong}, Z.X.; {Yin}, Z.Q.; {Quan}, H.T.; {Yin}, X.; {Zhang}, P.;
  {Duan}, L.M.; {Zhang}, X.
\newblock {Space-Time Crystals of Trapped Ions}.
\newblock {\em Phys. Rev. Lett.} {\bf 2012}, {\em 109},~163001,
  \href{http://xxx.lanl.gov/abs/1206.4772}{{\normalfont
  [arXiv:quant-ph/1206.4772]}}.
\newblock {\url{https://doi.org/10.1103/PhysRevLett.109.163001}}.

\bibitem[{Smits} \em{et~al.}(2018){Smits}, {Liao}, {Stoof}, and {van der
  Straten}]{PhysRevLett.121.185301}
{Smits}, J.; {Liao}, L.; {Stoof}, H.T.C.; {van der Straten}, P.
\newblock {Observation of a Space-Time Crystal in a Superfluid Quantum Gas}.
\newblock {\em Phys. Rev. Lett.} {\bf 2018}, {\em 121},~185301,
  \href{http://xxx.lanl.gov/abs/1807.05904}{{\normalfont
  [arXiv:cond-mat.quant-gas/1807.05904]}}.
\newblock {\url{https://doi.org/10.1103/PhysRevLett.121.185301}}.

\bibitem[Loll(1998)]{Loll1998}
Loll, R.
\newblock {Discrete approaches to quantum gravity in four-dimensions}.
\newblock {\em Living Rev. Rel.} {\bf 1998}, {\em 1},~13,
  \href{http://xxx.lanl.gov/abs/gr-qc/9805049}{{\normalfont [gr-qc/9805049]}}.
\newblock {\url{https://doi.org/10.12942/lrr-1998-13}}.

\bibitem[Heide \em{et~al.}(2018)Heide, Higuchi, Weber, and
  Hommelhoff]{lasergraphenePRL2018}
Heide, C.; Higuchi, T.; Weber, H.B.; Hommelhoff, P.
\newblock Coherent Electron Trajectory Control in Graphene.
\newblock {\em Phys. Rev. Lett.} {\bf 2018}, {\em 121},~207401.
\newblock {\url{https://doi.org/10.1103/PhysRevLett.121.207401}}.

\bibitem[{Higuchi} \em{et~al.}(2017){Higuchi}, {Heide}, {Ullmann}, {Weber}, and
  {Hommelhoff}]{natureLaserGraphene}
{Higuchi}, T.; {Heide}, C.; {Ullmann}, K.; {Weber}, H.B.; {Hommelhoff}, P.
\newblock {Light-field-driven currents in graphene}.
\newblock {\em Nature} {\bf 2017}, {\em 550},~224,
  \href{http://xxx.lanl.gov/abs/1607.04198}{{\normalfont
  [arXiv:physics.optics/1607.04198]}}.
\newblock {\url{https://doi.org/10.1038/nature23900}}.

\bibitem[Marino(1993)]{Marino:1992xi}
Marino, E.C.
\newblock {Quantum electrodynamics of particles on a plane and the Chern-Simons
  theory}.
\newblock {\em Nucl. Phys. B} {\bf 1993}, {\em 408},~551,
  \href{http://xxx.lanl.gov/abs/hep-th/9301034}{{\normalfont
  [hep-th/9301034]}}.
\newblock {\url{https://doi.org/10.1016/0550-3213(93)90379-4}}.

\bibitem[Gorbar \em{et~al.}(2001)Gorbar, Gusynin, and Miransky]{Gorbar:2001qt}
Gorbar, E.V.; Gusynin, V.P.; Miransky, V.A.
\newblock {Dynamical chiral symmetry breaking on a brane in reduced QED}.
\newblock {\em Phys. Rev. D} {\bf 2001}, {\em 64},~105028,
  \href{http://xxx.lanl.gov/abs/hep-ph/0105059}{{\normalfont
  [hep-ph/0105059]}}.
\newblock {\url{https://doi.org/10.1103/PhysRevD.64.105028}}.

\bibitem[Dudal \em{et~al.}(2018)Dudal, Mizher, and Pais]{Dudal:2018mms}
Dudal, D.; Mizher, A.J.; Pais, P.
\newblock {Remarks on the Chern-Simons photon term in the QED description of
  graphene}.
\newblock {\em Phys. Rev. D} {\bf 2018}, {\em 98},~065008,
  \href{http://xxx.lanl.gov/abs/1801.08853}{{\normalfont
  [arXiv:hep-th/1801.08853]}}.
\newblock {\url{https://doi.org/10.1103/PhysRevD.98.065008}}.

\bibitem[Dudal \em{et~al.}(2019)Dudal, Mizher, and Pais]{Dudal:2018pta}
Dudal, D.; Mizher, A.J.; Pais, P.
\newblock {Exact quantum scale invariance of three-dimensional reduced QED
  theories}.
\newblock {\em Phys. Rev. D} {\bf 2019}, {\em 99},~045017,
  \href{http://xxx.lanl.gov/abs/1808.04709}{{\normalfont
  [arXiv:hep-th/1808.04709]}}.
\newblock {\url{https://doi.org/10.1103/PhysRevD.99.045017}}.

\bibitem[Iorio and Pais(2015)]{ipfirst}
Iorio, A.; Pais, P.
\newblock Revisiting the gauge fields of strained graphene.
\newblock {\em Phys. Rev. D} {\bf 2015}, {\em 92},~125005.
\newblock {\url{https://doi.org/10.1103/PhysRevD.92.125005}}.

\bibitem[Andrianopoli \em{et~al.}(2020)Andrianopoli, Cerchiai, D'Auria,
  Gallerati, Noris, Trigiante, and Zanelli]{DauriaZanelli2019}
Andrianopoli, L.; Cerchiai, B.L.; D'Auria, R.; Gallerati, A.; Noris, R.;
  Trigiante, M.; Zanelli, J.
\newblock {$\mathcal{N}$-extended $D = 4$ supergravity, unconventional SUSY and
  graphene}.
\newblock {\em JHEP} {\bf 2020}, {\em 01},~084,
  \href{http://xxx.lanl.gov/abs/1910.03508}{{\normalfont
  [arXiv:hep-th/1910.03508]}}.
\newblock {\url{https://doi.org/10.1007/JHEP01(2020)084}}.

\bibitem[Peskin and Schroeder(1995)]{Peskin}
Peskin, M.E.; Schroeder, D.V.
\newblock {\em {An Introduction to quantum field theory}}; Addison-Wesley:
  Reading, USA,  1995.

\bibitem[{Kruchinin} \em{et~al.}(2018){Kruchinin}, {Krausz}, and
  {Yakovlev}]{StanislavRevModPhys}
{Kruchinin}, S.Y.; {Krausz}, F.; {Yakovlev}, V.S.
\newblock {Colloquium: Strong-field phenomena in periodic systems}.
\newblock {\em Rev. Mod. Phys.} {\bf 2018}, {\em 90},~021002,
  \href{http://xxx.lanl.gov/abs/1712.05685}{{\normalfont
  [arXiv:quant-ph/1712.05685]}}.
\newblock {\url{https://doi.org/10.1103/RevModPhys.90.021002}}.

\bibitem[Castro~Neto \em{et~al.}(2009)Castro~Neto, Guinea, Peres, Novoselov,
  and Geim]{pacoreview2009}
Castro~Neto, A.H.; Guinea, F.; Peres, N.M.R.; Novoselov, K.S.; Geim, A.K.
\newblock The electronic properties of graphene.
\newblock {\em Rev. Mod. Phys.} {\bf 2009}, {\em 81},~109--162.
\newblock {\url{https://doi.org/10.1103/RevModPhys.81.109}}.

\bibitem[Gusynin \em{et~al.}(2007)Gusynin, Sharapov, and Carbotte]{Gusynin}
Gusynin, V.P.; Sharapov, S.G.; Carbotte, J.P.
\newblock AC conductivity of graphene: from light-binding model to 2 +
  1-dimensional quantum electrodynamics.
\newblock {\em International Journal of Modern Physics B} {\bf 2007}, {\em
  21},~4611--4658,
  \href{http://xxx.lanl.gov/abs/https://doi.org/10.1142/S0217979207038022}{{\normalfont
  [https://doi.org/10.1142/S0217979207038022]}}.
\newblock {\url{https://doi.org/10.1142/S0217979207038022}}.

\bibitem[Gonzalez \em{et~al.}(1993)Gonzalez, Guinea, and
  Vozmediano]{GONZALEZ1993771}
Gonzalez, J.; Guinea, F.; Vozmediano, M.A.H.
\newblock {The Electronic spectrum of fullerenes from the Dirac equation}.
\newblock {\em Nucl. Phys. B} {\bf 1993}, {\em 406},~771,
  \href{http://xxx.lanl.gov/abs/cond-mat/9208004}{{\normalfont
  [cond-mat/9208004]}}.
\newblock {\url{https://doi.org/10.1016/0550-3213(93)90009-E}}.

\bibitem[Yazyev and Chen(2014)]{Yazyev2014}
Yazyev, O.V.; Chen, Y.P.
\newblock Polycrystalline graphene and other two-dimensional materials.
\newblock {\em Nature nanotechnology} {\bf 2014}, {\em 9},~755--767.

\bibitem[Yazyev and Louie(2010)]{Yazyev2010}
Yazyev, O.V.; Louie, S.G.
\newblock Topological defects in graphene: Dislocations and grain boundaries.
\newblock {\em Phys. Rev. B} {\bf 2010}, {\em 81},~195420,
  \href{http://xxx.lanl.gov/abs/1004.2031}{{\normalfont
  [arXiv:cond-mat/1004.2031]}}.

\bibitem[Hirth and Lothe(1967)]{hirth1967theory}
Hirth, J.; Lothe, J.
\newblock {\em Theory of Dislocations}; McGraw-Hill series in electrical
  engineering: Electronics and electronic circuits, McGraw-Hill,  1967.

\bibitem[Zhang \em{et~al.}(2015)Zhang, Xu, Yuan, Xin, and Ding]{C5NR04960A}
Zhang, X.; Xu, Z.; Yuan, Q.; Xin, J.; Ding, F.
\newblock The favourable large misorientation angle grain boundaries in
  graphene.
\newblock {\em Nanoscale} {\bf 2015}, {\em 7},~20082--20088.
\newblock {\url{https://doi.org/10.1039/C5NR04960A}}.

\bibitem[Freedman \em{et~al.}(1976)Freedman, van Nieuwenhuizen, and
  Ferrara]{n=1sugra1}
Freedman, D.Z.; van Nieuwenhuizen, P.; Ferrara, S.
\newblock {Progress Toward a Theory of Supergravity}.
\newblock {\em Phys. Rev. D} {\bf 1976}, {\em 13},~3214--3218.
\newblock {\url{https://doi.org/10.1103/PhysRevD.13.3214}}.

\bibitem[Ferrara and van Nieuwenhuizen(1979)]{n=1sugra1b}
Ferrara, S.; van Nieuwenhuizen, P.
\newblock {Simplifications of Einstein Supergravity}.
\newblock {\em Phys. Rev. D} {\bf 1979}, {\em 20},~2079.
\newblock {\url{https://doi.org/10.1103/PhysRevD.20.2079}}.

\bibitem[Van~Nieuwenhuizen(1981)]{n=1sugra2}
Van~Nieuwenhuizen, P.
\newblock {Supergravity}.
\newblock {\em Phys. Rept.} {\bf 1981}, {\em 68},~189--398.
\newblock {\url{https://doi.org/10.1016/0370-1573(81)90157-5}}.

\bibitem[Nilles(1984)]{Nilles}
Nilles, H.P.
\newblock {Supersymmetry, Supergravity and Particle Physics}.
\newblock {\em Phys. Rept.} {\bf 1984}, {\em 110},~1--162.
\newblock {\url{https://doi.org/10.1016/0370-1573(84)90008-5}}.

\bibitem[Alexandre \em{et~al.}(2013)Alexandre, Houston, and
  Mavromatos]{Alexandre1}
Alexandre, J.; Houston, N.; Mavromatos, N.E.
\newblock {Dynamical Supergravity Breaking via the Super-Higgs Effect
  Revisited}.
\newblock {\em Phys. Rev. D} {\bf 2013}, {\em 88},~125017,
  \href{http://xxx.lanl.gov/abs/1310.4122}{{\normalfont
  [arXiv:hep-th/1310.4122]}}.
\newblock {\url{https://doi.org/10.1103/PhysRevD.88.125017}}.

\bibitem[Alexandre \em{et~al.}(2015)Alexandre, Houston, and
  Mavromatos]{Alexandre2}
Alexandre, J.; Houston, N.; Mavromatos, N.E.
\newblock {Inflation via Gravitino Condensation in Dynamically Broken
  Supergravity}.
\newblock {\em Int. J. Mod. Phys. D} {\bf 2015}, {\em 24},~1541004,
  \href{http://xxx.lanl.gov/abs/1409.3183}{{\normalfont
  [arXiv:gr-qc/1409.3183]}}.
\newblock {\url{https://doi.org/10.1142/S0218271815410047}}.

\bibitem[Deser and Zumino(1977)]{deser}
Deser, S.; Zumino, B.
\newblock {Broken Supersymmetry and Supergravity}.
\newblock {\em Phys. Rev. Lett.} {\bf 1977}, {\em 38},~1433--1436.
\newblock {\url{https://doi.org/10.1103/PhysRevLett.38.1433}}.

\bibitem[Volkov and Akulov(1972)]{volkov}
Volkov, D.V.; Akulov, V.P.
\newblock {Possible universal neutrino interaction}.
\newblock {\em JETP Lett.} {\bf 1972}, {\em 16},~438--440.

\bibitem[Ellis and Mavromatos(2013)]{hilltopEllis}
Ellis, J.; Mavromatos, N.E.
\newblock {Inflation induced by gravitino condensation in supergravity}.
\newblock {\em Phys. Rev. D} {\bf 2013}, {\em 88},~085029,
  \href{http://xxx.lanl.gov/abs/1308.1906}{{\normalfont
  [arXiv:hep-th/1308.1906]}}.
\newblock {\url{https://doi.org/10.1103/PhysRevD.88.085029}}.

\bibitem[Alvarez \em{et~al.}(2012)Alvarez, Valenzuela, and Zanelli]{AVZ}
Alvarez, P.D.; Valenzuela, M.; Zanelli, J.
\newblock {Supersymmetry of a different kind}.
\newblock {\em JHEP} {\bf 2012}, {\em 1204},~058,
  \href{http://xxx.lanl.gov/abs/1109.3944}{{\normalfont
  [arXiv:hep-th/1109.3944]}}.
\newblock {\url{https://doi.org/10.1007/JHEP04(2012)058}}.

\bibitem[Guevara \em{et~al.}(2016)Guevara, Pais, and Zanelli]{GPZ}
Guevara, A.; Pais, P.; Zanelli, J.
\newblock {Dynamical Contents of Unconventional Supersymmetry}.
\newblock {\em JHEP} {\bf 2016}, {\em 08},~085,
  \href{http://xxx.lanl.gov/abs/1606.05239}{{\normalfont
  [arXiv:hep-th/1606.05239]}}.
\newblock {\url{https://doi.org/10.1007/JHEP08(2016)085}}.

\bibitem[Alvarez \em{et~al.}(2015)Alvarez, Pais, Rodr\'\i{}guez,
  Salgado-Rebolledo, and Zanelli]{USUSYSU2}
Alvarez, P.D.; Pais, P.; Rodr\'\i{}guez, E.; Salgado-Rebolledo, P.; Zanelli, J.
\newblock {Supersymmetric 3D model for gravity with $SU(2)$ gauge symmetry,
  mass generation and effective cosmological constant}.
\newblock {\em Class. Quant. Grav.} {\bf 2015}, {\em 32},~175014,
  \href{http://xxx.lanl.gov/abs/1505.03834}{{\normalfont
  [arXiv:hep-th/1505.03834]}}.
\newblock {\url{https://doi.org/10.1088/0264-9381/32/17/175014}}.

\bibitem[Ba\~nados \em{et~al.}(1992)Ba\~nados, Teitelboim, and
  Zanelli]{BTZ1992}
Ba\~nados, M.; Teitelboim, C.; Zanelli, J.
\newblock Black hole in three-dimensional spacetime.
\newblock {\em Phys. Rev. Lett.} {\bf 1992}, {\em 69},~1849--1851.
\newblock {\url{https://doi.org/10.1103/PhysRevLett.69.1849}}.

\bibitem[Alvarez \em{et~al.}(2014)Alvarez, Pais, Rodr\'\i{}guez,
  Salgado-Rebolledo, and Zanelli]{BTZLorentzflat}
Alvarez, P.D.; Pais, P.; Rodr\'\i{}guez, E.; Salgado-Rebolledo, P.; Zanelli, J.
\newblock {The BTZ black hole as a Lorentz-flat geometry}.
\newblock {\em Phys. Lett. B} {\bf 2014}, {\em 738},~134--135,
  \href{http://xxx.lanl.gov/abs/1405.6657}{{\normalfont
  [arXiv:gr-qc/1405.6657]}}.
\newblock {\url{https://doi.org/10.1016/j.physletb.2014.09.032}}.

\bibitem[Miskovic and Zanelli(2009)]{Miskovic2009}
Miskovic, O.; Zanelli, J.
\newblock {On the negative spectrum of the 2+1 black hole}.
\newblock {\em Phys. Rev. D} {\bf 2009}, {\em 79},~105011,
  \href{http://xxx.lanl.gov/abs/0904.0475}{{\normalfont
  [arXiv:hep-th/0904.0475]}}.
\newblock {\url{https://doi.org/10.1103/PhysRevD.79.105011}}.

\bibitem[Iorio \em{et~al.}(2020)Iorio, Lambiase, Pais, and Scardigli]{GUPBTZ}
Iorio, A.; Lambiase, G.; Pais, P.; Scardigli, F.
\newblock Generalized uncertainty principle in three-dimensional gravity and
  the BTZ black hole.
\newblock {\em Phys. Rev. D} {\bf 2020}, {\em 101},~105002.
\newblock {\url{https://doi.org/10.1103/PhysRevD.101.105002}}.

\bibitem[Iorio and Pais(2019)]{ip5}
Iorio, A.; Pais, P.
\newblock {Generalized uncertainty principle in graphene}.
\newblock {\em J. Phys. Conf. Ser.} {\bf 2019}, {\em 1275},~012061,
  \href{http://xxx.lanl.gov/abs/1902.00116}{{\normalfont
  [arXiv:hep-th/1902.00116]}}.
\newblock {\url{https://doi.org/10.1088/1742-6596/1275/1/012061}}.

\bibitem[Iorio and Lambiase(2012)]{Iorio:2011yz}
Iorio, A.; Lambiase, G.
\newblock {The Hawking-Unruh phenomenon on graphene}.
\newblock {\em Phys. Lett.} {\bf 2012}, {\em B716},~334--337,
  \href{http://xxx.lanl.gov/abs/1108.2340}{{\normalfont
  [arXiv:cond-mat.mtrl-sci/1108.2340]}}.
\newblock {\url{https://doi.org/10.1016/j.physletb.2012.08.023}}.

\bibitem[Aghanim \em{et~al.}(2020)Aghanim et~al.]{Planck}
Aghanim, N.;  et~al.
\newblock {Planck 2018 results. VI. Cosmological parameters}.
\newblock {\em Astron. Astrophys.} {\bf 2020}, {\em 641},~A6,
  \href{http://xxx.lanl.gov/abs/1807.06209}{{\normalfont
  [arXiv:astro-ph.CO/1807.06209]}}.
\newblock [Erratum: Astron.Astrophys. 652, C4 (2021)],
  {\url{https://doi.org/10.1051/0004-6361/201833910}}.

\bibitem[Verde \em{et~al.}(2019)Verde, Treu, and Riess]{tensions1}
Verde, L.; Treu, T.; Riess, A.G.
\newblock {Tensions between the Early and the Late Universe}.
\newblock {\em Nature Astron.} {\bf 2019}, {\em 3},~891,
  \href{http://xxx.lanl.gov/abs/1907.10625}{{\normalfont
  [arXiv:astro-ph.CO/1907.10625]}}.
\newblock {\url{https://doi.org/10.1038/s41550-019-0902-0}}.

\bibitem[Perivolaropoulos and Skara(2022)]{tensions2}
Perivolaropoulos, L.; Skara, F.
\newblock {Challenges for \ensuremath{\Lambda}CDM: An update}.
\newblock {\em New Astron. Rev.} {\bf 2022}, {\em 95},~101659,
  \href{http://xxx.lanl.gov/abs/2105.05208}{{\normalfont
  [arXiv:astro-ph.CO/2105.05208]}}.
\newblock {\url{https://doi.org/10.1016/j.newar.2022.101659}}.

\bibitem[Abdalla \em{et~al.}(2022)Abdalla et~al.]{tensions3}
Abdalla, E.;  et~al.
\newblock {Cosmology intertwined: A review of the particle physics,
  astrophysics, and cosmology associated with the cosmological tensions and
  anomalies}.
\newblock {\em JHEAp} {\bf 2022}, {\em 34},~49--211,
  \href{http://xxx.lanl.gov/abs/2203.06142}{{\normalfont
  [arXiv:astro-ph.CO/2203.06142]}}.
\newblock {\url{https://doi.org/10.1016/j.jheap.2022.04.002}}.

\bibitem[Freedman(2017)]{Freedman}
Freedman, W.L.
\newblock {Cosmology at a Crossroads}.
\newblock {\em Nature Astron.} {\bf 2017}, {\em 1},~0121,
  \href{http://xxx.lanl.gov/abs/1706.02739}{{\normalfont
  [arXiv:astro-ph.CO/1706.02739]}}.
\newblock {\url{https://doi.org/10.1038/s41550-017-0121}}.

\bibitem[Green \em{et~al.}(2012{\natexlab{a}})Green, Schwarz, and
  Witten]{string1}
Green, M.B.; Schwarz, J.H.; Witten, E.
\newblock {\em {Superstring Theory Vol. 1}: {25th Anniversary Edition}};
  Cambridge Monographs on Mathematical Physics, Cambridge University Press,
  2012.
\newblock {\url{https://doi.org/10.1017/CBO9781139248563}}.

\bibitem[Green \em{et~al.}(2012{\natexlab{b}})Green, Schwarz, and
  Witten]{string2}
Green, M.B.; Schwarz, J.H.; Witten, E.
\newblock {\em {Superstring Theory Vol. 2}: {25th Anniversary Edition}};
  Cambridge Monographs on Mathematical Physics, Cambridge University Press,
  2012.
\newblock {\url{https://doi.org/10.1017/CBO9781139248570}}.

\bibitem[Polchinski(2007)]{Polch}
Polchinski, J.
\newblock {\em {String theory. Vol. 2: Superstring theory and beyond}};
  Cambridge Monographs on Mathematical Physics, Cambridge University Press,
  2007.
\newblock {\url{https://doi.org/10.1017/CBO9780511618123}}.

\bibitem[Hellerman \em{et~al.}(2001)Hellerman, Kaloper, and Susskind]{smatrix1}
Hellerman, S.; Kaloper, N.; Susskind, L.
\newblock {String theory and quintessence}.
\newblock {\em JHEP} {\bf 2001}, {\em 06},~003,
  \href{http://xxx.lanl.gov/abs/hep-th/0104180}{{\normalfont
  [hep-th/0104180]}}.
\newblock {\url{https://doi.org/10.1088/1126-6708/2001/06/003}}.

\bibitem[Fischler \em{et~al.}(2001)Fischler, Kashani-Poor, McNees, and
  Paban]{smatrix2}
Fischler, W.; Kashani-Poor, A.; McNees, R.; Paban, S.
\newblock {The Acceleration of the universe, a challenge for string theory}.
\newblock {\em JHEP} {\bf 2001}, {\em 07},~003,
  \href{http://xxx.lanl.gov/abs/hep-th/0104181}{{\normalfont
  [hep-th/0104181]}}.
\newblock {\url{https://doi.org/10.1088/1126-6708/2001/07/003}}.

\bibitem[Palti(2019)]{Palti1}
Palti, E.
\newblock {The Swampland: Introduction and Review}.
\newblock {\em Fortsch. Phys.} {\bf 2019}, {\em 67},~1900037,
  \href{http://xxx.lanl.gov/abs/1903.06239}{{\normalfont
  [arXiv:hep-th/1903.06239]}}.
\newblock {\url{https://doi.org/10.1002/prop.201900037}}.

\bibitem[Palti(2022)]{Palti2}
Palti, E.
\newblock {The swampland and string theory}.
\newblock {\em Contemp. Phys.} {\bf 2022}, {\em 62},~165--179.
\newblock {\url{https://doi.org/10.1080/00107514.2022.2103275}}.

\bibitem[Obied \em{et~al.}(2018)Obied, Ooguri, Spodyneiko, and Vafa]{swamp1}
Obied, G.; Ooguri, H.; Spodyneiko, L.; Vafa, C.
\newblock {De Sitter Space and the Swampland} {\bf 2018}.
\newblock  \href{http://xxx.lanl.gov/abs/1806.08362}{{\normalfont
  [arXiv:hep-th/1806.08362]}}.

\bibitem[Agrawal \em{et~al.}(2018)Agrawal, Obied, Steinhardt, and Vafa]{swamp2}
Agrawal, P.; Obied, G.; Steinhardt, P.J.; Vafa, C.
\newblock {On the Cosmological Implications of the String Swampland}.
\newblock {\em Phys. Lett. B} {\bf 2018}, {\em 784},~271--276,
  \href{http://xxx.lanl.gov/abs/1806.09718}{{\normalfont
  [arXiv:hep-th/1806.09718]}}.
\newblock {\url{https://doi.org/10.1016/j.physletb.2018.07.040}}.

\bibitem[Garg and Krishnan(2019)]{swamp3}
Garg, S.K.; Krishnan, C.
\newblock {Bounds on Slow Roll and the de Sitter Swampland}.
\newblock {\em JHEP} {\bf 2019}, {\em 11},~075,
  \href{http://xxx.lanl.gov/abs/1807.05193}{{\normalfont
  [arXiv:hep-th/1807.05193]}}.
\newblock {\url{https://doi.org/10.1007/JHEP11(2019)075}}.

\bibitem[Ooguri \em{et~al.}(2019)Ooguri, Palti, Shiu, and Vafa]{swamp4}
Ooguri, H.; Palti, E.; Shiu, G.; Vafa, C.
\newblock {Distance and de Sitter Conjectures on the Swampland}.
\newblock {\em Phys. Lett. B} {\bf 2019}, {\em 788},~180--184,
  \href{http://xxx.lanl.gov/abs/1810.05506}{{\normalfont
  [arXiv:hep-th/1810.05506]}}.
\newblock {\url{https://doi.org/10.1016/j.physletb.2018.11.018}}.

\bibitem[Mohayaee \em{et~al.}(2021)Mohayaee, Rameez, and Sarkar]{subir1}
Mohayaee, R.; Rameez, M.; Sarkar, S.
\newblock {Do supernovae indicate an accelerating universe?}
\newblock {\em Eur. Phys. J. ST} {\bf 2021}, {\em 230},~2067--2076,
  \href{http://xxx.lanl.gov/abs/2106.03119}{{\normalfont
  [arXiv:astro-ph.CO/2106.03119]}}.
\newblock {\url{https://doi.org/10.1140/epjs/s11734-021-00199-6}}.

\bibitem[Secrest \em{et~al.}(2022)Secrest, von Hausegger, Rameez, Mohayaee, and
  Sarkar]{subir2}
Secrest, N.J.; von Hausegger, S.; Rameez, M.; Mohayaee, R.; Sarkar, S.
\newblock {A Challenge to the Standard Cosmological Model}.
\newblock {\em Astrophys. J. Lett.} {\bf 2022}, {\em 937},~L31,
  \href{http://xxx.lanl.gov/abs/2206.05624}{{\normalfont
  [arXiv:astro-ph.CO/2206.05624]}}.
\newblock {\url{https://doi.org/10.3847/2041-8213/ac88c0}}.

\bibitem[Di~Valentino \em{et~al.}(2021)Di~Valentino, Mena, Pan, Visinelli,
  Yang, Melchiorri, Mota, Riess, and Silk]{Tensol1}
Di~Valentino, E.; Mena, O.; Pan, S.; Visinelli, L.; Yang, W.; Melchiorri, A.;
  Mota, D.F.; Riess, A.G.; Silk, J.
\newblock {In the realm of the Hubble tension\textemdash{}a review of
  solutions}.
\newblock {\em Class. Quant. Grav.} {\bf 2021}, {\em 38},~153001,
  \href{http://xxx.lanl.gov/abs/2103.01183}{{\normalfont
  [arXiv:astro-ph.CO/2103.01183]}}.
\newblock {\url{https://doi.org/10.1088/1361-6382/ac086d}}.

\bibitem[Mavromatos(2022)]{anommav}
Mavromatos, N.E.
\newblock {Anomalies, the Dark Universe and Matter-Antimatter asymmetry}.
\newblock In Proceedings of the {DICE 2022: Spacetime, Matter, Quantum
  Mechanics},  2022,  \href{http://xxx.lanl.gov/abs/2212.13437}{{\normalfont
  [arXiv:hep-th/2212.13437]}}.

\bibitem[Basilakos \em{et~al.}(2020{\natexlab{a}})Basilakos, Mavromatos, and
  Sol\`a~Peracaula]{bms1}
Basilakos, S.; Mavromatos, N.E.; Sol\`a~Peracaula, J.
\newblock {Gravitational and Chiral Anomalies in the Running Vacuum Universe
  and Matter-Antimatter Asymmetry}.
\newblock {\em Phys. Rev. D} {\bf 2020}, {\em 101},~045001,
  \href{http://xxx.lanl.gov/abs/1907.04890}{{\normalfont
  [arXiv:hep-ph/1907.04890]}}.
\newblock {\url{https://doi.org/10.1103/PhysRevD.101.045001}}.

\bibitem[Basilakos \em{et~al.}(2020{\natexlab{b}})Basilakos, Mavromatos, and
  Sol\`a~Peracaula]{bms2}
Basilakos, S.; Mavromatos, N.E.; Sol\`a~Peracaula, J.
\newblock {Quantum Anomalies in String-Inspired Running Vacuum Universe:
  Inflation and Axion Dark Matter}.
\newblock {\em Phys. Lett. B} {\bf 2020}, {\em 803},~135342,
  \href{http://xxx.lanl.gov/abs/2001.03465}{{\normalfont
  [arXiv:gr-qc/2001.03465]}}.
\newblock {\url{https://doi.org/10.1016/j.physletb.2020.135342}}.

\bibitem[Mavromatos and Sol\`a~Peracaula(2021{\natexlab{a}})]{ms1}
Mavromatos, N.E.; Sol\`a~Peracaula, J.
\newblock {Stringy-running-vacuum-model inflation: from primordial
  gravitational waves and stiff axion matter to dynamical dark energy}.
\newblock {\em Eur. Phys. J. ST} {\bf 2021}, {\em 230},~2077--2110,
  \href{http://xxx.lanl.gov/abs/2012.07971}{{\normalfont
  [arXiv:hep-ph/2012.07971]}}.
\newblock {\url{https://doi.org/10.1140/epjs/s11734-021-00197-8}}.

\bibitem[Mavromatos and Sol\`a~Peracaula(2021{\natexlab{b}})]{ms2}
Mavromatos, N.E.; Sol\`a~Peracaula, J.
\newblock {Inflationary physics and trans-Planckian conjecture in the stringy
  running vacuum model: from the phantom vacuum to the true vacuum}.
\newblock {\em Eur. Phys. J. Plus} {\bf 2021}, {\em 136},~1152,
  \href{http://xxx.lanl.gov/abs/2105.02659}{{\normalfont
  [arXiv:hep-th/2105.02659]}}.
\newblock {\url{https://doi.org/10.1140/epjp/s13360-021-02149-6}}.

\bibitem[Gross and Sloan(1987)]{gross}
Gross, D.J.; Sloan, J.H.
\newblock {The Quartic Effective Action for the Heterotic String}.
\newblock {\em Nucl. Phys. B} {\bf 1987}, {\em 291},~41--89.
\newblock {\url{https://doi.org/10.1016/0550-3213(87)90465-2}}.

\bibitem[Metsaev and Tseytlin(1987)]{MT}
Metsaev, R.R.; Tseytlin, A.A.
\newblock {Order alpha-prime (Two Loop) Equivalence of the String Equations of
  Motion and the Sigma Model Weyl Invariance Conditions: Dependence on the
  Dilaton and the Antisymmetric Tensor}.
\newblock {\em Nucl. Phys. B} {\bf 1987}, {\em 293},~385--419.
\newblock {\url{https://doi.org/10.1016/0550-3213(87)90077-0}}.

\bibitem[Bento and Mavromatos(1987)]{bento}
Bento, M.C.; Mavromatos, N.E.
\newblock {Ambiguities in the Low-energy Effective Actions of String Theories
  With the Inclusion of Antisymmetric Tensor and Dilaton Fields}.
\newblock {\em Phys. Lett. B} {\bf 1987}, {\em 190},~105--109.
\newblock {\url{https://doi.org/10.1016/0370-2693(87)90847-1}}.

\bibitem[Green and Schwarz(1984)]{gs}
Green, M.B.; Schwarz, J.H.
\newblock {Anomaly Cancellation in Supersymmetric D=10 Gauge Theory and
  Superstring Theory}.
\newblock {\em Phys. Lett. B} {\bf 1984}, {\em 149},~117--122.
\newblock {\url{https://doi.org/10.1016/0370-2693(84)91565-X}}.

\bibitem[Svrcek and Witten(2006)]{svrcek}
Svrcek, P.; Witten, E.
\newblock {Axions In String Theory}.
\newblock {\em JHEP} {\bf 2006}, {\em 06},~051,
  \href{http://xxx.lanl.gov/abs/hep-th/0605206}{{\normalfont
  [hep-th/0605206]}}.
\newblock {\url{https://doi.org/10.1088/1126-6708/2006/06/051}}.

\bibitem[Arvanitaki \em{et~al.}(2010)Arvanitaki, Dimopoulos, Dubovsky, Kaloper,
  and March-Russell]{arvanitaki}
Arvanitaki, A.; Dimopoulos, S.; Dubovsky, S.; Kaloper, N.; March-Russell, J.
\newblock {String Axiverse}.
\newblock {\em Phys. Rev. D} {\bf 2010}, {\em 81},~123530,
  \href{http://xxx.lanl.gov/abs/0905.4720}{{\normalfont
  [arXiv:hep-th/0905.4720]}}.
\newblock {\url{https://doi.org/10.1103/PhysRevD.81.123530}}.

\bibitem[Marsh(2016)]{marsh}
Marsh, D.J.E.
\newblock {Axion Cosmology}.
\newblock {\em Phys. Rept.} {\bf 2016}, {\em 643},~1--79,
  \href{http://xxx.lanl.gov/abs/1510.07633}{{\normalfont
  [arXiv:astro-ph.CO/1510.07633]}}.
\newblock {\url{https://doi.org/10.1016/j.physrep.2016.06.005}}.

\bibitem[Alexander \em{et~al.}(2006)Alexander, Peskin, and
  Sheikh-Jabbari]{stephon}
Alexander, S.H.S.; Peskin, M.E.; Sheikh-Jabbari, M.M.
\newblock {Leptogenesis from gravity waves in models of inflation}.
\newblock {\em Phys. Rev. Lett.} {\bf 2006}, {\em 96},~081301,
  \href{http://xxx.lanl.gov/abs/hep-th/0403069}{{\normalfont
  [hep-th/0403069]}}.
\newblock {\url{https://doi.org/10.1103/PhysRevLett.96.081301}}.

\bibitem[Lyth \em{et~al.}(2005)Lyth, Quimbay, and Rodriguez]{lyth}
Lyth, D.H.; Quimbay, C.; Rodriguez, Y.
\newblock {Leptogenesis and tensor polarisation from a gravitational
  Chern-Simons term}.
\newblock {\em JHEP} {\bf 2005}, {\em 03},~016,
  \href{http://xxx.lanl.gov/abs/hep-th/0501153}{{\normalfont
  [hep-th/0501153]}}.
\newblock {\url{https://doi.org/10.1088/1126-6708/2005/03/016}}.

\bibitem[Mavromatos(2022)]{mavrosourcegw}
Mavromatos, N.E.
\newblock {Lorentz Symmetry Violation in String-Inspired Effective Modified
  Gravity Theories}.
\newblock In Proceedings of the {740. WE-Heraeus-Seminar}: {Experimental Tests
  and Signatures of Modified and Quantum Gravity Workshop},  2022,
  \href{http://xxx.lanl.gov/abs/2205.07044}{{\normalfont
  [arXiv:hep-th/2205.07044]}}.

\bibitem[Lalak \em{et~al.}(1995)Lalak, Lola, Ovrut, and Ross]{ross}
Lalak, Z.; Lola, S.; Ovrut, B.A.; Ross, G.G.
\newblock {Large scale structure from biased nonequilibrium phase transitions:
  Percolation theory picture}.
\newblock {\em Nucl. Phys. B} {\bf 1995}, {\em 434},~675--696,
  \href{http://xxx.lanl.gov/abs/hep-ph/9404218}{{\normalfont
  [hep-ph/9404218]}}.
\newblock {\url{https://doi.org/10.1016/0550-3213(94)00557-U}}.

\bibitem[Shapiro and Sola(2002)]{sola1}
Shapiro, I.L.; Sola, J.
\newblock {Scaling behavior of the cosmological constant: Interface between
  quantum field theory and cosmology}.
\newblock {\em JHEP} {\bf 2002}, {\em 02},~006,
  \href{http://xxx.lanl.gov/abs/hep-th/0012227}{{\normalfont
  [hep-th/0012227]}}.
\newblock {\url{https://doi.org/10.1088/1126-6708/2002/02/006}}.

\bibitem[Shapiro and Sola(2009)]{sola2}
Shapiro, I.L.; Sola, J.
\newblock {On the possible running of the cosmological 'constant'}.
\newblock {\em Phys. Lett. B} {\bf 2009}, {\em 682},~105--113,
  \href{http://xxx.lanl.gov/abs/0910.4925}{{\normalfont
  [arXiv:hep-th/0910.4925]}}.
\newblock {\url{https://doi.org/10.1016/j.physletb.2009.10.073}}.

\bibitem[Shapiro and Sola(2004)]{sola3}
Shapiro, I.L.; Sola, J.
\newblock {Cosmological constant, renormalization group and Planck scale
  physics}.
\newblock {\em Nucl. Phys. B Proc. Suppl.} {\bf 2004}, {\em 127},~71--76,
  \href{http://xxx.lanl.gov/abs/hep-ph/0305279}{{\normalfont
  [hep-ph/0305279]}}.
\newblock {\url{https://doi.org/10.1016/S0920-5632(03)02402-2}}.

\bibitem[Perico \em{et~al.}(2013)Perico, Lima, Basilakos, and Sola]{lima1}
Perico, E.L.D.; Lima, J.A.S.; Basilakos, S.; Sola, J.
\newblock {Complete Cosmic History with a dynamical $\Lambda=\Lambda(H)$ term}.
\newblock {\em Phys. Rev. D} {\bf 2013}, {\em 88},~063531,
  \href{http://xxx.lanl.gov/abs/1306.0591}{{\normalfont
  [arXiv:astro-ph.CO/1306.0591]}}.
\newblock {\url{https://doi.org/10.1103/PhysRevD.88.063531}}.

\bibitem[Lima \em{et~al.}(2013)Lima, Basilakos, and Sola]{lima2}
Lima, J.A.S.; Basilakos, S.; Sola, J.
\newblock {Expansion History with Decaying Vacuum: A Complete Cosmological
  Scenario}.
\newblock {\em Mon. Not. Roy. Astron. Soc.} {\bf 2013}, {\em 431},~923--929,
  \href{http://xxx.lanl.gov/abs/1209.2802}{{\normalfont
  [arXiv:gr-qc/1209.2802]}}.
\newblock {\url{https://doi.org/10.1093/mnras/stt220}}.

\bibitem[Sola~Peracaula(2022)]{sola4}
Sola~Peracaula, J.
\newblock {The cosmological constant problem and running vacuum in the
  expanding universe}.
\newblock {\em Phil. Trans. Roy. Soc. Lond. A} {\bf 2022}, {\em 380},~20210182,
   \href{http://xxx.lanl.gov/abs/2203.13757}{{\normalfont
  [arXiv:gr-qc/2203.13757]}}.
\newblock {\url{https://doi.org/10.1098/rsta.2021.0182}}.

\bibitem[Kanti \em{et~al.}(1996)Kanti, Mavromatos, Rizos, Tamvakis, and
  Winstanley]{kanti}
Kanti, P.; Mavromatos, N.E.; Rizos, J.; Tamvakis, K.; Winstanley, E.
\newblock {Dilatonic black holes in higher curvature string gravity}.
\newblock {\em Phys. Rev. D} {\bf 1996}, {\em 54},~5049--5058,
  \href{http://xxx.lanl.gov/abs/hep-th/9511071}{{\normalfont
  [hep-th/9511071]}}.
\newblock {\url{https://doi.org/10.1103/PhysRevD.54.5049}}.

\bibitem[Moreno-Pulido and Sola(2020)]{pul1}
Moreno-Pulido, C.; Sola, J.
\newblock {Running vacuum in quantum field theory in curved spacetime:
  renormalizing $\rho_{vac}$ without $\sim m^4$ terms}.
\newblock {\em Eur. Phys. J. C} {\bf 2020}, {\em 80},~692,
  \href{http://xxx.lanl.gov/abs/2005.03164}{{\normalfont
  [arXiv:gr-qc/2005.03164]}}.
\newblock {\url{https://doi.org/10.1140/epjc/s10052-020-8238-6}}.

\bibitem[Moreno-Pulido and Sola~Peracaula(2022{\natexlab{a}})]{pul2}
Moreno-Pulido, C.; Sola~Peracaula, J.
\newblock {Renormalizing the vacuum energy in cosmological spacetime:
  implications for the cosmological constant problem}.
\newblock {\em Eur. Phys. J. C} {\bf 2022}, {\em 82},~551,
  \href{http://xxx.lanl.gov/abs/2201.05827}{{\normalfont
  [arXiv:gr-qc/2201.05827]}}.
\newblock {\url{https://doi.org/10.1140/epjc/s10052-022-10484-w}}.

\bibitem[Moreno-Pulido and Sola~Peracaula(2022{\natexlab{b}})]{pul3}
Moreno-Pulido, C.; Sola~Peracaula, J.
\newblock {Equation of state of the running vacuum}.
\newblock {\em Eur. Phys. J. C} {\bf 2022}, {\em 82},~1137,
  \href{http://xxx.lanl.gov/abs/2207.07111}{{\normalfont
  [arXiv:gr-qc/2207.07111]}}.
\newblock {\url{https://doi.org/10.1140/epjc/s10052-022-11117-y}}.

\bibitem[Moreno-Pulido \em{et~al.}(2023)Moreno-Pulido, Sola~Peracaula, and
  Cheraghchi]{pul4}
Moreno-Pulido, C.; Sola~Peracaula, J.; Cheraghchi, S.
\newblock {Running vacuum in QFT in FLRW spacetime: The dynamics of $\rho_{\rm
  vac}(H)$ from the quantized matter fields} {\bf 2023}.
\newblock  \href{http://xxx.lanl.gov/abs/2301.05205}{{\normalfont
  [arXiv:gr-qc/2301.05205]}}.

\bibitem[Bossingham \em{et~al.}(2018)Bossingham, Mavromatos, and
  Sarkar]{lepto1}
Bossingham, T.; Mavromatos, N.E.; Sarkar, S.
\newblock {Leptogenesis from Heavy Right-Handed Neutrinos in CPT Violating
  Backgrounds}.
\newblock {\em Eur. Phys. J. C} {\bf 2018}, {\em 78},~113,
  \href{http://xxx.lanl.gov/abs/1712.03312}{{\normalfont
  [arXiv:hep-ph/1712.03312]}}.
\newblock {\url{https://doi.org/10.1140/epjc/s10052-018-5587-5}}.

\bibitem[Bossingham \em{et~al.}(2019)Bossingham, Mavromatos, and
  Sarkar]{lepto2}
Bossingham, T.; Mavromatos, N.E.; Sarkar, S.
\newblock {The role of temperature dependent string-inspired CPT violating
  backgrounds in leptogenesis and the chiral magnetic effect}.
\newblock {\em Eur. Phys. J. C} {\bf 2019}, {\em 79},~50,
  \href{http://xxx.lanl.gov/abs/1810.13384}{{\normalfont
  [arXiv:hep-ph/1810.13384]}}.
\newblock {\url{https://doi.org/10.1140/epjc/s10052-019-6564-3}}.

\bibitem[Mavromatos and Sarkar(2020)]{sarkarbaryo}
Mavromatos, N.E.; Sarkar, S.
\newblock {Curvature and thermal corrections in tree-level CPT-Violating
  Leptogenesis}.
\newblock {\em Eur. Phys. J. C} {\bf 2020}, {\em 80},~558,
  \href{http://xxx.lanl.gov/abs/2004.10628}{{\normalfont
  [arXiv:hep-ph/2004.10628]}}.
\newblock {\url{https://doi.org/10.1140/epjc/s10052-020-8109-1}}.

\bibitem[Kostelecky and Russell(2011)]{smebounds}
Kostelecky, V.A.; Russell, N.
\newblock {Data Tables for Lorentz and CPT Violation}.
\newblock {\em Rev. Mod. Phys.} {\bf 2011}, {\em 83},~11--31,
  \href{http://xxx.lanl.gov/abs/0801.0287}{{\normalfont
  [arXiv:hep-ph/0801.0287]}}.
\newblock {\url{https://doi.org/10.1103/RevModPhys.83.11}}.

\bibitem[Capanelli \em{et~al.}(2023)Capanelli, Jenks, Kolb, and
  McDonough]{Capanelli:2023uwv}
Capanelli, C.; Jenks, L.; Kolb, E.W.; McDonough, E.
\newblock {Cosmological Implications of Kalb-Ramond-Like-Particles} {\bf 2023}.
\newblock  \href{http://xxx.lanl.gov/abs/2309.02485}{{\normalfont
  [arXiv:hep-ph/2309.02485]}}.

\bibitem[Pop\l{}awski(2010)]{poplinfl}
Pop\l{}awski, N.J.
\newblock {Cosmology with torsion: An alternative to cosmic inflation}.
\newblock {\em Phys. Lett. B} {\bf 2010}, {\em 694},~181--185,
  \href{http://xxx.lanl.gov/abs/1007.0587}{{\normalfont
  [arXiv:astro-ph.CO/1007.0587]}}.
\newblock [Erratum: Phys.Lett.B 701, 672--672 (2011)],
  {\url{https://doi.org/10.1016/j.physletb.2010.09.056}}.

\bibitem[Poplawski(2011)]{poplcosm}
Poplawski, N.J.
\newblock {Cosmological constant from quarks and torsion}.
\newblock {\em Annalen Phys.} {\bf 2011}, {\em 523},~291--295,
  \href{http://xxx.lanl.gov/abs/1005.0893}{{\normalfont
  [arXiv:gr-qc/1005.0893]}}.
\newblock {\url{https://doi.org/10.1002/andp.201000162}}.

\bibitem[Magueijo \em{et~al.}(2013)Magueijo, Zlosnik, and Kibble]{kibble}
Magueijo, J.a.; Zlosnik, T.G.; Kibble, T.W.B.
\newblock {Cosmology with a spin}.
\newblock {\em Phys. Rev. D} {\bf 2013}, {\em 87},~063504,
  \href{http://xxx.lanl.gov/abs/1212.0585}{{\normalfont
  [arXiv:astro-ph.CO/1212.0585]}}.
\newblock {\url{https://doi.org/10.1103/PhysRevD.87.063504}}.

\bibitem[Poplawski(2012)]{poplbounc}
Poplawski, N.J.
\newblock {Nonsingular, big-bounce cosmology from spinor-torsion coupling}.
\newblock {\em Phys. Rev. D} {\bf 2012}, {\em 85},~107502,
  \href{http://xxx.lanl.gov/abs/1111.4595}{{\normalfont
  [arXiv:gr-qc/1111.4595]}}.
\newblock {\url{https://doi.org/10.1103/PhysRevD.85.107502}}.

\bibitem[Giacosa \em{et~al.}(2008)Giacosa, Hofmann, and Neubert]{neubert}
Giacosa, F.; Hofmann, R.; Neubert, M.
\newblock {A model for the very early Universe}.
\newblock {\em JHEP} {\bf 2008}, {\em 02},~077,
  \href{http://xxx.lanl.gov/abs/0801.0197}{{\normalfont
  [arXiv:hep-th/0801.0197]}}.
\newblock {\url{https://doi.org/10.1088/1126-6708/2008/02/077}}.

\bibitem[Kostelecky \em{et~al.}(2008)Kostelecky, Russell, and
  Tasson]{Kostelecky:2007kx}
Kostelecky, V.A.; Russell, N.; Tasson, J.
\newblock {New Constraints on Torsion from Lorentz Violation}.
\newblock {\em Phys. Rev. Lett.} {\bf 2008}, {\em 100},~111102,
  \href{http://xxx.lanl.gov/abs/0712.4393}{{\normalfont
  [arXiv:gr-qc/0712.4393]}}.
\newblock {\url{https://doi.org/10.1103/PhysRevLett.100.111102}}.

\bibitem[Bolejko \em{et~al.}(2020)Bolejko, Cinus, and Roukema]{Bolejko:2020nbw}
Bolejko, K.; Cinus, M.; Roukema, B.F.
\newblock {Cosmological signatures of torsion and how to distinguish torsion
  from the dark sector}.
\newblock {\em Phys. Rev. D} {\bf 2020}, {\em 101},~104046,
  \href{http://xxx.lanl.gov/abs/2003.06528}{{\normalfont
  [arXiv:astro-ph.CO/2003.06528]}}.
\newblock {\url{https://doi.org/10.1103/PhysRevD.101.104046}}.

\bibitem[Aluri \em{et~al.}(2023)Aluri et~al.]{Aluri:2022hzs}
Aluri, P.K.;  et~al.
\newblock {Is the observable Universe consistent with the cosmological
  principle?}
\newblock {\em Class. Quant. Grav.} {\bf 2023}, {\em 40},~094001,
  \href{http://xxx.lanl.gov/abs/2207.05765}{{\normalfont
  [arXiv:astro-ph.CO/2207.05765]}}.
\newblock {\url{https://doi.org/10.1088/1361-6382/acbefc}}.

\bibitem[G\'omez-Valent \em{et~al.}(2023)G\'omez-Valent, Mavromatos, and
  Sol\`a~Peracaula]{Gomez-Valent:2023hov}
G\'omez-Valent, A.; Mavromatos, N.E.; Sol\`a~Peracaula, J.
\newblock {Stringy Running Vacuum Model and current Tensions in Cosmology} {\bf
  2023}.
\newblock  \href{http://xxx.lanl.gov/abs/2305.15774}{{\normalfont
  [arXiv:gr-qc/2305.15774]}}.

\bibitem[Garcia~de Andrade(2011)]{GarciadeAndrade:2011zzc}
Garcia~de Andrade, L.C.
\newblock {Torsion bounds from CP violation alpha(2)-dynamo in axion-photon
  cosmic plasma}.
\newblock {\em Mod. Phys. Lett. A} {\bf 2011}, {\em 26},~2863--2868.
\newblock {\url{https://doi.org/10.1142/S0217732311037182}}.

\bibitem[Campanelli and Giannotti(2005)]{Campanelli:2005ye}
Campanelli, L.; Giannotti, M.
\newblock {Magnetic helicity generation from the cosmic axion field}.
\newblock {\em Phys. Rev. D} {\bf 2005}, {\em 72},~123001,
  \href{http://xxx.lanl.gov/abs/astro-ph/0508653}{{\normalfont
  [astro-ph/0508653]}}.
\newblock {\url{https://doi.org/10.1103/PhysRevD.72.123001}}.

\bibitem[Cai \em{et~al.}(2016)Cai, Capozziello, De~Laurentis, and
  Saridakis]{telep}
Cai, Y.F.; Capozziello, S.; De~Laurentis, M.; Saridakis, E.N.
\newblock {f(T) teleparallel gravity and cosmology}.
\newblock {\em Rept. Prog. Phys.} {\bf 2016}, {\em 79},~106901, and references
  therein.,  \href{http://xxx.lanl.gov/abs/1511.07586}{{\normalfont
  [arXiv:gr-qc/1511.07586]}}.
\newblock {\url{https://doi.org/10.1088/0034-4885/79/10/106901}}.

\bibitem[D'Ambrosio \em{et~al.}(2022)D'Ambrosio, Fell, Heisenberg, and
  Kuhn]{qgrav}
D'Ambrosio, F.; Fell, S.D.B.; Heisenberg, L.; Kuhn, S.
\newblock {Black holes in f(Q) gravity}.
\newblock {\em Phys. Rev. D} {\bf 2022}, {\em 105},~024042,
  \href{http://xxx.lanl.gov/abs/2109.03174}{{\normalfont
  [arXiv:gr-qc/2109.03174]}}.
\newblock {\url{https://doi.org/10.1103/PhysRevD.105.024042}}.

\bibitem[Fern\'andez \em{et~al.}(2023)Fern\'andez, Pujol, Sol\'\i{}s, and
  Vargas]{Fernandez:2023kwc}
Fern\'andez, N.; Pujol, P.; Sol\'\i{}s, M.; Vargas, T.
\newblock {Revisiting the electronic properties of disclinated graphene
  sheets}.
\newblock {\em Eur. Phys. J. B} {\bf 2023}, {\em 96},~68,
  \href{http://xxx.lanl.gov/abs/2302.04804}{{\normalfont
  [arXiv:cond-mat.str-el/2302.04804]}}.
\newblock {\url{https://doi.org/10.1140/epjb/s10051-023-00542-x}}.

\bibitem[Nissinen and Volovik(2019)]{volovik1}
Nissinen, J.; Volovik, G.E.
\newblock {On thermal Nieh\textendash{}Yan anomaly in topological Weyl
  materials}.
\newblock {\em Pisma Zh. Eksp. Teor. Fiz.} {\bf 2019}, {\em 110},~797--798,
  \href{http://xxx.lanl.gov/abs/1911.03382}{{\normalfont
  [arXiv:cond-mat.str-el/1911.03382]}}.
\newblock {\url{https://doi.org/10.1134/S0021364019240020}}.

\bibitem[Nissinen and Volovik(2020)]{volovik2}
Nissinen, J.; Volovik, G.E.
\newblock {Thermal Nieh-Yan anomaly in Weyl superfluids}.
\newblock {\em Phys. Rev. Res.} {\bf 2020}, {\em 2},~033269,
  \href{http://xxx.lanl.gov/abs/1909.08936}{{\normalfont
  [arXiv:cond-mat.str-el/1909.08936]}}.
\newblock {\url{https://doi.org/10.1103/PhysRevResearch.2.033269}}.

\bibitem[Bombacigno \em{et~al.}(2016)Bombacigno, Cianfrani, and
  Montani]{Bombacigno}
Bombacigno, F.; Cianfrani, F.; Montani, G.
\newblock {Big-Bounce cosmology in the presence of Immirzi field}.
\newblock {\em Phys. Rev. D} {\bf 2016}, {\em 94},~064021,
  \href{http://xxx.lanl.gov/abs/1607.00910}{{\normalfont
  [arXiv:gr-qc/1607.00910]}}.
\newblock {\url{https://doi.org/10.1103/PhysRevD.94.064021}}.

\bibitem[Bombacigno and Montani(2019)]{Bombacigno1}
Bombacigno, F.; Montani, G.
\newblock {Big bounce cosmology for Palatini $R^2$ gravity with a
  Nieh\textendash{}Yan term}.
\newblock {\em Eur. Phys. J. C} {\bf 2019}, {\em 79},~405,
  \href{http://xxx.lanl.gov/abs/1809.07563}{{\normalfont
  [arXiv:gr-qc/1809.07563]}}.
\newblock {\url{https://doi.org/10.1140/epjc/s10052-019-6918-x}}.

\bibitem[Bombacigno \em{et~al.}(2021)Bombacigno, Boudet, and
  Montani]{Bombacigno2}
Bombacigno, F.; Boudet, S.; Montani, G.
\newblock {Generalized Ashtekar variables for Palatini $f(\mathcal {R})$
  models}.
\newblock {\em Nucl. Phys. B} {\bf 2021}, {\em 963},~115281,
  \href{http://xxx.lanl.gov/abs/1911.09066}{{\normalfont
  [arXiv:gr-qc/1911.09066]}}.
\newblock {\url{https://doi.org/10.1016/j.nuclphysb.2020.115281}}.

\bibitem[Boudet \em{et~al.}(2021)Boudet, Bombacigno, Montani, and
  Rinaldi]{Bombacigno3}
Boudet, S.; Bombacigno, F.; Montani, G.; Rinaldi, M.
\newblock {Superentropic black hole with Immirzi hair}.
\newblock {\em Phys. Rev. D} {\bf 2021}, {\em 103},~084034,
  \href{http://xxx.lanl.gov/abs/2012.02700}{{\normalfont
  [arXiv:gr-qc/2012.02700]}}.
\newblock {\url{https://doi.org/10.1103/PhysRevD.103.084034}}.

\bibitem[Bombacigno \em{et~al.}(2021)Bombacigno, Boudet, Olmo, and
  Montani]{Bombacigno4}
Bombacigno, F.; Boudet, S.; Olmo, G.J.; Montani, G.
\newblock {Big bounce and future time singularity resolution in Bianchi I
  cosmologies: The projective invariant Nieh-Yan case}.
\newblock {\em Phys. Rev. D} {\bf 2021}, {\em 103},~124031,
  \href{http://xxx.lanl.gov/abs/2105.06870}{{\normalfont
  [arXiv:gr-qc/2105.06870]}}.
\newblock {\url{https://doi.org/10.1103/PhysRevD.103.124031}}.

\bibitem[Elizalde \em{et~al.}(2019)Elizalde, Odintsov, Paul, and
  S\'aez-Chill\'on~G\'omez]{tan1}
Elizalde, E.; Odintsov, S.D.; Paul, T.; S\'aez-Chill\'on~G\'omez, D.
\newblock {Inflationary universe in $F(R)$ gravity with antisymmetric tensor
  fields and their suppression during its evolution}.
\newblock {\em Phys. Rev. D} {\bf 2019}, {\em 99},~063506,
  \href{http://xxx.lanl.gov/abs/1811.02960}{{\normalfont
  [arXiv:gr-qc/1811.02960]}}.
\newblock {\url{https://doi.org/10.1103/PhysRevD.99.063506}}.

\bibitem[Paul and Banerjee(2020)]{tan2}
Paul, T.; Banerjee, N.
\newblock {Cosmological quantum entanglement: a possible testbed for the
  existence of Kalb\textendash{}Ramond field}.
\newblock {\em Class. Quant. Grav.} {\bf 2020}, {\em 37},~135013,
  \href{http://xxx.lanl.gov/abs/2004.10111}{{\normalfont
  [arXiv:gr-qc/2004.10111]}}.
\newblock {\url{https://doi.org/10.1088/1361-6382/ab8bb9}}.

\bibitem[Paul(2020)]{tan3}
Paul, T.
\newblock {Antisymmetric tensor fields in modified gravity: a summary}.
\newblock {\em Symmetry} {\bf 2020}, {\em 12},~1573,
  \href{http://xxx.lanl.gov/abs/2009.07732}{{\normalfont
  [arXiv:gr-qc/2009.07732]}}.
\newblock {\url{https://doi.org/10.3390/sym12091573}}.

\bibitem[Paul and SenGupta(2019)]{tan4}
Paul, T.; SenGupta, S.
\newblock {Dynamical suppression of spacetime torsion}.
\newblock {\em Eur. Phys. J. C} {\bf 2019}, {\em 79},~591,
  \href{http://xxx.lanl.gov/abs/1808.00172}{{\normalfont
  [arXiv:gr-qc/1808.00172]}}.
\newblock {\url{https://doi.org/10.1140/epjc/s10052-019-7109-5}}.

\bibitem[Das \em{et~al.}(2018)Das, Paul, and Sengupta]{tan5}
Das, A.; Paul, T.; Sengupta, S.
\newblock {Invisibility of antisymmetric tensor fields in the light of $F(R)$
  gravity}.
\newblock {\em Phys. Rev. D} {\bf 2018}, {\em 98},~104002,
  \href{http://xxx.lanl.gov/abs/1804.06602}{{\normalfont
  [arXiv:hep-th/1804.06602]}}.
\newblock {\url{https://doi.org/10.1103/PhysRevD.98.104002}}.

\bibitem[Nascimento \em{et~al.}(2022)Nascimento, Petrov, and
  Porf\'\i{}rio]{lvt}
Nascimento, J.R.; Petrov, A.Y.; Porf\'\i{}rio, P.J.
\newblock {Induced gravitational topological term and the Einstein-Cartan
  modified theory}.
\newblock {\em Phys. Rev. D} {\bf 2022}, {\em 105},~044053,
  \href{http://xxx.lanl.gov/abs/2108.05705}{{\normalfont
  [arXiv:gr-qc/2108.05705]}}.
\newblock {\url{https://doi.org/10.1103/PhysRevD.105.044053}}.

\bibitem[Battista and De~Falco(2021)]{falco1}
Battista, E.; De~Falco, V.
\newblock {First post-Newtonian generation of gravitational waves in
  Einstein-Cartan theory}.
\newblock {\em Phys. Rev. D} {\bf 2021}, {\em 104},~084067,
  \href{http://xxx.lanl.gov/abs/2109.01384}{{\normalfont
  [arXiv:gr-qc/2109.01384]}}.
\newblock {\url{https://doi.org/10.1103/PhysRevD.104.084067}}.

\bibitem[Battista and De~Falco(2022{\natexlab{a}})]{falco2}
Battista, E.; De~Falco, V.
\newblock {Gravitational waves at the first post-Newtonian order with the
  Weyssenhoff fluid in Einstein\textendash{}Cartan theory}.
\newblock {\em Eur. Phys. J. C} {\bf 2022}, {\em 82},~628,
  \href{http://xxx.lanl.gov/abs/2206.12907}{{\normalfont
  [arXiv:gr-qc/2206.12907]}}.
\newblock {\url{https://doi.org/10.1140/epjc/s10052-022-10558-9}}.

\bibitem[Battista and De~Falco(2022{\natexlab{b}})]{falco3}
Battista, E.; De~Falco, V.
\newblock {First post-Newtonian N-body problem in Einstein\textendash{}Cartan
  theory with the Weyssenhoff fluid: equations of motion}.
\newblock {\em Eur. Phys. J. C} {\bf 2022}, {\em 82},~782,
  \href{http://xxx.lanl.gov/abs/2208.09839}{{\normalfont
  [arXiv:gr-qc/2208.09839]}}.
\newblock {\url{https://doi.org/10.1140/epjc/s10052-022-10746-7}}.

\bibitem[De~Falco \em{et~al.}(2023)De~Falco, Battista, and Antoniadis]{falco4}
De~Falco, V.; Battista, E.; Antoniadis, J.
\newblock {Analytical coordinate time at first post-Newtonian order}.
\newblock {\em EPL} {\bf 2023}, {\em 141},~29002,
  \href{http://xxx.lanl.gov/abs/2301.02472}{{\normalfont
  [arXiv:gr-qc/2301.02472]}}.
\newblock {\url{https://doi.org/10.1209/0295-5075/acb07e}}.

\bibitem[Battista \em{et~al.}(2023)Battista, De~Falco, and Usseglio]{falco5}
Battista, E.; De~Falco, V.; Usseglio, D.
\newblock {First post-Newtonian N-body problem in Einstein\textendash{}Cartan
  theory with the Weyssenhoff fluid: Lagrangian and first integrals}.
\newblock {\em Eur. Phys. J. C} {\bf 2023}, {\em 83},~112,
  \href{http://xxx.lanl.gov/abs/2301.08954}{{\normalfont
  [arXiv:gr-qc/2301.08954]}}.
\newblock {\url{https://doi.org/10.1140/epjc/s10052-023-11249-9}}.

\bibitem[De~Falco and Battista(2023)]{falco6}
De~Falco, V.; Battista, E.
\newblock {Analytical results for binary dynamics at the first post-Newtonian
  order in Einstein-Cartan theory with the Weyssenhoff fluid}.
\newblock {\em Phys. Rev. D} {\bf 2023}, {\em 108},~064032,
  \href{http://xxx.lanl.gov/abs/2309.00319}{{\normalfont
  [arXiv:gr-qc/2309.00319]}}.
\newblock {\url{https://doi.org/10.1103/PhysRevD.108.064032}}.

\bibitem[Mondal and Chakraborty(2023)]{chak1}
Mondal, V.; Chakraborty, S.
\newblock {Lorentzian quantum cosmology with torsion} {\bf 2023}.
\newblock  \href{http://xxx.lanl.gov/abs/2305.01690}{{\normalfont
  [arXiv:gr-qc/2305.01690]}}.

\bibitem[Chakraborty and Dey(2018)]{chak2}
Chakraborty, S.; Dey, R.
\newblock {Noether Current, Black Hole Entropy and Spacetime Torsion}.
\newblock {\em Phys. Lett. B} {\bf 2018}, {\em 786},~432--441,
  \href{http://xxx.lanl.gov/abs/1806.05840}{{\normalfont
  [arXiv:gr-qc/1806.05840]}}.
\newblock {\url{https://doi.org/10.1016/j.physletb.2018.10.027}}.

\bibitem[Banerjee \em{et~al.}(2018)Banerjee, Chakraborty, and Mukherjee]{chak3}
Banerjee, R.; Chakraborty, S.; Mukherjee, P.
\newblock {Late-time acceleration driven by shift-symmetric Galileon in the
  presence of torsion}.
\newblock {\em Phys. Rev. D} {\bf 2018}, {\em 98},~083506,
  \href{http://xxx.lanl.gov/abs/1802.04150}{{\normalfont
  [arXiv:gr-qc/1802.04150]}}.
\newblock {\url{https://doi.org/10.1103/PhysRevD.98.083506}}.

\bibitem[Sharma and Sur(2021)]{Sharma:2021fou}
Sharma, M.K.; Sur, S.
\newblock {Growth of matter perturbations in an interacting dark energy
  scenario emerging from metric-scalar-torsion couplings} {\bf 2021}.
\newblock  \href{http://xxx.lanl.gov/abs/2102.01525}{{\normalfont
  [arXiv:gr-qc/2102.01525]}}.

\bibitem[Boos and Hehl(2017)]{length1}
Boos, J.; Hehl, F.W.
\newblock {Gravity-induced four-fermion contact interaction implies
  gravitational intermediate W and Z type gauge bosons}.
\newblock {\em Int. J. Theor. Phys.} {\bf 2017}, {\em 56},~751--756,
  \href{http://xxx.lanl.gov/abs/1606.09273}{{\normalfont
  [arXiv:gr-qc/1606.09273]}}.
\newblock {\url{https://doi.org/10.1007/s10773-016-3216-3}}.

\bibitem[Gialamas and Veerm\"ae(2023)]{Gialamas:2023emn}
Gialamas, I.D.; Veerm\"ae, H.
\newblock {Electroweak vacuum decay in metric-affine gravity}.
\newblock {\em Phys. Lett. B} {\bf 2023}, {\em 844},~138109,
  \href{http://xxx.lanl.gov/abs/2305.07693}{{\normalfont
  [arXiv:hep-th/2305.07693]}}.
\newblock {\url{https://doi.org/10.1016/j.physletb.2023.138109}}.

\bibitem[Pal \em{et~al.}(2023)Pal, Pal, and Sarkar]{Pal:2022szb}
Pal, K.; Pal, K.; Sarkar, T.
\newblock {Conformal Fisher information metric with torsion}.
\newblock {\em J. Phys. A} {\bf 2023}, {\em 56},~335001,
  \href{http://xxx.lanl.gov/abs/2210.04759}{{\normalfont
  [arXiv:physics.class-ph/2210.04759]}}.
\newblock {\url{https://doi.org/10.1088/1751-8121/ace74b}}.

\bibitem[Gallegos \em{et~al.}(2021)Gallegos, G\"ursoy, and
  Yarom]{Gallegos:2021bzp}
Gallegos, A.D.; G\"ursoy, U.; Yarom, A.
\newblock {Hydrodynamics of spin currents}.
\newblock {\em SciPost Phys.} {\bf 2021}, {\em 11},~041,
  \href{http://xxx.lanl.gov/abs/2101.04759}{{\normalfont
  [arXiv:hep-th/2101.04759]}}.
\newblock {\url{https://doi.org/10.21468/SciPostPhys.11.2.041}}.

\bibitem[Adamczyk \em{et~al.}(2017)Adamczyk et~al.]{STAR:2017ckg}
Adamczyk, L.;  et~al.
\newblock {Global $\Lambda$ hyperon polarization in nuclear collisions:
  evidence for the most vortical fluid}.
\newblock {\em Nature} {\bf 2017}, {\em 548},~62--65,
  \href{http://xxx.lanl.gov/abs/1701.06657}{{\normalfont
  [arXiv:nucl-ex/1701.06657]}}.
\newblock {\url{https://doi.org/10.1038/nature23004}}.

\bibitem[Adam \em{et~al.}(2018)Adam et~al.]{STAR:2018gyt}
Adam, J.;  et~al.
\newblock {Global polarization of $\Lambda$ hyperons in Au+Au collisions at
  $\sqrt{s_{_{NN}}}$ = 200 GeV}.
\newblock {\em Phys. Rev. C} {\bf 2018}, {\em 98},~014910,
  \href{http://xxx.lanl.gov/abs/1805.04400}{{\normalfont
  [arXiv:nucl-ex/1805.04400]}}.
\newblock {\url{https://doi.org/10.1103/PhysRevC.98.014910}}.

\bibitem[Takahashi(2015)]{takahashi}
Takahashi, R.
\newblock {Spin hydrodynamic generation}.
\newblock {\em Nature Physics} {\bf 2015}, {\em 12},~52.

\end{thebibliography}

%=====================================
% References, variant B: internal bibliography
%=====================================

% If authors have biography, please use the format below
%\section*{Short Biography of Authors}
%\bio
%{\raisebox{-0.35cm}{\includegraphics[width=3.5cm,height=5.3cm,clip,keepaspectratio]{Definitions/author1.pdf}}}
%
%{\textbf{Firstname Lastname} Biography of first author}
%
%\bio

%{\raisebox{-0.35cm}{\includegraphics[width=3.5cm,height=5.3cm,clip,keepaspectratio]{Definitions/author2.jpg}}}

%{\textbf{Firstname Lastname} Biography of second author}

% For the MDPI journals use author-date citation, please follow the formatting guidelines on http://www.mdpi.com/authors/references
% To cite two works by the same author: \citeauthor{ref-journal-1a} (\citeyear{ref-journal-1a}, \citeyear{ref-journal-1b}). This produces: Whittaker (1967, 1975)
% To cite two works by the same author with specific pages: \citeauthor{ref-journal-3a} (\citeyear{ref-journal-3a}, p. 328; \citeyear{ref-journal-3b}, p.475). This produces: Wong (1999, p. 328; 2000, p. 475)

%%%%%%%%%%%%%%%%%%%%%%%%%%%%%%%%%%%%%%%%%%
%% for journal Sci
%\reviewreports{\\
%Reviewer 1 comments and authors’ response\\
%Reviewer 2 comments and authors’ response\\
%Reviewer 3 comments and authors’ response
%}
%%%%%%%%%%%%%%%%%%%%%%%%%%%%%%%%%%%%%%%%%%
\end{adjustwidth}
\end{document}